\pdfoutput=1
\documentclass[3p,times,twocolumn]{elsarticle}

\usepackage{ecrc}

\volume{-- preprint}

\firstpage{1}

\journalname{Nuclear Physics B Proceedings Supplement}

\runauth{L.\ von Smekal}

\jid{nuphbp}

\jnltitlelogo{Nuclear Physics B Proceedings Supplement}

\usepackage{amsmath}
\usepackage{amsfonts}
\usepackage{bbm}
\usepackage{dsfont}
\usepackage{xspace}
\usepackage{slashed}
\usepackage{widetext}

\usepackage{amssymb}

\biboptions{sort&compress}

\usepackage[figuresright]{rotating}

\begin{document}

\newcommand{\qctd}{QC\ensuremath{_2}D\xspace}
\newcommand{\Tr}{{\text{Tr}}}
\newcommand{\tr}{{\text{tr}}}
\newcommand{\imag}{i}
\newcommand{\smallfrac}[2]{\mbox{\small ${\displaystyle \frac{#1}{#2}}$}}
\renewcommand{\log}{\ln}

\newcommand{\Ur}{\ensuremath{U_{k,\rho}}}
\newcommand{\Urr}{\ensuremath{U_{k,\rho\rho}}}
\newcommand{\Ud}{\ensuremath{U_{k,d}}}
\newcommand{\Up}{\ensuremath{U_{k,\phi}}}
\newcommand{\Upp}{\ensuremath{U_{k,\phi\phi}}}
\newcommand{\Udd}{\ensuremath{U_{k,dd}}}
\newcommand{\Urd}{\ensuremath{U_{k,\rho d}}} \def\eq#1{eq.~(\ref{#1})}

\def\Eq#1{Eq.~(\ref{#1})}
\def\Fig#1{Fig.~\ref{#1}} 
\def\Tab#1{Tab.~\ref{#1}}
\def\Eqs#1{Eqs.~(\ref{#1})}

\def\Quadrat#1#2{{\vcenter{\hrule height #2
  \hbox{\vrule width #2 height #1 \kern#1
    \vrule width #2}
  \hrule height #2}}}
\def\dAlemb{\mathop{\kern 1pt\hbox{$\Quadrat{5pt}{0.4pt}$} \kern1pt}}
\def\dAlember{\mathop{\kern 1pt\raise-2pt\hbox{$\Quadrat{3pt}{0.4pt}$} \kern1pt}}



\begin{frontmatter}



\dochead{}

\title{Universal Aspects of QCD-like Theories\tnoteref{SWS}}
\tnotetext[SWS]{For ``Physics at all scales: The Renormalization
  Group,''\\ The $49^\mathrm{th}$ Schladming Winter School on Theoretical
  Physics.} 


\author{Lorenz von Smekal}

\address{Technische Universit\"at Darmstadt, Institut f\"ur Kernphysik, Theoriezentrum, Schlossgartenstra\ss e 2, 64289 Darmstadt, Germany}

\begin{abstract}
In these lectures I review some basic examples of how the concepts of
universality and scaling can be used to study aspects of the chiral
and the deconfinement transition, if not in QCD directly but in
QCD-like theories. As an example for flavor dynamics I discuss a
quark-hadron model to describe the phase diagram of two-color
QCD with the functional renormalization group. Universal
aspects of deconfinement are illustrated mainly in the $2+1$
dimensional SU$(N)$ gauge theories with second order transition where
many exact results from spin models can be exploited.    
\end{abstract}


\end{frontmatter}


\section{Introduction}
\label{intro}

Strongly interacting matter fuels the stars and makes up almost the
entire mass of the luminous universe. The underlying theory of quarks
and gluons, Quantum Chromodynamics (QCD), completely specifies the
interactions. However, these are so complex and non-linear that they
have yet to be fully understood. Indeed, it is these strong
interactions which under normal conditions confine quarks and gluons
into the interior of hadrons. Understanding the generation of their
masses, the confinement of quarks and gluons, the different phases of
QCD at extreme temperatures or densities and the transitions between
them are some of the great challenges in physics. At temperatures in
familiar units close to $2 \times 10^{12}$ K $\sim 170$ MeV, for
example, confined hadronic matter undergoes a transition into an
unconfined state which is a nearly perfect fluid, the quark-gluon
plasma. Degenerate neutron matter presumably exists in neutron stars
at similarly extreme pressures of above $1.6 \times 10^{33}$~Pa $\sim
10$~MeV$/$fm$^3$. 
Compared to the conditions around us, both are of course beyond any
imagination. The conditions in the finite temperature deconfinement
transition are almost 10 orders of magnitude hotter than the surface
of the Earth and still more than 5 orders of magnitude hotter than the
center of the Sun. They bring us back in the evolution of the early
universe within 10 $\mu$s after the Big Bang, and they are being
recreated in the heavy-ion collision experiments at
RHIC and LHC \cite{BraunMunzinger:2009zz}. The pressure inside neutron
stars for comparison corresponds to the weight of about 100 solar masses
pressing on one square meter. Such conditions at comparatively
moderate temperatures require heavy-ion collisions at much lower beam
energies, corresponding to center-of-mass energies per colliding
nucleon pair in the GeV range instead of TeV at the LHC. 
This energy range was pioneered at the AGS and it will be the goal of
the HADES and CBM experiments at FAIR where penetrating probes such as
lepton pairs will be used to study the ultradense conditions in the early
stages of the collisions, in particular \cite{Friman:2011zz,Senger:2012}.

\begin{figure}[t]
\leftline{\hskip -.4cm
\includegraphics[width=\linewidth]{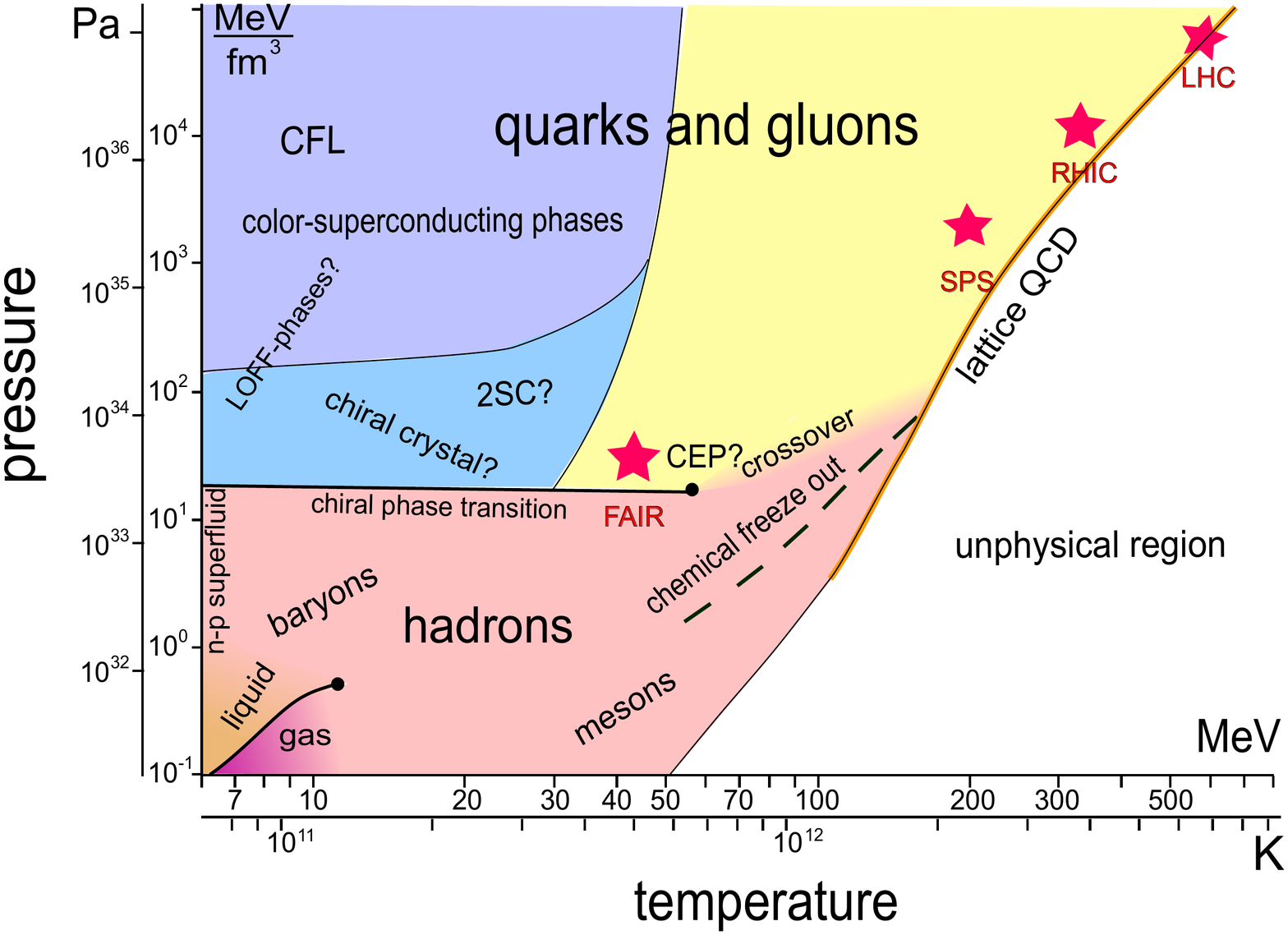}}
\vspace{-.2cm}
\caption{Sketch of the QCD phase diagram from \cite{Wambach:2011bj}. 
\vspace{-.2cm} }
\label{fig:pdsketch}
\end{figure}

Nevertheless, the same characteristic features are being discussed in
the QCD phase diagram as those, {\it e.g.}, in the phase diagram of
water under more normal conditions of pressure and temperature.
Conventionally one plots the QCD phase diagram in the plane of
temperature over baryon chemical potential. An unconventionally
conventional pressure versus temperature phase diagram for strongly
interacting matter is shown in the sketch of Fig.~\ref{fig:pdsketch}.    
A number of color superconducting and other ordered condensed matter phases
are expected to exist in the high pressure region perhaps comparable
to the sixteen or so known crystalline phases of water. There is the
liquid-gas transition to nuclear matter with a critical endpoint at a
temperature of around 15 MeV, and just as in the phase diagram of
water there may be a second critical endpoint at a higher pressure
where the conjectured first-order line for chiral symmetry restoration
ends. But also the possibility of an at least approximate triple point
in the QCD phase diagram has been discussed
\cite{McLerran:2007qj}, which is a point of three-phase coexistence as
in water at the low pressure end of the freezing temperature around 610 Pa.

In order to understand these main characteristic
features in the phase digram of strongly interacting matter,
it has proven to be very useful to deform QCD by not only varying the
individual quarks' masses but also the numbers of their different
flavors and colors. Such deformations are particularly useful when
they lead to second-order phase transitions. Then we can apply the
powerful concepts of universality, scaling and finite-size scaling
which provide by now standard tools that are straightforward applications
of the renormalization group from statistical physics. The classic
example is to consider chiral symmetry restoration with only two flavors of
(nearly) massless quarks. In the chiral limit, the low
temperature phase is characterized by chiral symmetry breaking with
the quark condensate $\langle \bar q q \rangle_T $ as the order
parameter analogous to the spontaneous magnetization in a ferromagnet.
Just as the magnetization, the condensate melts with temperature due
to thermal fluctuations. For low temperatures this is very well
described by chiral effective field theory \cite{Gerber:1988tt} based
on a {\em non-linear} realization of chiral symmetry in terms of the
(would-be) Goldstone bosons which is determined entirely by the
geometry of the coset $G/H$ or vacuum manifold of the symmetry
breaking $G\to H$. With massless quarks  at two-loop, for example,
\begin{equation}
 \langle \bar qq\rangle_T =  \langle \bar qq\rangle_0\,  \Big[ 1 -
 \frac{T^2}{8f_\pi^2} - \frac{1}{6} \Bigg(\frac{T^2}{8f_\pi^2}\Bigg)^2 +
 \mathcal O (T^6)\,\Big] \; . 
\end{equation}

As temperature increases, however, such asymptotic
expansions necessarily break down eventually. In contrast, the
restoration of chiral symmetry at the analogue of the Curie
temperature in the ferromagnet is very well described by a {\em linear}
sigma model. The global chiral symmetry is $G =$ SU$(2)_R \times $
SU$(2)_L$, corresponding to independent flavor rotations of the right and
left-handed components of the two massless quarks. This symmetry is
locally the same as SO$(4)$, the group of rotations in 4 Euclidean
dimensions. It is broken spontaneously in the ordered phase down to
the diagonal, vector-like isospin symmetry $H =$ SU$(2)_V \simeq $
SO$(3)$. Therefore, scaling and universality predict that the
singularities of the thermodynamic observables at this critical point,
the endpoint of the first-order line in the plane of external field
and temperature, are all described by the critical exponents in the
class of the O$(4)$ Heisenberg ferromagnet in three dimensions
\cite{Rajagopal:1992qz}. In particular, with the analogue of the
external magnetic field being the explicitly symmetry breaking
quark mass $m_q $, and $t= T/T_c - 1$ the reduced temperature, 
this implies that 
\begin{equation}
\begin{split}
 \langle \bar q q \rangle_T \, &\sim  (-t)^\beta , \;\; \mbox{for } t\to 0^-
 \;\;  \mbox{at } m_q=0\; , \;\;\mbox{and} \\
 \langle \bar q q \rangle_{T_c} &\sim  \, m_q^{1/\delta}   , \;\;
 \mbox{for } m_q\to 0 \; , \;\;\mbox{where} \\
 & \beta = 0.38(1)\, ,  \;\; \delta = 4.82(1) \;.
\end{split}
\end{equation}  
Scaling and hyperscaling relations which hold 
below the upper critical dimension $d=4$ then determine all other
critical exponents, which is referred to as two-exponent scaling. More
generally, the behavior of the thermodynamic observables in the
vicinity of the critical point is governed by universal scaling
functions. For example, the magnetic equation of state here is of the
form 
\begin{equation}  
   \langle \bar q q\rangle_T \propto m_q^{1/\delta} E\big(t \,
   m_q^{-1/\delta\beta}\big) \; ,   
\end{equation}
with $E(0) =1$ and $E(y) \sim (-y)^\beta$ for $y\to -\infty$
\cite{Pelissetto:2000ek}. 

The study of this O$(4)$ universality and scaling in the quark-meson
model with the functional renormalization group (FRG) by now also has
a long history \cite{Berges:1997eu,Schaefer:1999em}. There is recent
evidence, however, that the true scaling window might actually be very
small, and that it might not include the region of physical pion masses in
which only an apparent scaling may have been observed \cite{Braun:2010vd}. 
Finite volume effects were also investigated in the two-flavor
quark-meson model and found to be under very good control with 
finite-size scaling \cite{Braun:2010vd,Klein:2011zk}.   

\begin{figure}[t]
\leftline{\hskip .4cm
\includegraphics[width=.8\linewidth]{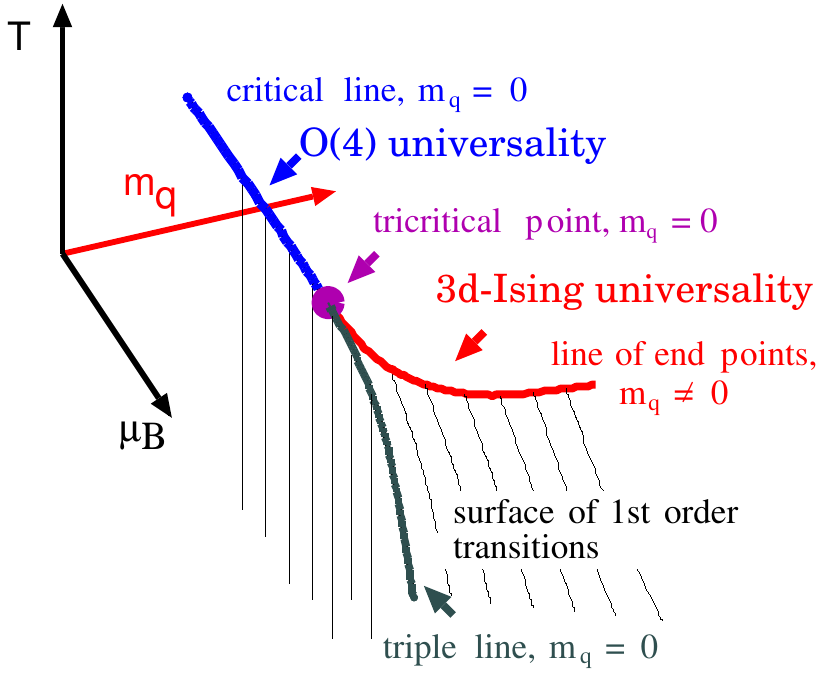}}
\vspace{-.2cm}
\caption{Sketch of a three-dimensional phase diagram with two light quarks
  in the vicinity of a tricritical point at finite baryon chemical
  potential $\mu_B = 3\mu$, by courtesy of the Director of this Winter
  School. 
  \vspace{-.2cm} }
\label{fig:wings}
\end{figure}

When a third, strange quark is included, the situation changes
depending on its mass. With three massless or nearly massless quarks,
the transition must be of first order because of the axial U$_A(1)$
anomaly \cite{Pisarski:1983ms}. Near its physical value, there is
nowadays good evidence from lattice simulations for O$(N)$ scaling
\cite{Ejiri:2009ac,Bazavov:2012ty}, or perhaps at least apparent
scaling. If the critical endpoint exists in the QCD phase diagram,
{\it i.e.}, for quark masses at the physical point, then it seems
certainly no less likely to exist also when the two light quark masses
are sent to zero, in which limit it will inevitably turn into a
tricritical point separating the second-order O$(4)$ line from a
triple line at $m_q=0$, see Fig.~\ref{fig:wings}. With
increasing $\mp m_q$, wing lines of critical endpoints would emanate
from such a tricritical point where three coexisting phases become
indistinguishable. This is a very common view but it is seriously being
challenged from studies of the corresponding critical surface in the
three-dimensional phase diagram depending on $m_q$, $m_s$ and
$(\mu/T)^2$ through lattice simulations at imaginary chemical 
potential, with $-(3/\pi)^2 \le (\mu/T)^2 \le 0 $ \cite{Bonati:2012pe}.

Other important ways to change QCD are to vary the number of colors $N_c$,
to change the quarks' color representation or to replace the gauge group
altogether.

For example, the limit of infinitely many colors $N_c$ has inspired many
qualitative descriptions of the QCD phase diagram 
\cite{McLerran:2007qj,Hidaka:2008yy,Andronic:2009gj}.  One interesting
aspect of this limit is that the baryon density becomes an order
parameter for $N_c\to\infty$, in particular, also when the number of
flavors $N_f$ grows along with $N_c$, {\it i.e.} with the ratio $N_f/N_c$ held
fixed. In the next section we will discuss two-color QCD as another
example, which perhaps surprisingly shares some aspects with the
large-$N_c$ limit, however with $N_c=N_f=2$. Therefore, if the phases
of many-color QCD with  $N_c\sim N_f \to\infty$ might be reflected in
some approximate sense in the real world, it  
seems quite worthwhile to also consider $N_c=N_f=2$.  

When changing the quarks' color representation or the gauge group as a
whole one often exploits the fact that a special role in various ways
is being played by the center of the group, that is the set of all
elements which commute with the whole group.   
The center of SU$(N)$ is Z$_N = \{ \exp(2\pi i \,n/N) :\;
n=0,\,1,\,\dots N-1 \}$, the set of the $N$th roots of unity. The way
the center is represented in any given irreducible representation
defines its $N$-ality $k = 0,\,1,\,\dots N-1 $. If $k$ is coprime with
$N$, then the center is represented faithfully. This is the case in
the fundamental representation and all representations with $k=1$. The
other extreme are the real representations with $N$-ality $k = 0$ such as the
adjoint representation in which the center is represented trivially,
{\it i.e.} in a fully degenerate way by the unit element. Another
special case worth mentioning are representations with $k=N/2$ for
even $N$ in which the center is represented by the real Z$_2 =
\{1,-1\}$. These representations are called pseudo-real because they
are isomorphic to their complex conjugates. One special interest in
quarks in real or pseudo-real color representations derives from the
fact that the fermion determinant remains positive or real,
respectively, even at finite chemical potential. This means that such
theories can be simulated at finite baryon density without fermion-sign
problem on the lattice (using pairs of degenerate flavors, if
necessary in the worst case).

For example, the center of the SU$(2)$ gauge group of two-color QCD is
Z$_2 = \{1,-1\}$. Its integer representations, {\it i.e.} the proper
rotations of SO$(3)$, have $N$-ality $k = 0$ and are real while the
half-odd integer representations with $k=1$ such as the fundamental
representation are pseudo-real, and this exhausts all possibilities in
this case. We will have much more to say about the consequences in the next
section. Here it suffices to make one general observation, and that is
that color singlets in product representations can occur only for even
powers of pseudo-real representations. Therefore, theories with quarks
in pseudo-real color representations can contain only bosonic but no
fermionic baryons. Theories with fermionic baryons require singlets in
odd powers of the quarks' representations which singles out the real
ones if the sign problem is to be avoided. Adjoint fermions in SU$(N)$
gauge theories are the most obvious candidates. Other important
examples include the compact exceptional Lie
Groups $G_2$, $F_4$ and $E_8$ which all have a trivial center. The
smallest one of these is $G_2$. It has rank 2 and dimension 14, it is
simply connected and it can conveniently accommodate SU$(3)$ as a
subgroup. All its representations are real, so there is no
fermion-sign problem even for a single flavor and lattice simulations
at finite baryon density are possible \cite{Maas:2012wr}.  Moreover,
if the $G_2$ gauge symmetry can be broken down to SU$(3)$
via the Higgs mechanism in a controlled way, one can study how the
sign problem gradually reemerges. 

We should eventually be able to
completely understand the phase diagrams of QCD's closest relatives
without the fermion-sign problem, and this should be a worthwhile
exercise also for QCD itself. QCD-like theories can be used to study
universal aspects but also as benchmarks for functional methods which
work for QCD as well but which rely on truncations or model
assumptions. While two-color QCD with its bosonic baryons has been
studied extensively in the past, as will be reviewed in the next
section, studies of the phase diagram of $G_2$ gauge theory are still in
their infancy. They will be an important next step in understanding a
sign-problem-free variant of QCD with fermionic baryons.   

The other special role that is being played by the center of the gauge
group is via the global center symmetry and its spontaneous breaking
in the deconfinement transition of the pure gauge theory. Again, if we
change the number of colors or the dimension of space from three to
two, this transition becomes second order and we can gain a very
precise understanding of it from universality, scaling and finite-size
scaling. This will be discussed in Section 3.


\section{Quark-meson-diquark model of two-color QCD}
\label{QMDmodel}

Quantum Chromodynamics with two colors (\qctd) has been well studied
for many years within chiral effective field theory and random matrix theory
\cite{Kogut:1999iv, Kogut:2000ek, Splittorff:2000mm,
  Splittorff:2001fy, Dunne:2002vb, Brauner:2006dv, Kanazawa:2009ks,
  Kanazawa:2009en,Kanazawa:2011tt}, in lattice simulations
\cite{Nakamura:1984uz,Hands:1999md, Hands:2000ei, Muroya:2002jj,
  Chandrasekharan:2006tz, Hands:2006ve, Hands:2010gd, Hands:2011ye}, and the
Nambu--Jona-Lasinio model
\cite{Kondratyuk:1991hf, Kondratyuk:1992he, Rapp:1997zu, Ratti:2004ra,
  Sun:2007fc, Brauner:2009gu, Andersen:2010vu, Harada:2010vy, Zhang:2010kn, 
  He:2010nb}. In this section we focus on the
(Polyakov-)quark-meson-diquark model for studying the phase diagram of
\qctd with the functional renormalization group
\cite{Strodthoff:2011tz}, including fluctuations 
due to collective baryonic excitations.

The most important differences between two and three colors both follow
from the property of the SU$(2)$ gauge group of \qctd that
its representations are either real or pseudo-real.
This leads to an anti\-unitary symmetry in the Dirac operator
\cite{Kogut:2000ek}. As one consequence, the fermion determinant remains
real for non-vanishing baryon or quark chemical potential,
$\mu\not=0$, as it does for adjoint quarks in any-color QCD, or in the
$G_2$ gauge theory with fundamental fermions. Thus,
at least for an even number of degenerate fundamental quark flavors in
\qctd there is no fermion-sign problem and the phase diagram is
amenable to Monte-Carlo simulations. 

Another consequence of the reality or pseudo-reality of the quarks' color
representation is a Pauli-G\"ursey symmetry which allows to combine
quarks and charge-conjugated antiquarks into enlarged flavor multiplets. As a
result, for vanishing chemical potential and quark mass, $\mu=m_q=0$,
the usual $\mathrm{SU}(N_f)\times \mathrm{SU}(N_f) \times \mathrm
U(1)_B$ chiral and baryon number symmetries are replaced by an
extended $\mathrm{SU}(2N_f)$ flavor symmetry. A similarly extended
flavor symmetry is also known from (even numbers of) fermions in $2+1$
dimensions with U$(1)$ or SU$(N)$ gauge fields with $N\ge 3$. There it
is due to the fact that there is no physical helicity and that the
4-dimensional spinor representation of Dirac fermions is reducible in
the 3-dimensional space-time. The resulting extended U$(2N_f)$ flavor
symmetry is relevant for the expected  semimetal-insulator transition in QED$_3$
with not too many flavors, estimated for $N_f\lesssim 4$, or for the description of the electronic excitations
around the Dirac points of graphene at half-filling which are also
described by $N_f =2$ such 4-spinors of 3-dimensional Dirac fermions
in a single layer or $N_f=4$ in a double layer, for example
\cite{Gusynin:2007ix}. With (pseudo-)real
gauge fields such as SU$(2)$ or $G_2$ in 3 dimensions one obtains
an even larger extended flavor symmetry \cite{Dunne:2002vb}.   

As usual, the extended flavor symmetry of our 4-dimensional QCD-like theories 
is (spontaneously) broken by a (dynamical) Dirac mass. The breaking
patterns are somewhat different from the usual, however. In the
pseudo-real case of the fundamental two-color quarks the
(non-anomalous) extended SU$(2N_f)$ gets broken down to the $(2N_f+1)N_f$
dimensional compact symplectic group Sp$(N_f)$. In the real cases of
adjoint SU$(N)$ or fundamental $G_2$ quarks, for example, the
corresponding breaking would be SU$(2N_f) \to $ Spin$(2N_f)$, the
double cover of the proper rotation group SO$(2N_f)$.   

For $N_f=2$ the extended SU$(4)$ flavor symmetry group and
its Sp$(2)$ subgroup combining the isospin and baryon number
symmetries of \qctd are locally isomorphic to the
rotation groups SO$(6)$ and SO$(5)$, respectively. The coset is given
by $S^5$, the unit sphere in six dimensions, and a spontaneously generated 
Dirac mass will lead to five instead of the usual three Goldstone
bosons in this case, the three pions plus a scalar diquark-antidiquark
pair. 

Moreover, for $N_c=2$ these color-singlet scalar diquarks play a dual
role, as would-be-Goldstone bosons and bosonic baryons at the same
time. While this thus represents the perhaps most important difference
as compared to the real world, it also makes it much easier to
investigate the effects of baryonic degrees of freedom on the phase
diagram in functional approaches. In that sense the
quark-meson-diquark  (QMD) model can be considered as a first 
step towards their inclusion in a `quark-meson-baryon' model for real
QCD with  $N_c=3$.

For the same reason the QMD model of \qctd provides a relativistic
analogue of the BEC-BCS crossover in ultracold fermionic
quantum-gases, which has also been described successfully with
functional renormalization group methods
\cite{Boettcher:2012cm,Diehl:2009ma,Scherer:2010sv}.  
In contrast to non-relativistic models of the BEC-BCS crossover, 
an interesting additional constraint thereby arises from the Silver
Blaze property \cite{Cohen:2003kd}: When a relativistic chemical
potential $\mu$ is coupled to degrees of freedom with a mass gap
$\Delta$, as temperature approaches zero, the partition function and hence
thermodynamic observables must actually become independent of the
chemical potential as long as $\mu < \Delta$. In general, it is not
guaranteed that this constraint is automatically satisfied in
non-perturbative approaches such as functional renormalization group
studies which rely on truncations, but it is yet another example of
valuable extra information to devise intelligent 
truncation schemes \cite{Strodthoff:2011tz}.
 
One main virtue of the QMD model, however, is to explicitly demonstrate the
impact of baryonic degrees of freedom on the phase diagram by
comparing it with the corresponding purely mesonic model, as
representative of typical three-color QCD model calculations.
For this comparison it is more appropriate to think of
the vacuum diquark mass as the baryon mass $m_B$ rather than the pion
mass $m_\pi$.  In \qctd with its extended flavor symmetry they are
the same, but the essential aspect of this assignment is that a
continuous phase transition at zero temperature occurs at a critical
quark chemical potential $\mu_c = m_B/N_c$. Except for the scale
separation between $m_\pi/2$ and $m_B/3$ in the real world, this
transition is then to be compared to the liquid-gas
transition of nuclear matter in QCD with three colors which is of
first order, involves the binding energy, and thus occurs somewhat
below $\mu = m_B/N_c$.

As temperature increases the liquid gas transition ends, turning into
a crossover with continuously varying but nevertheless probably still
relatively abruptly increasing baryon density along some narrow
region. This rapid increase is generally expected to lead to the
strong chemical-potential dependence of the chemical freeze-out line
observed in heavy ion collisions at center-of-mass energies below
about 10 GeV per nucleon pair, the baryonic freeze-out
\cite{BraunMunzinger:2007zz,Andronic:2008gu}. 
One might conclude that the phase transition line for diquark 
condensation, where a rapidly increasing baryon density  
develops, would be the origin of a corresponding baryonic freeze-out line 
in two-color QCD, with $N_c=N_f =2$ arguably not
  necessarily further from reality than the large $N_c $ limits.  
As in the latter, one might then even identify a two-color version of
quarkyonic matter \cite{McLerran:2007qj, Hidaka:2008yy, Hands:2010gd,
  Brauner:2009gu}.

Finally, it is worth noting that the model, the functional
renormalization group equations and the techniques to solve them have
a broad scope of applications beyond two-color QCD. One example is QCD
with two light flavors at finite isospin chemical potential, which has
been studied with the NJL model in mean-field plus
random phase approximation (RPA) \cite{He:2005nk, Xiong:2009zz}. There
is a precise equivalence between the corresponding quark-meson model
with isospin chemical potential and 
the quark-meson-diquark model of two-color QCD discussed here. Besides  
changing $N_c$ this merely involves reducing the number of
would-be-Goldstone bosons from five to three again, retaining only one
of the three degenerate pions for the neutral one, and reinterpreting the
diquark-antidiquark pair of \qctd as the charged pions of QCD with
isospin chemical potential \cite{KazuhikoEtAlInPrep}.  Similar models
are also studied in the context of color superconductivity
\cite{Alford:2007xm,Steiner:2002gx,Huang:2003xd}. The capacity to 
numerically solve functional renormalization group equations 
on higher dimensional grids in field space is generally useful for
competing symmetries, as in a quark-meson model study of the axial
anomaly with scale dependent 't~Hooft couplings, for example
\cite{Mario2011}.  

\medskip

\subsection{Extended flavor symmetry and model Lagrangian}
\label{subsec:symm}

\medskip

As all half-odd integer representations of SU$(2)$, its fundamental
representation is pseudo-real because it is isomorphic to its
complex conjugate representation with the isometry given explicitly 
by $S = \imag\sigma_2$, $S^2 = -1$. 
Therefore, charge conjugation of the gauge fields in \qctd can be
undone by the constant SU$(2)$ gauge transformation $S =
\imag\sigma_2$. With $T^a = \sigma^a/2$ for the
fundamental color generators in SU$(2)$, one has 
\begin{equation}
\label{pseudo-real}
{T^a}^T = {T^a}^* = - S T^a S^{-1} \; .
\end{equation}
We will reserve $\sigma_i$ ($\tau_i$) for the Pauli matrices in spinor
(flavor) space from now on. Together with the charge conjugation
matrix $C$ in spinor space, likewise with $C^2= -1$, and complex
conjugation denoted by $K$ one then defines an antiunitary symmetry
$T= SCK$ with $T^2 = +1$ as time-reversal invariance in quantum
mechanics which leaves the Dirac operator invariant. 

For comparison, the irreducible representations of the proper
rotations, as the adjoint representations of all SU$(N)$, are examples
of real representations with hermitian generators that satisfy  
\begin{equation}
{T^a}^T = {T^a}^* = - T^a \; .
\end{equation}
More generally this corresponds to an isometry $S$ for complex
conjugation as in (\ref{pseudo-real}) but now with $S^2 = +1$. 
For all real color representations, 
as those of the adjoint groups SU$(N)/Z_N$ or the exceptional Lie
groups mentioned in the introduction such as $G_2$, one thus has $T^2=
-1$, correspondingly. The Dirac operator then has an antiunitary
symplectic symmetry which results in a two-fold degeneracy of its
eigenstates and a positive fermion determinant even for a single
flavor.  This leads to the classification of the Dirac
operator by the Dyson index $\beta$ of random matrix theory
\cite{Kogut:1999iv,Kogut:2000ek}, with $\beta = 1$ for fermions in
pseudo-real color representations corresponding to the Gaussian orthogonal
ensemble, $\beta = 4$ in real color representations corresponding to
the Gaussian symplectic, and $\beta = 2$ corresponding to the Gaussian unitary  
ensemble otherwise. The Dirac operator in the fundamental color
representation of \qctd (and in the $N$-ality $N/2$ representations of
the SU$(N)$'s with even $N$) falls in the first class ($\beta = 1$). Those
of $G_2$, $F_4$ and $E_8$ (and the $N$-ality zero representations of
SU$(N)$) fall in the second ($\beta =4 $).\footnote{Without the extra
  minus sign from the charge conjugation matrix $C$ after spin
  diagonalization, the antiunitary symmetries of staggered fermion
  operators for SU$(2)$ are opposite to those of the continuum or
  other lattice Dirac operators, $\beta =4$ for fundamental 
  (pseudo-real, half-odd integer) and $\beta = 1$ for adjoint (real,
  integer colored) quarks.}    
  
Following \cite{Kogut:2000ek}, let's start from the 
standard kinetic part of the Euclidean \qctd Lagrangian, in the chiral basis,  
\begin{equation}
\mathcal{L}_\text{kin}= \bar{\psi}\slashed{D}\psi  =  \psi_L^\dagger \imag
\sigma_\mu D_\mu \psi_L - \psi_R^\dagger \imag \sigma_\mu^\dagger D_\mu
\psi_R \; , \label{eq:2}
\end{equation}
with hermitian $\gamma$-matrices, $\sigma_\mu = (-\imag, \vec\sigma
)$, and $\psi_{R/L}$, $\psi_{R/L}^*$ as independent Grassmann
variables with $\psi_{R/L}^\dagger\equiv \psi_{R/L}^{*\, T}$. 
The covariant derivative is $D_\mu=\partial_\mu+\imag A_\mu$, and the
coupling is absorbed in the gauge fields $A_\mu=A_\mu^aT^a$. 

The two terms in (\ref{eq:2}) get interchanged under the anti\-unitary
symmetry $T$. One can apply a corresponding transformation to only one
of the two terms to change its sign, however. Using $(-\imag
\sigma_2)$ for the chiral $R$-component of the charge conjugation
matrix $C$ in the second term, say, by changing variables to
$\tilde\psi_R = -\imag\sigma_2 S \psi_R^*$ and $\tilde\psi_R^* =
-\imag\sigma_2 S \psi_R$, one can therefore re-express  
\begin{equation}
\mathcal{L}_\text{kin}=\Psi^\dagger\imag \sigma^\mu D_\mu \Psi \label{eq:3}
\end{equation}
in terms of $2N_f$ $4$-dimensional spinors
$\Psi=(\psi_L,\tilde\psi_R)^T$ and
$\Psi^\dagger=(\psi_L^\dagger,\tilde\psi_R^\dagger)$. Because
$\mathcal{L}_\text{kin}$  is now block diagonal, the SU$(2N_f)$
symmetry in the space combining flavor and transformed chiral
components is manifest in this form. With the same transformation of
variables the quarks' Dirac-mass term becomes   
\begin{equation} 
m \bar{\psi}\psi = \smallfrac{m}{2} \big( \Psi^T \imag\sigma_2 S
\Sigma_0 \Psi \, -  {\Psi^*}^T\imag\sigma_2  S  \Sigma_0\Psi^*  \big)\, ,
\label{eq:4}
\end{equation}
where the symplectic matrix
\begin{equation} 
\Sigma_0=\begin{pmatrix}0& \mathds 1_{N_f}\\-\mathds 1_{N_f}&0\end{pmatrix}
\end{equation}
acts in the $2N_f$-dimensional extended flavor space and transforms as
$\Sigma_0 \to U^T\Sigma_0 U$, because it is antisymmetric according to
the $N_f(2N_f -1)$-dimensional antisymmetric rank 2 tensor
representation of SU$(2N_f)$. The invariance 
group of $\Sigma_0$ as bilinear form on complex $2N_f$-vectors is the
compact symplectic group Sp$(N_f)$ which is the intersection of the
unitary U$(2N_f)$ and the symplectic Sp$(2N_f,\mathds C)$, therefore
sometimes also referred to as USp$(2N_f)$. An explicit(dynamical)
Dirac mass thus explicitly(spontaneously) breaks the original
SU$(2N_f)$ down to Sp$(N_f)$. 

For real color representations ($\beta = 4$), going through the same
steps with replacing $S\to \mathds 1$  \cite{Kogut:2000ek}, the mass
term is a symmetric color singlet and the corresponding flavor matrix
is therefore symmetric, likewise. It then belongs to the
$N_f(2N_f+1)$-dimensional symmetric rank 2 tensor 
representation of SU$(2N_f)$, and the invariance group, generated by the 
$N_f(2N_f -1)$ antisymmetric hermitian $2N_f\times 2N_f$ matrices, is
SO$(2N_f)$ or its double cover Spin$(2N_f)$ for fermionic states.

For $N_f=2$ flavors the enlarged flavor symmetry group is 
SU$(4)$, not $U(4)$ because of the axial anomaly, it replaces the
usual chiral and baryon number symmetries $\mathrm{SU}(2)_L\times
\mathrm{SU(2)}_R\times \mathrm U(1)_B$. Just as this extended flavor
SU$(4)$ shares its 15 dimensional Lie algebra with the group of
rotations in 6 dimensions, SO$(6)$, its Sp$(2)$ subgroup leaving the
($\beta =1$) Dirac-mass term invariant has the 10 dimensional Lie
algebra of SO$(5)$ (in fact they are both the universal covers of the
respective rotation groups). So the symmetry breaking patterns by
Dirac mass terms are locally the same as SO$(6)\to $ SO$(5)$  and
SO$(6)\to $ SO$(4)$ for $\beta = 1$ and $\beta=4$, respectively.

On the other hand, at finite chemical potential $\mu $, but still with
massless quarks, $m=0$, the SU$(2N_f)$ symmetry is broken explicitly
by \cite{Kogut:2000ek}    
\begin{equation}
\mu \bar\psi\gamma_0\psi =  \mu \Psi^\dagger B_0\Psi \; , \;\;
\mbox{with} \;\; 
B_0 = \begin{pmatrix} \mathds 1_{N_f}&0 \\0& -\mathds 1_{N_f}
\end{pmatrix} \; ,
\end{equation}
down to $\mathrm{SU}(N_f)_L\times
\mathrm{SU}(N_f)_R\times \mathrm U(1)_B$. 
For $N_f=2$, in terms of the rotation groups, this symmetry breaking
pattern is locally the same as $\mathrm{SO}(6) \to
\mathrm{SO}(4)\times \mathrm{SO}(2)$. 
  
When both $\mu$ and $m$ are non-zero, the unbroken flavor symmetry is
of course given by the common subgroup  $\mathrm{SU}(2)_V \times
\mathrm U(1)_B$ of the two limiting cases: (i) $\mu\to 0$ at finite
$m$ with either Sp$(2)$ ($\beta=1$) or Spin$(4)$ ($\beta = 4$)
symmetry, and 
(ii) $m\to 0$ at finite $\mu $ with $\mathrm{SU}(N_f)_L\times
\mathrm{SU}(N_f)_R\times \mathrm U(1)_B$ in either case, as discussed
above. 

Whether the resulting $\mathrm{SU}(2)_V \times \mathrm U(1)_B$
is actually closer to the combined isospin plus baryon-number
symmetries or the standard chiral symmetry, naturally depends on the
relative sizes of the quarks' Dirac mass $m$ and their chemical
potential $\mu$. More precisely, 
for quark chemical potential $\mu < m_\pi/2 $ 
the chiral symmetry breaking pattern essentially remains the $\mu
= 0$ one and the vacuum alignment is said to be $\langle \bar q
q\rangle$-like. There is only an explicit breaking of the combined
isospin/baryon-number symmetry proportional to $\mu$: 
according to $\mathrm{Sp}(2) \to \mathrm{SU}(2)_V \times \mathrm U(1)_B$
which is the same as $\mathrm{SO}(5) \to \mathrm{SO}(3) \times
\mathrm{SO}(2) $ for the bosonic states of \qctd with its pseudo-real
fundamental quarks ($\beta=1$), or according to Spin$(4) =
\mathrm{SU}(2) \times \mathrm{SU}(2) \to  \mathrm{SU}(2)_V \times
\mathrm U(1)_B$ for $G_2$ with real fundamental quarks or other theories with
quarks in real color representations ($\beta =4$). If the chemical potential is
small, the respective enlarged isospin/baryon-number symmetries are
approximately realized. The vacuum is unchanged and the degeneracies
in the spectrum simply split up by quark number proportional to
$\mu$. For example, among the would-be-Goldstone bosons the masses
$m_{\Delta^\pm}$ of diquark and antidiquark in \qctd split from those
of the pion as  $m_{\Delta^\pm}=m_\pi \pm 2\mu$.

When $m_{\Delta^-}=0$, {\it i.e.},  for $\mu \ge m_\pi/2 $, a diquark condensate
develops at zero temperature and the vacuum alignment starts rotating
from being $\langle \bar q q\rangle$-like to becoming more and more  $\langle q
q\rangle$-like as $\mu$ is further increased. The chiral condensate
then rapidly decreases and chiral symmetry gets restored to 
the approximate $\mathrm{SU}(2)_L\times \mathrm{SU}(2)_R \simeq
\mathrm{SO}(4)$. In this phase, the
baryon number U$(1)_B$ is spontaneously broken and the remaining
isospin SU$(2)_V$ changes from the approximately realized enlarged 
isospin/baryon-number Sp$(2)$ symmetry 
to becoming an approximate standard chiral  
 $\mathrm{SU}(2)_L\times \mathrm{SU}(2)_R $ symmetry.\footnote{In our
  example with $N_f=2$ quarks in a real color representation it might
  seem that there is no change in the approximate symmetry, it is
  Spin$(4) = \mathrm{SU}(2)\times \mathrm{SU}(2) $ for small chemical
  potential below the crossover and $\mathrm{SU}(2)_L\times
  \mathrm{SU}(2)_R $ above. They may look the same but they are two
  different subgroups of the original SU$(4)$, the first one contains
  U$(1)_B$ as a subgroup and the second one does not.}
Both are only approximate symmetries, the first is weakly broken by
the small $\mu$ (but there is a large dynamical quark mass), while the
second is weakly broken by the small current quark mass $m$ (at large
$\mu$). They change in a crossover. In fact, the dynamical
contribution to the quark mass changes in this crossover from being
predominantly the original Dirac mass, with a condensate of tightly
bound light diquarks, to becoming a dynamical Majorana
mass as in BCS theory. This vacuum realignment in the
diquark-condensation phase with superfluidity of the bosonic baryons
in \qctd is the analogue of the BEC-BCS crossover in ultracold
fermionic quantum gases.      

At zero temperature, for $2\mu <
m_\pi $, below the onset of the condensation of diquarks as bosonic
baryons,  the baryon density remains zero and the thermodynamic
observables must be independent of $\mu$. Because this is far from obvious to verify explicitly in actual
calculations, it has been named the Silver Blaze Problem
\cite{Cohen:2003kd}. In order to be able to excite any states at zero
temperature, and with a gap in the spectrum, the relativistic chemical
potential needs to be increased beyond the mass gap in the
correlations to which it couples. Here, with a continuous
zero-temperature quantum phase transition at $\mu_B \equiv 2\mu =
m_\pi$ this gap is simply given by the lightest baryon mass in vacuum which
because of the extended flavor symmetry in \qctd coincides with the
pion mass, $m_B = m_\pi$.   
This latter property is of course special to $N_c=2$ or other theories
with quarks in pseudo-real (without fermionic baryons) or real color
representations (with fermionic baryons, but where the lightest baryon
with mass $m_B = m_\pi$ is still bosonic). The Silver Blaze property
must hold as it does here, however, also when there are no bosonic
baryons as in QCD up to a quark chemical potential of the order of
$m_B/N_c$ (reduced by $1/N_c$ of the binding energy per nucleon when
the transition is of first order).

\begin{figure}[t]

\vspace*{.1cm}

\centering
\includegraphics[width=0.36\textwidth]{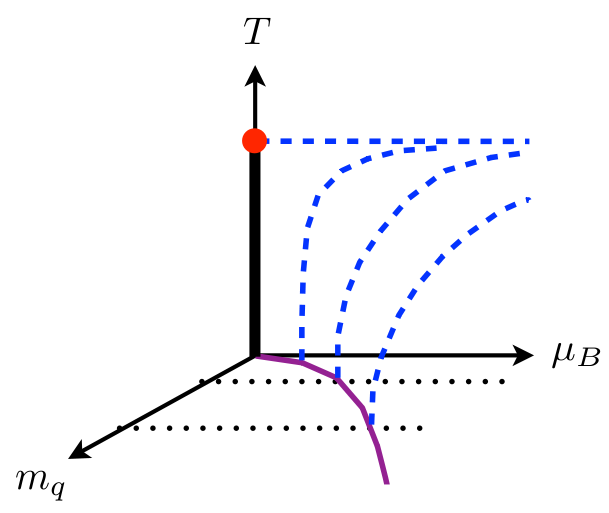}

\caption{Schematic phase diagram for \qctd  in the parameter space of
  temperature $T$, quark mass $m_q$ and baryon chemical potential
  $\mu_B$. 
\vspace{.1cm} 
}
\label{fig:pdtcsketch}
\end{figure}

At finite temperature, a qualitative picture emerges for the phase
diagram of \qctd as
sketched in Fig.~\ref{fig:pdtcsketch}. The solid line in the $T=0$ plane
represents the continuous zero-temperature 
transition with diquark condensation which is of mean-field
type. Because the quark mass $m_q$ scales quadratically with the pion
mass, it will occur along a parabola $m_q\propto \mu_B^2$. The thick
dashed lines represent the corresponding second-order transitions at finite
temperature in fixed $m_q$ planes of the O$(2)$ universality. The thick
line along the temperature axis is the magnetic first-order transition in the
 $\mu_B= 0$ plane which might end in a multicritical point.
When viewed in the $\mu_B= 0$ plane, this is the critical endpoint in
the O$(6)$ universality class for the chiral phase transition in \qctd 
with its extended SU$(4)$ flavor symmetry. In the $m_q = 0$ plane, the
vacuum alignment will always be $\langle q q \rangle$-like, for
no-matter-how-small $\mu>0$. Therefore, in this plane one only has the
second-order O$(2)$ line which, if it ends in the same point, would make it
multicritical.  

The construction of the QMD model for  \qctd starts from the flavor structure
of the standard chiral condensate and the quark mass term which is
of the form $\Psi^T \Sigma_0 \Psi$. It therefore transforms under the
full flavor SU$(4)$ according to the six-dimensional antisymmetric
representation in the decomposition $4\otimes 4= 6 \oplus 10$.

The other components belonging to the same multiplet are obtained from
transformations 
\begin{equation}
\Psi\to U\Psi \; , \quad
 U =\exp(\imag \theta^a X^a) \, \in \, \mathrm{SU}(4)/\mathrm{Sp}(2)
 \; .
\end{equation}
Then, $\Psi^T \Sigma_0 \Psi \to \Psi^T \Sigma \Psi$, where, from
Cartan's immersion theorem, the whole coset SU$(4)/$Sp$(2) \cong S^5$ 
is obtained in this way via $\Sigma \equiv U^T \Sigma_0 U$.  The coset
elements $\Sigma$ are in turn parameterized by six-dimensional unit
vectors $\vec n$ as $\Sigma = \vec n \vec\Sigma$, with
$\Sigma_i^\dagger \Sigma_j + \Sigma_j^\dagger \Sigma_i= 2\delta_{ij}$
and $\vec \Sigma=(\Sigma_0,  \imag \Sigma_0 X^{a})$ such that 
$X^a$,  $a=1\dots 5$, form a basis for the coset generators
\cite{Brauner:2006dv}. Thus, one verifies explicitly that the
vector $\Psi^T \vec \Sigma \Psi $ transforms as a (complex)
six-dimensional vector under SO$(6)$.

The QMD model for \qctd is
therefore defined by coupling the real SO$(6)$ vector of quark
bilinears $(\Psi^T\vec \Sigma\Psi+\text{h.c.})$ to a vector of mesonic
fields $\vec\phi=(\sigma,\vec{\pi},\text{Re}\,\Delta,\text{Im}\,\Delta)^T$
formed by the scalar $\sigma$ meson, the pseudoscalar pions $\vec\pi$
and the scalar diquark-antidiquark pair $\Delta$. This yields the
Lagrangian, 
\begin{equation}
\begin{split}
\label{eq:sigmamodel}
\mathcal{L}_\text{QMD}=&\Psi^\dagger\imag \sigma^\mu
\partial_\mu\Psi+\frac{g}{2}\big(\Psi^T \imag \sigma_2
S\vec \Sigma\Psi \\
& - {\Psi^*}^T \imag \sigma_2
S\vec \Sigma\Psi^*\big)\cdot \vec \phi 
+\frac{1}{2} (\partial_\mu \vec \phi)^2 +V(\vec \phi),
\end{split}
\end{equation}
where $V(\vec \phi)$ is the meson and diquark potential. A
non-vanishing chemical potential not only couples to the quarks but
also to the bosonic diquarks. Rewriting \Eq{eq:sigmamodel} in terms of
the original variables, one obtains
\begin{equation}
\label{eq:sigmamodeloriginal}
\begin{split} \mathcal{L}_\text{QMD}=&\bar{\psi}\left(\slashed{\partial}+g(\sigma+\imag\gamma^5\vec{\pi}\vec{\tau})-\mu
    \gamma^0\right)\psi\\
  &+\frac{g}{2}\left(\Delta^* (\psi^T C \gamma^5\tau_2 S \psi)+
    \Delta(\psi^\dagger C \gamma^5\tau_2 S\psi^*)\right)\\
  &+\frac{1}{2} (\partial_\mu \sigma)^2+\frac{1}{2} (\partial_\mu \vec
  \pi)^2  +
  V(\vec \phi)\\
  &+\frac{1}{2} \big((\partial_\mu-2\mu \,
  \delta_{\mu}^{0})\Delta\big) 
 (\partial_\mu+2\mu\, \delta_\mu^0)\Delta^*\ ,
\end{split}
\raisetag{\baselineskip}
\end{equation}
with $C=\gamma^2\gamma^0$ and a flavor- and color-blind
Yukawa coupling $g$. With 
\begin{equation} 
 V(\vec\phi) = \frac{\lambda}{4} (\vec\phi^2 - v^2 )^2 - c\sigma \, ,
\label{eq:Vlinsig}
\end{equation} 
one obtains the corresponding O$(6)$ linear sigma model; and 
in the limit $\lambda\to\infty$, the bosonic part of $
\mathcal{L}_\text{QMD}$ is then equivalent to the leading-order $\chi$PT
Lagrangian of Refs.~\cite{Kogut:2000ek} with the identifications $v=
f_\pi= 2F$ and $c=f_\pi m_\pi^2 = 2Fm_\pi^2$. The coefficient of the
leading term in $\mu$ of the $\chi$PT Lagrangian, which is 
$\mu^2 \mbox{tr}(\Sigma B^T\Sigma^\dagger B) $ with $B = 
UB_0U^\dagger$, was fixed from gauging the flavor SU$(4)$ in
\cite{Kogut:1999iv}. Here it simply follows from $-2\mu^2 |\Delta|^2$ as
part of the kinetic term of the complex scalar diquark field $\Delta$
with chemical potential $\mu_B=2\mu$. This implies in particular, that
the meson/diquark potential $V(\vec\phi)$ itself, up to the explicit breaking by
$-c\sigma$, which in principle needs to be only SO$(4) \times
\mathrm{SO}(2)$ invariant at finite $\mu$, must remain SO$(6)$
invariant, however, at this leading order, $\mathcal O(\mu^2)$, and
therefore at $\mathcal O(\phi^2)$ in the fields, likewise.  We can thus only
have an SO$(6)$ invariant mass term in $V(\vec\phi)$.

Gauge field dynamics and confinement effects can be modeled 
also in \qctd by including a constant Polyakov-loop variable as a background
field as in the NJL model \cite{Brauner:2009gu}, and analogous to what
is commonly done in the so-called Polyakov-loop-extended quark-meson
models of three-color QCD \cite{Schaefer:2007pw,Schaefer:2009ui,  
  Herbst:2010rf}. To this end one introduces a constant temporal
background gauge field $A_\mu=A_0\delta_{\mu 0}$ which is furthermore
assumed to be in the Cartan subalgebra as in the Polyakov gauge, {\it
  i.e.}, for the color SU$(2)$ of \qctd simply given by $A_0=T^3
2a_0$. This leads to the Polyakov loop variable 
\begin{equation}
\label{eq:polyvar}
\Phi\equiv\frac{1}{2}\Tr_c e^{\imag \beta A_0}=\cos(\beta a_0),
\end{equation}
to model a thermal expectation value of the color-traced Polyakov loop 
at an inverse temperature $\beta = 1/T$, as an order parameter for the
deconfinement transition at vanishing chemical potential.  
With the covariant derivative
$D_\mu=\partial_\mu-\imag \delta_{\mu 0} A_0$ replacing the 
ordinary one in the fermionic part of the QMD model in
\Eq{eq:sigmamodeloriginal}, this leads to a 
contribution of the form $-\imag\bar{\psi}\gamma^0 T^3 2a_0 \psi$ for 
a Polyakov-loop-extended quark-meson-diquark model (PQMD) Lagrangian.

It is often convenient to rewrite the fermionic part of the Lagrangian
in yet another form, $ \mathcal{L}_q = \bar{\Psi}_\mathrm{G}
S^{-1}\Psi_\mathrm{G} $, 
 in terms of eight-component Nambu-Gorkov spinors
$\Psi_\mathrm G =\left(\begin{smallmatrix}\psi_r\\
    \tau_2\psi_g^C \end{smallmatrix}\right)$, where $\psi_r$
($\psi_g$) explicitly denote the red (green) color components of $\psi$ and
$\psi^C\equiv C\bar{\psi}^T$ per flavor, and 
\begin{equation}
\label{eq:S0-1}
S^{-1}=\left(\begin{smallmatrix}
    \slashed{\partial}-\gamma^0(i
    a_0 + \mu) +g(\sigma+\imag\gamma^5\vec{\pi}\vec{\tau})
    &  g\gamma^5 \Delta \\
 - g\gamma^5\Delta^*
    &
    \slashed{\partial}-\gamma^0(ia_0
    - \mu)
    +g(\sigma-\imag\gamma^5\vec{\pi}\vec{\tau}) \end{smallmatrix}\right). 
\end{equation}
The PQMD model Lagrangian then becomes
\begin{equation}
\label{eq:LagrangianPQM}
\begin{split}
\mathcal{L}_\text{PQMD} =&  \, \bar{\Psi}_\mathrm G S^{-1}\Psi_\mathrm G
+\mathcal{U}_\text{Pol}(\Phi) +\frac{1}{2}
(\partial_\mu \sigma)^2+\frac{1}{2} (\partial_\mu \vec  \pi)^2 \\   
&+\frac{1}{2} \big((\partial_\mu-2\mu \,
  \delta_{\mu}^{0})\Delta\big) 
 (\partial_\mu+2\mu\, \delta_\mu^0)\Delta^* + V(\vec\phi)\ ,
\end{split}
\end{equation}
where $\mathcal{U}_\text{Pol}(\Phi)$ is the Polyakov-loop potential
\cite{Brauner:2009gu} which is commonly fitted to lattice results, but
which can also be computed with functional methods
\cite{Braun:2007bx,Marhauser:2008fz}. In contrast to the three-color case the 
Polyakov-loop potential is a function of one single real variable
$\Phi$ here, even in the presence of a diquark condensate.

\medskip

\subsection{Mesonic and baryonic fluctuations with the\\ 
 functional renormalization group}

\medskip

The phases of \qctd and the competing dynamics of the fluctuations of
the order parameters due to collective mesonic and baryonic excitations  
are most conveniently described within the framework of the functional
renormalization group (FRG)  
\cite{Litim:1998nf,Berges:2000ew,Polonyi:2001se,Pawlowski:2005xe, Gies:2006wv,
  Schaefer:2006sr,Braun:2011pp}.
The central object in the Wilsonian RG flow equation as pioneered by
Wetterich \cite{Wetterich:1992yh},   
is the scale $k$ dependent effective average action $\Gamma_k[\Phi]$,
where $\Phi$ generically represents the set of all local fields in
the theory.  This scale-dependent effective average action 
interpolates between the microscopic classical action at some ultraviolet 
(UV) cutoff scale $k=\Lambda$, at which fluctuations of essentially all  
momentum modes are suppressed, and the effective action of the  
full quantum theory in the infrared (IR), for $k\to0$, which then
includes all quantum and thermal fluctuations.  The scale-dependence
is described by the Wetterich flow equation,  
\begin{equation}
\label{eq:Wettericheq}
\partial_t \Gamma_k\equiv k\partial_k
\Gamma_k[\Phi]=\frac{1}{2}\text{Tr}\left\{\,\partial_t R_k
  (\Gamma^{(2)}_k+R_k)^{-1}\right\}\ ,
\end{equation}
which involves a momentum- and scale-dependent regulator $R_k$, whose 
precise form is not fixed but leaves a considerable flexibility. The
role of the regulator $R_k$ is to suppress the fluctuations of modes
with momenta below the renormalization scale $k$, and the flow
equation is UV as well as IR finite.  $\Gamma^{(2)}_k[\Phi]$ 
are the second functional derivatives of the effective average action
with respect to all the fields at scale $k$.  The functional trace
represents a one-loop integration typically evaluated in momentum
space and includes the sum over all fields and their internal and
space-time indices as well, with standard modifications for fermionic fields.  
It contains the full field and $k$-dependent
propagators of the regulated theory with cutoff $R_k$, the inverse of  
$\Gamma^{(2)}_k[\Phi ] + R_k$.  
In order to solve the flow equation an initial microscopic action
$S=\Gamma_{k=\Lambda}$ at some UV scale $\Lambda$ has to be specified.
Truncating the effective action to a specific form,  
the functional equation can be converted into a closed set of
(integro-)differential equations, but will in general also introduce
some regulator dependence in the flow. The choice of an
optimized regulator minimizes this regulator dependence for physical
observables. As bosonic (fermionic) regulators $R_{k,B}$ ($R_{k,F}$) we
choose 

\vspace{-.1cm}

\begin{equation}
\label{eq:regulators}
\begin{split}
R_{k,B}(\vec p)&=(k^2-\vec p^2)\theta(k^2-\vec p^2),\\
R_{k,F}(\vec p)&=-\imag \vec{p}\cdot\vec\gamma
\left(\sqrt{\frac{k^2}{\vec p^2}} -1\right)\theta(k^2-\vec p^2),
\end{split}
\end{equation} 
which are three-momentum analogues of the optimized Litim regulators
\cite{Litim:2001up}. With this choice the three-momentum integration
becomes trivial and the remaining Matsubara sums can be evaluated
analytically. Furthermore, this choice leaves the semilocal
U$(1)$-symmetry of the Lagrangian unaffected, analogous to
\cite{Diehl:2009ma}, where the chemical potential acts like the
zero-component of an Abelian gauge field. 

At leading-order in a derivative expansion, all wave-function
renormalization factors are neglected and only the scale-dependent
effective potential $U_k$ is taken into account. The ansatz for the
effective average action then simply reads 
$\Gamma_k=\left.\int d^4x \, \mathcal{L}_\text{PQMD}\right|_{V+c\sigma
  \to U_k}$.
This means that in $\mathcal{L}_\text{PQMD}$ from
\Eq{eq:LagrangianPQM} the
meson/diquark potential $V(\vec\phi)$ of the O$(6)$ linear sigma model
from \Eq{eq:Vlinsig} is replaced by $U_k-c\sigma$. The explicit symmetry
breaking term $-c\sigma $ does not affect the flow and is thus not
part of $U_k$ but added after the RG evolution to the full effective
potential again. At $\mu=0$, the scale-dependent $U_k$ then only depends on
the modulus of
$\vec\phi=(\sigma,\vec{\pi},\text{Re}\,\Delta,\text{Im}\,\Delta)^T$. At   
non-vanishing chemical potential, however, we only have
$\mathrm{SO}(4)\times \mathrm{SO}(2)$ symmetry and must therefore
allow it to depend on two invariants, {\it i.e.}, $U_k\equiv
U_k(\rho^2,d^2)$ where $\rho^2 = \sigma^2+\vec\pi^2$, and $d^2 =
|\Delta|^2$ as before. For $\mu\to 0$ we recover the full extended
SO$(6)$ invariance, of course, so that $U_k$ then  
depends only on the combination $\phi^2 = \rho^2 + d^2$ again.

Working out the second functional derivatives of the effective action,
one obtains in momentum space with constant fields, and coordinates
such that  $\sigma=\rho$, $\vec\pi=\vec 0$, $\text{Re}\, \Delta =  d$,
$\text{Im}\,\Delta=0$, the inverse bosonic propagator
$\Gamma^{(2)}_{k,B} + R_{k,B}$, with   

\vspace*{-.6cm}

\begin{widetext}

\vspace*{-.5cm}

\begin{eqnarray}
\label{eq:gamma2b}
\hskip -1.6cm \Gamma^{(2)}_{k,B} &=& \\[4pt] 
&& \hskip -1cm \left(\begin{array}{cccccc}
\hspace*{-4pt} p^2+2\Ur &0 &0 &0 &0&0\\
0& \hspace*{-4pt}   p^2+2\Ur &0 &0 &0 &0\\
0& 0&  \hspace*{-4pt}  p^2+2\Ur &0 &0 &0\\
0& 0& 0& \hspace*{-4pt} p^2+2 \Ur+4 \rho^2 \Urr&4 \rho d \Urd&0\\
0& 0& 0& 4 \rho d \Urd& \hspace*{-4pt}  p^2+2\Ud+4 d^2 \Udd-4\mu^2& -4 \mu p_0\\
0& 0& 0& 0&\hspace*{-4pt}  4 \mu p_0&p^2+2\Ud-4 \mu^2 \hspace*{-4pt} 
\end{array}\right)\, , \nonumber
\end{eqnarray}


\end{widetext}

\noindent
where short-hand index notations for the
derivatives  of the potential with respect to the fields are 
defined as 
\[
\Ud \equiv \frac{\partial U_k}{\partial d^2} \; , \;\; \Ur \equiv
\frac{\partial U_k}{\partial \rho^2}\; , \; \mbox{and} \;\;
\Urd \equiv \frac{\partial^2
U_k}{\partial \rho^2\partial d^2} \; ,\;\mbox{etc.}
\]
The inverse fermion-propagator can essentially be read off from
\Eq{eq:S0-1}. With the above replacements for constant fields 
it is diagonal in flavor and given by,
$\Gamma^{(2)}_{k,F} +  R_{k,F}$, with  
\begin{equation}
\label{eq:Gamma2F}
  \Gamma^{(2)}_{k,F}=\left(\begin{smallmatrix}-\imag \slashed p-\imag a_0 \gamma^0 + g \rho -\gamma^0 \mu &g\gamma^5 d\\
      -g\gamma^5 d &-\imag \slashed p -\imag a_0 \gamma^0+ g\rho
      +\gamma^0\mu\end{smallmatrix}\right) \; ,
\end{equation}
for each flavor. With the regulators from \Eqs{eq:regulators} the
$k$-dependent propagators are obtained upon inverting these expressions for 
$\Gamma^{(2)}_{k,B/F} + R_{k,B/F} $. When these propagators are  inserted into 
the  Wetterich equation, \Eq{eq:Wettericheq}, the integration over 
the spatial momentum components in the explicit loop, with the
three-dimensional  optimized regulators in \Eqs{eq:regulators}, can be
performed rather straightforwardly. For example, for the bosonic
contribution to the flow, with bosonic Matsubara frequencies $p_0 =
\omega_n= 2\pi T n$ for periodic boundary conditions in the
zero-component $p_0$ of the loop momentum at finite temperature we
have in the subspace of the three pions,
\[
 \Big(\Gamma^{(2)}_{k,B}(p) + R_{k,B}(\vec p)\Big)_{ij} =
\Big(\omega_n^2 + \max\{\vec p^2,k^2\}+2\Ur \Big) \, \delta_{ij}  
\]
with $i,j = 1,\dots 3$ for the upper-left $3\times 3 $ 
block of $\Gamma^{(2)}_{k,B} $ in \Eq{eq:gamma2b}. 
In the subspace of the $\sigma $ meson and the diquark-antidiquark pair,
which mix when the diquark condensate $d = |\Delta|$  is non-zero, one
analogously obtains a $3\times 3 $ matrix corresponding to the
lower-right block in \Eq{eq:gamma2b} with the same replacements, $p_0
\to \omega_n$ and $\vec p^2 \to \max\{\vec p^2,k^2\} $. If we call the
inverse of this   $3\times 3 $ submatrix 
\begin{equation} 
A(\omega_n,\max\{\vec p^2,k^2\}) \equiv \Big(\Gamma^{(2)}_{k,B} +
R_{k,B} \Big)^{-1}_{\sigma,\Delta} \, ,
\end{equation}
the total bosonic contribution to the flow equation for the effective
potential $U_k$ (up to a 4-dimensional spacetime-volume factor $V$ the same
as that for the effective action in this truncation), 
 is then given by the following Matsubara sum
\begin{align} 
\partial_t U_{k,B} &= \frac{1}{2V}\text{Tr}\left\{\,\partial_t
  R_{k,B} \,  \Big(\Gamma^{(2)}_{k,B}+R_{k,B}\Big)^{-1}\right\} \\
 &= \frac{1}{2} T\sum_{n=-\infty}^\infty \int_0^k \frac{p^2 dp}{2\pi^2} \, 2 k^2\, \bigg( 
\frac{3}{\omega_n^2 + k^2 + 2\Ur} \notag \\
& \hskip 4.5cm + \tr\, A(\omega_n,k^2)\bigg)  \notag \\
&= \frac{k^5 T}{6\pi^2} \sum_{n=-\infty}^\infty \bigg(\frac{3}{\omega_n^2 +
  k^2 + 2\Ur}  + \tr\, A(\omega_n,k^2)\bigg)  \; .  \notag
\end{align}
The trace of $A$ is the ratio of the sum of three $2\times 2$
determinants of submatrices of  $\big(\Gamma^{(2)}_{k,B/F} + 
R_{k,B/F}\big)_{\sigma,\Delta} $ over the full determinant of the
$3\times 3$ block $\big(\Gamma^{(2)}_{k,B/F} + R_{k,B/F}
\big)_{\sigma,\Delta}$ corresponding to the sigma and
diquark-antidiquark directions in field space  in \Eq{eq:gamma2b}.  
Therefore, the numerator of $\tr\, A$ is a quadratic polynomial in
$\omega_n^2$ and the denominator a cubic one, 
\begin{equation}
\tr\, A(\omega_n,k^2) =  \frac{3 (\omega_n^2)^2+\alpha_1
  \omega_n^2+\alpha_0}{(\omega_n^2)^3+\beta_2 (\omega_n^2)^2+\beta_1
  \omega_n^2+\beta_0}  
\end{equation}
The coefficients of these polynomials are probably not very
illuminating. They are worked out to be
\cite{Strodthoff:2011tz},
\begin{align}
\alpha_0 =&\, 3 k^4\\ 
      &\hskip -.2cm + 4 k^2 \big(-4 \mu^2 + 2 U_{k,d} + 2 d^2 U_{k,dd}
      + U_{k,\rho} + 2       \rho^2 U_{k,\rho\rho}\big)\notag\\  
&\hskip -.2cm  + 4 \Big(4 \mu^4 + U_{k,d}^2 + 2 U_{k,d} \big(d^2
U_{k,dd} + U_{k,\rho}   + 2 \rho^2 U_{k,\rho\rho}\big)\notag\\
&\qquad - 4 \mu^2 \big(U_{k,d} + d^2 U_{k,dd} + U_{k,\rho} + 2 \rho^2
U_{k,\rho\rho}\big) \notag\\
&\qquad + 2 d^2 \big(U_{k,dd} U_{k,\rho} - 2 \rho^2 U_{k,\rho d}^2 + 2
\rho^2 U_{k,dd} U_{k,\rho\rho}\big)\Big)\notag\\ 
\alpha_1 =& \, 6 k^2 + 8 U_{k,d} + 8 d^2 U_{k,dd} + 4 U_{k,\rho} + 8
\rho^2 U_{k,\rho\rho} \notag \\
\beta_0 =& \, \big(k^2 - 4 \mu^2 + 2 U_{k,d}\big) \Big(k^4 \\
&\hskip -.2cm + 2 k^2 \big(-2 \mu^2 + U_{k,d} + 2 d^2 U_{k,dd} +
U_{k,\rho} + 2 \rho^2 U_{k,\rho\rho}\big)\notag\\  
&\hskip -.2cm + 4 \big(-2 \mu^2 U_{k,\rho} + U_{k,d} U_{k,\rho} + 2
d^2 U_{k,dd} U_{k,\rho} \notag\\
&\hskip -.2cm  - 4 d^2 \rho^2 U_{k,\rho d}^2 + 2 \rho^2 (-2
\mu^2 + U_{k,d} + 2 d^2 U_{k,dd}) U_{k,\rho\rho}\big)\Big)\notag\\ 
\beta_1 =& \, 3 k^4 + 4 k^2 (2 U_{k,d} + 2 d^2 U_{k,dd} + U_{k,\rho} +
2 \rho^2 U_{k,\rho\rho})\notag\\ 
 &\hskip -.2cm + 4 \Big(4 \mu^4 + U_{k,d}^2 - 4 \mu^2 \big(U_{k,d} +
 d^2 U_{k,dd} - U_{k,\rho} \notag\\
&\; - 2 \rho^2 U_{k,\rho\rho}\big) +  2 U_{k,d} \big(d^2
U_{k,dd} + U_{k,\rho} + 2 \rho^2 U_{k,\rho\rho}\big)\notag\\ 
 &\quad + 2 d^2 \big(U_{k,dd} U_{k,\rho} - 2 \rho^2 U_{k,\rho d}^2 +
   2 \rho^2 U_{k,dd} U_{k,\rho\rho}\big)\Big)\notag \\ 
\beta_2 =& \, 3 k^2 + 8 \mu^2 + 4 U_{k,d} + 4 d^2 U_{k,dd} + 2
U_{k,\rho} + 4 \rho^2 U_{k,\rho\rho} \notag 
\end{align}
These coefficient functions $\alpha_{i}$ and  $\beta_{i}$ depend on
the renormalization scale, the chemical potential, the fields, and the
derivatives of the potential. This mixing of the sigma with the
diquark sector in the diquark condensation phase, where $d=|\Delta
|\not= 0 $ and baryon number is no-longer conserved, is what makes the
equations somewhat more complicated than usual. But the structure
behind these lengthy expressions is actually not as bad as it might
at first appear. With a partial-fraction decomposition of $\tr\, A$,
for example, the Matsubara sum can still be computed analytically from
the residue theorem in the standard way \cite{Das:1997gg,LeBellac}.  
With the roots of the polynomial in the denominator denoted as
$\omega_{n,0}^2=-z_i^2$, $i=1,...,3$, this yields for 
the bosonic flow
\begin{align}
\label{eq:bosonicflowfinal}
  \partial_t U_{k,B}=&\,
  \frac{k^5}{12\pi^2}\, \Bigg\{\, \frac{3}{E^\pi_k}\,
  \coth\left(\frac{E^\pi_k}{2T}\right)\\    
 &    +\sum_{i=1}^3  \,\frac{\alpha_2 z_i^4-\alpha_1
   z_i^2+\alpha_0} {(z_{i+1}^2-z_i^2) (z_{i+2}^2-z_i^2)}
  \,  \frac{1}{z_i}\, \coth\left(\frac{z_i}{2T}\right)\, \Bigg\} \; ,
\notag
\end{align}
where  $E^\pi_k=\sqrt{k^2+2 \Ur}\,$.

The fermionic contribution to the flow can be worked out
analogously. The derivative of the regulator  $R_{k,F}$ in
\Eqs{eq:regulators} cuts off the spatial loop-momentum integration
at $\vec p^2 = k^2$, and for $\vec p^2 < k^2$ the fermionic two-point
function $\Gamma^{(2)}_{k,F} + R_{k,F} $ is obtained from
\Eq{eq:Gamma2F} upon replacing $ -i \vec p \cdot \vec\gamma \to -i (\vec p
\cdot \vec\gamma/|\vec p|)\, k$ therein. Inverting the resulting 
fermion matrix, this then yields for the trace over Dirac (and flavor)
indices for $\vec p^2 < k^2$, 
\begin{align}
\tr\Big( -ik\, \frac{\vec p\cdot\vec\gamma}{|\vec p|}
\,\big(\Gamma^{(2)}_{k,F} + R_{k,F}  \big)^{-1} \Big) &=\\
&\hskip -2.5cm 16 k^2 \, \frac{(p_0+a_0)^2 -\mu^2 + k^2 + g^2 (d^2 +\rho^2)}{\big((p_0+a_0)^2 +
  {E_k^+}^2\big)\big((p_0+a_0)^2 + {E_k^-}^2\big)} \; . \notag
\end{align}
One factor of $N_f= 2$ hereby arises from the sum of the two flavors.
The square of the denominator is the determinant of the 
fermion matrix $\Gamma^{(2)}_{k,F} + R_{k,F} $ per flavor,
whose eigenvalues are given by the eight combinations of different signs
of $\mp i(p_0+a_0) \pm E^\pm_k$, with  
\begin{equation}
{E}^\pm_k  =\sqrt{g^2 d^2 + (\epsilon_k \pm \mu)^2} \; ,\;
\mbox{and} \;\; \epsilon_k =\sqrt{k^2+ g^2\rho^2} \; .
\end{equation}
For $\vec p^2 > k^2 $ one would simply have to replace $k^2$ by $\vec
p^2$ in these expressions again, as one would for $\Gamma^{(2)}_{F}$ without
the regulator, {\it e.g.}, for a mean-field analysis of the model,
see Ref.~\cite{Strodthoff:2011tz}. These momenta are not needed for
the optimized fermionic flow either, however, which then readily
follows to be given by, 
\begin{align} 
\label{eq:fermionicflow}
\partial_t U_{k,F} &= -\frac{1}{V}\text{Tr}\left\{\,\partial_t
  R_{k,F} \, \Big(\Gamma^{(2)}_{k,F}+R_{k,F}\Big)^{-1}\right\} \\
&\hskip -.8cm= -\frac{8k^5 T}{3\pi^2} \sum_{n=-\infty}^\infty 
\frac{(\nu_n+a_0)^2 -\mu^2 + k^2 + g^2 (d^2 +\rho^2)}{\big((\nu_n
  +a_0)^2 +   {E_k^+}^2\big)\big((\nu_n+a_0)^2 + {E_k^-}^2\big)} 
\; ,\notag
\end{align}
where the fermionic Matsubara frequencies $p_0 = \nu_n= (2 n+ 1) \pi T$
are used for antiperiodic boundary conditions in imaginary time. 
When their sum is evaluated, one finally obtains,
\begin{align} 
\label{eq:fermionicflowfinal}
\partial_t U_{k,F} &= \\
& \hskip -.8cm -\frac{k^5}{3\pi^2} 
\sum_\pm  \frac{2}{E^\pm_k}\Bigg(1 \pm
      \frac{\mu} {\sqrt{k^2+g^2\rho^2}}\Bigg) 
   \, \Big( 1 - 2 N_q(E_k^\pm/T,\Phi) \Big)\; . \notag 
\end{align}
Here, $N_q(E/T,\Phi)$ are Polyakov-loop enhanced quark occupation
numbers for \qctd,
\begin{equation}
\label{eq:PLenhancedoccnumbers}
N_q(E/T,\Phi)=\frac{1+\Phi e^{{E}/{T}}}{1+2\Phi e^{{E}/{T}}
  +e^{{2E}/{T}}}\; 
\end{equation}
which reduce to the Fermi-Dirac and Bose-Einstein distributions for $\Phi=1$
and  $\Phi= -1$, respectively. It is also common to use \cite{Brauner:2009gu},
\begin{equation}
\varphi_\Phi(E/T) \,\equiv  1 - 2 N_q(E/T,\Phi) =
\frac{\sinh(E/T)}{\Phi+\cosh(E/T)} \; ,
\end{equation}
which satisfies
\begin{equation}
\varphi_\Phi(x) =  \left\{ 
\begin{array}{lll} 
 \tanh(x/2) &=  1 - 2\rho_f(x) \, , & \Phi = 1\\[2pt] 
 \tanh(x)   &=  1 - 2\rho_f(2x)\, , & \Phi = 0\\[2pt] 
 \coth(x/2) &=  1 + 2\rho_b(x) \, , & \Phi = -1\\ 
\end{array} \right.
\end{equation}
\[
\hskip -1.2cm\mbox{where} \;\; \rho_f(x) = \frac{1}{e^x +1} \; , \;\;\mbox{and} \;\;
\rho_b(x) = \frac{1}{e^x - 1} \; .
\]
The SU$(2)$ center sector with $\Phi = -1 $ corresponds to using
periodic quarks in \qctd and leads to a bosonic excitation
spectrum. It is explicit in \Eq{eq:fermionicflow} that nothing
changes, if one changes to periodic quarks and changes the background
according to such a center flip $\Phi \to -\Phi$ at the same time. The
first amounts to replacing $\nu_n \to \omega_n$ and the second to $a_0
\to a_0 + \pi T $ which together leave the fermionic flow
invariant.\footnote{In \Eq{eq:fermionicflow} we can actually 
  continuously change $\nu_n \to \nu_n +c$ and $a_0\to a_0 -c$ without
  effect. It might appear that we can thus change  the quarks'
  boundary conditions by a general U$(1)$ phase as we could in QED
  with unbroken displacement symmetry. This is not the case, however,
  because  $\nu_n \to \nu_n +c$ would rotate the boundary conditions
  of the red quarks in the same way as those of the green antiquarks,
  {\it i.e.}, those of the red ones in the opposite direction of those
  of the green ones. Thus this U$(1)$ rotates about the
  $T^3$-direction of SU$(2)$ in color and there is no contradiction.}  
This is a manifestation of the so-called Roberge-Weiss
symmetry \cite{Roberge:1986mm}. For  $\Phi = 0 $, in a center-symmetric
background with $a_0 = \pi T/2\mod \pi$, the thermal excitation
energies are twice the quark energies which is occasionally
interpreted as modeling confinement.  
  

The full and final flow equation for the effective potential in the PQMD model
for \qctd \cite{Strodthoff:2011tz} in the leading order derivative
expansion is given by the sum of the bosonic and the
fermionic flows in \Eqs{eq:bosonicflowfinal} and
(\ref{eq:fermionicflowfinal}), 

\begin{equation} 
\label{eq:fullflowfinal}
\partial_t U_{k} \, =\,  \partial_t U_{k,B} \, +\, \partial_t U_{k,F}
\; . 
\end{equation}

\noindent
A significant complexity in this flow equation for the effective
potential $U(\rho,d)$ is introduced by the presence of two fields
$\rho$ and $ d$ corresponding to the standard chiral quark condensate
and the diquark condensate, respectively, which have a
mutual influence on one another with physically important implications.  

\medskip

\subsection{SO$(6)$ symmetric flow}

\medskip

In the normal hadronic phase without diquark condensation, we may set
$d = |\Delta| = 0$ to obtain an explicitly SO$(6)$ symmetric flow
for $U_k(\phi )$. If we set $\Up\equiv \Ur= \Ud$,
\Eq{eq:fullflowfinal} then reduces to a more familiar looking form,

\vspace*{-4cm}

\begin{widetext}
\begin{equation}
\label{eq:floweqdelta0}
\begin{split}
  \partial_t
  U_k=\frac{k^5}{12\pi^2}&
  \left\{\frac{3}{E_k^{\pi}}\coth\left(\frac{E_k^{\pi}}{2T}\right)
+\frac{1}{E_k^\sigma}\coth\left(\frac{E_k^\sigma}{2T}\right)
    +\frac{1}{E_k^{\pi}}\coth\left(\frac{E_k^{\pi}-2\mu}{2T}\right)
    +\frac{1}{E_k^{\pi}}\coth\left(\frac{E_k^{\pi}+2\mu}{2T}\right) \right.\\
  &\hspace{.5cm} \left. -\frac{16}{\epsilon_k} \,
  \Big(\left. 1-N_{q}\left(\epsilon_k-\mu;T,\Phi\right) 
    -N_{q}(\epsilon_k+\mu;T,\Phi)\right. \Big)\right\}, 
\end{split}
\end{equation}
\end{widetext}

\vspace*{-.2cm}

\noindent
with equal single-particle energies for mesons and diquarks,
\begin{equation}
E_k^{\pi}= E_k^\Delta = \sqrt{k^2+2\Up} \; , 
\end{equation}
while for the sigma meson one has 
\begin{equation}
E_k^\sigma=\sqrt{k^2+2 \Up    +4\phi^2 \Upp} \; ,
\end{equation}
and for the quarks $\epsilon_k =\sqrt{k^2+ g^2\phi^2} $ as above.
Except for the change in the number of active degrees of freedom  
contributing to this flow, and the isospin-like chemical potential
coupling to one pseudo-Goldstone boson pair, the SO$(6)$ symmetric flow
equation here is entirely analogous to the one of the PQM model for QCD 
with three colors, see {\it e.g.},
\cite{Herbst:2010rf,Skokov:2010wb,Skokov:2010uh}.   
In the PQM model for QCD with isospin chemical potential, however, one
must allow for pion condensation and then arrives at a flow equation
for two competing fields 
\cite{KazuhikoEtAlInPrep} analogous to \Eq{eq:fullflowfinal}, with
\Eqs{eq:bosonicflowfinal} and (\ref{eq:fermionicflowfinal}).  

A further simplification occurs for $\mu = 0$ and in a trivial
gauge-field background, $a_0 = 0$ corresponding to $\Phi = 1$,
\begin{equation}
\label{eq:flowmu0}
\begin{split}
\partial_t U_k= \frac{k^5}{12 \pi^2} & \left\{\frac{5}{E_k^\pi}
\coth\left(\frac{E_k^\pi}{2 T}\right) +\frac{1}{E_k^\sigma} \coth\left(
  \frac{E_k^\sigma}{2 T}\right) \right. \\ 
& \hspace{1.8cm} \left. 
-\frac{16}{\epsilon_k }
\tanh\left(\frac{\epsilon_k}{2 T} \right) \right\}\ .
\end{split}
\end{equation}
Because $\mu=0$ the diquarks are now fully degenerate with the pions
which leaves us with the $N_c=2$ analogue of the familiar three-color
QM model flow equation \cite{Braun:2003ii,Schaefer:2004en} except that
there are now five pseudo-Goldstone bosons instead of the usual three pions.

\begin{figure}[t]
\centering
\includegraphics[width=0.9\linewidth]{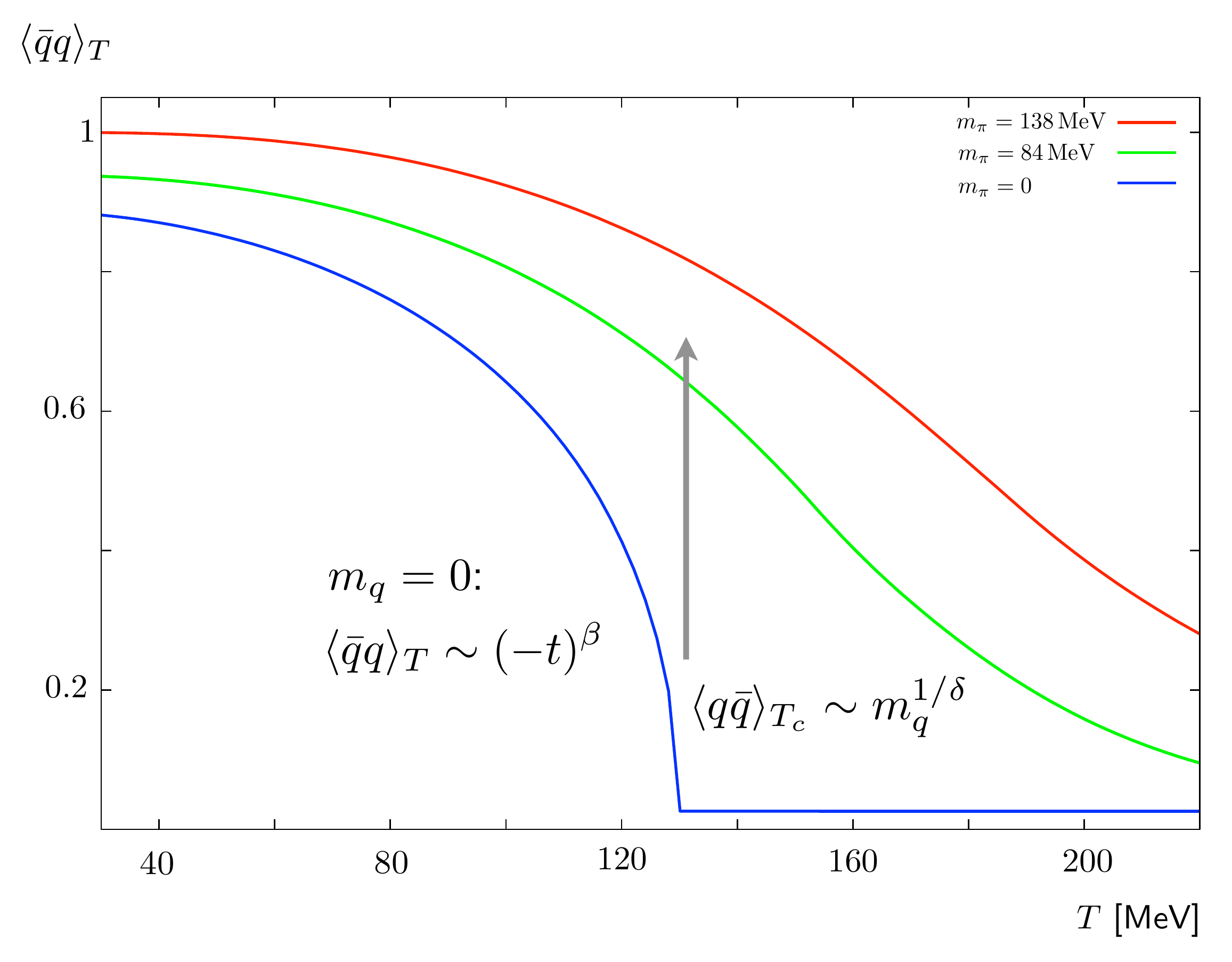}
\vspace{-.2cm}
\caption{The chiral condensate in the QMD model for \qctd  over 
  temperature $T$ from the SO$(6)$ symmetric RG flow for various quark
  masses $m_q$ corresponding to pion masses $m_\pi = 138$ MeV, $84$ MeV
  and in the chiral limit, relative to the $T=0$ condensate
  at $m_\pi = 138$ MeV. 
\vspace{-.2cm} }
\label{fig:finiteTscaling}
\end{figure}

As in the studies of O$(4)$ universality and scaling in the three-color QM model
\cite{Berges:1997eu,Schaefer:1999em,Bohr:2000gp,Stokic:2009uv,Braun:2010vd}, 
one can use this flow equation to analogously check the
symmetry breaking pattern discussed in Section
\ref{subsec:symm}, and the corresponding magnetic scaling, by
computing the two critical exponents $\beta$ and $\delta$ as mentioned 
in the Introduction. For $\mu=m=0$, the
SU$(4)\simeq \mathrm{SO}(6)$ dynamically breaks down to Sp$(2) \simeq
\mathrm{SO}(5)$ so that we expect a finite temperature phase
transition in the three-dimensional O$(6)$ universality class. The
critical exponent $\beta$ can be extracted from the dependence of the
chiral condensate on the reduced temperature $t=(T-T_c)/T_c$ in the
chiral limit, whereas the exponent $\delta$ governs the dependence of
the chiral condensate at $T_c$ on the quark mass $m_q$ or
correspondingly on the explicit symmetry-breaking parameter $c$, 
\begin{equation}
\langle \bar q q\rangle_T\sim(-t)^\beta , \quad \langle \bar q
q\rangle_{T_c}\sim c^{1/\delta} .
\end{equation}
With the usual 
two-exponent scaling all other critical
exponents are then obtained from these two.
A careful analysis of the behavior of the quark condensate from
solutions to the 1$d$ flow equation (\ref{eq:flowmu0}) such as those shown in
Fig.~\ref{fig:finiteTscaling} yields critical exponents
$\beta=0.4318(4)$ and $\delta=5.08(8)$ \cite{Strodthoff:2011tz}.
Literature values for these exponents obtained from Monte-Carlo
simulations are given by $\beta = 0.425(2)$ and $\delta = 4.77(2)$
\cite{Holtmann:2003he}. At leading order in the derivative expansion
one would not expect to reproduce these values exactly, however.
The more appropriate benchmark here should be the functional 
renormalization group result for the O$(6)$ model in the local
potential approximation \cite{Litim:2002cf}. 
In absence of wave-function renormalizations there is no 
anomalous dimension for the fields and their critical exponent
therefore vanishes, $\eta = 0$. Then the hyperscaling relations,  
\begin{equation}
\delta = \frac{d+2 -\eta}{d-2+\eta} \, , \quad \beta = \frac{\nu}{2}
\big( d-2+\eta \big) \, ,
\end{equation}
immediately entail that $\delta = 5$ and $\beta = \nu/2$ in $d=3$
dimensions. With the correlation-length critical exponent 
$\nu = 0.863076 $ from Ref.~\cite{Litim:2002cf} this corresponds to
$\beta = 0.4315 $, and both these values are in good agreement ({\it
  i.e.}, within the errors) with the two from \Eq{eq:flowmu0},
$\beta=0.4318(4)$ and $\delta=5.08(8)$ as quoted
above \cite{Strodthoff:2011tz}. 

\begin{figure}[t]
\centering
\includegraphics[width=0.9\linewidth]{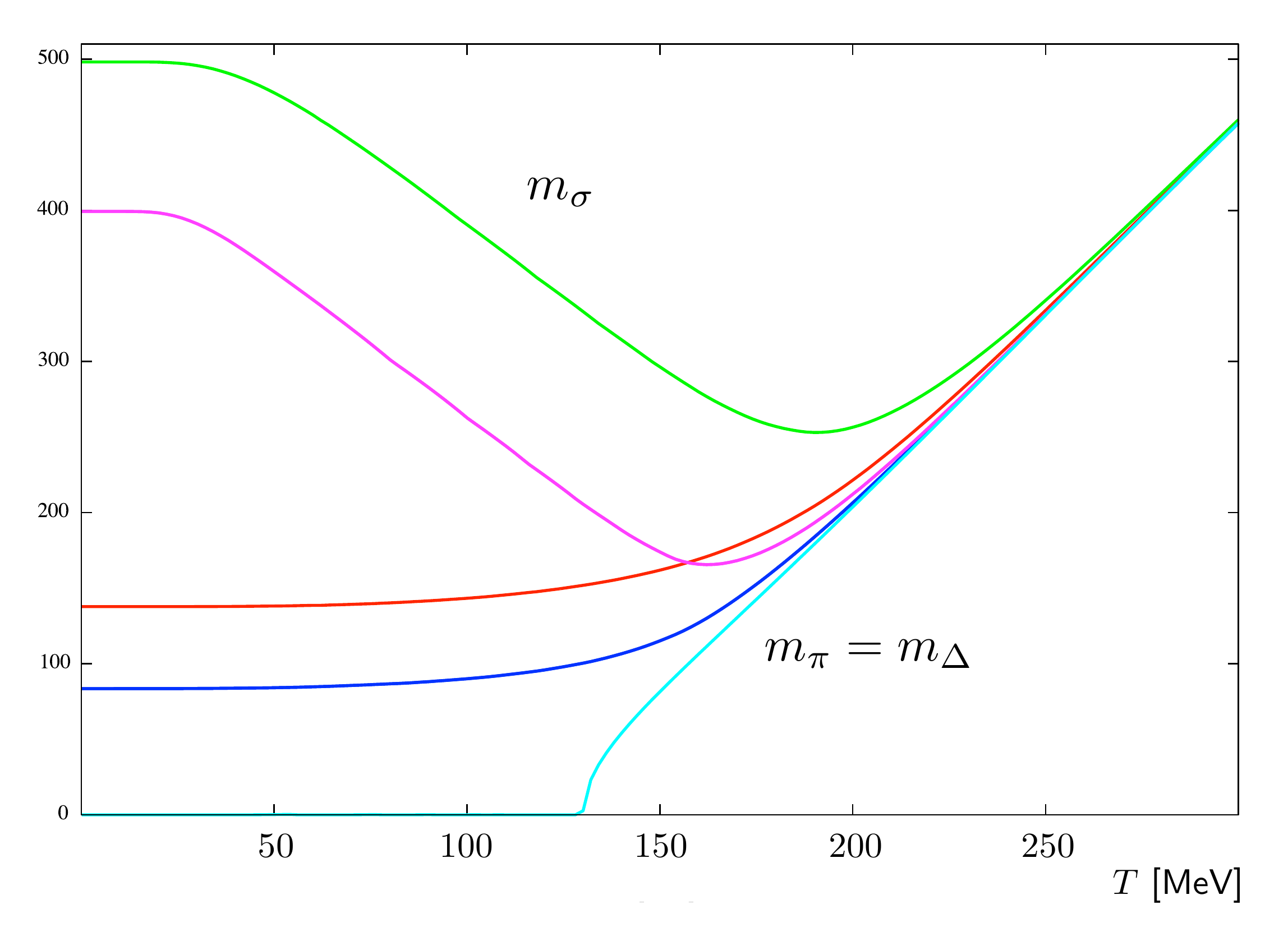}
\vspace{-.2cm}
\caption{The degenerate pion and diquark masses in comparison with
  that of the sigma meson over temperature $T$ from the SO$(6)$
  symmetric RG flow for the same model parameters as in
  Fig.~\ref{fig:finiteTscaling}.
\vspace{-.2cm} }
\label{fig:finiteTmasses}
\end{figure}

The corresponding temperature dependent screening masses as extracted
from the resulting effective potential also look qualitatively
exactly the same as in the QM model and are shown in
Fig.~\ref{fig:finiteTmasses} for the same set of zero-temperature  
pion masses as in Fig.~\ref{fig:finiteTscaling}.

\begin{figure}[t]
\centering
\includegraphics[width=\linewidth]{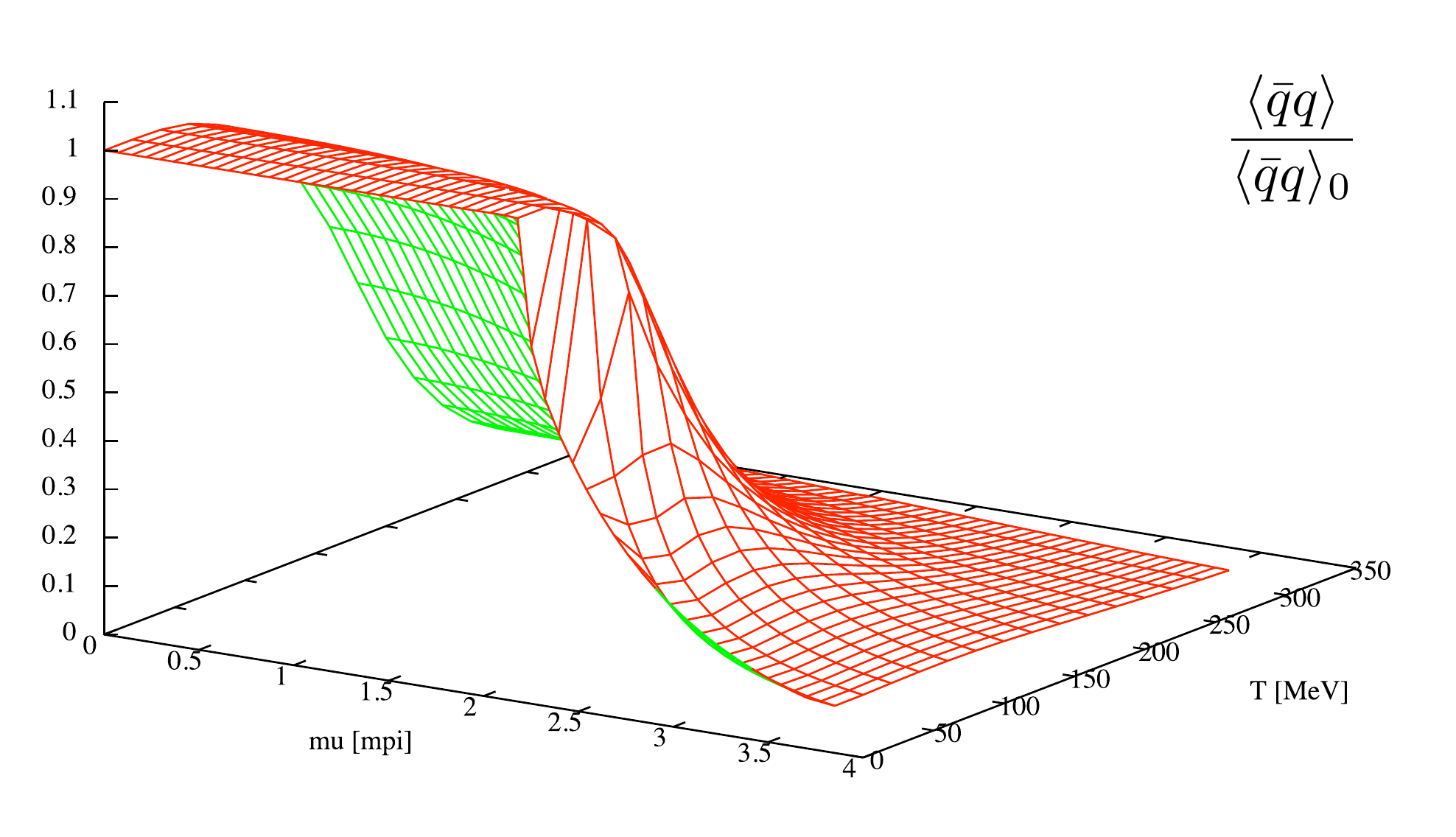}
\vspace{-.2cm}
\caption{The chiral condensate from the SO$(6)$ symmetric RG flow over 
  quark chemical potential and temperature. The $\mu=0 $ plane
  corresponds to the data shown in Fig.~\ref{fig:finiteTscaling} 
  for the $m_\pi = 138$ MeV pion mass.
\vspace{-.2cm} }
\label{fig:QCwoD}
\end{figure}

\begin{figure}[t]
\centering
\includegraphics[width=\linewidth]{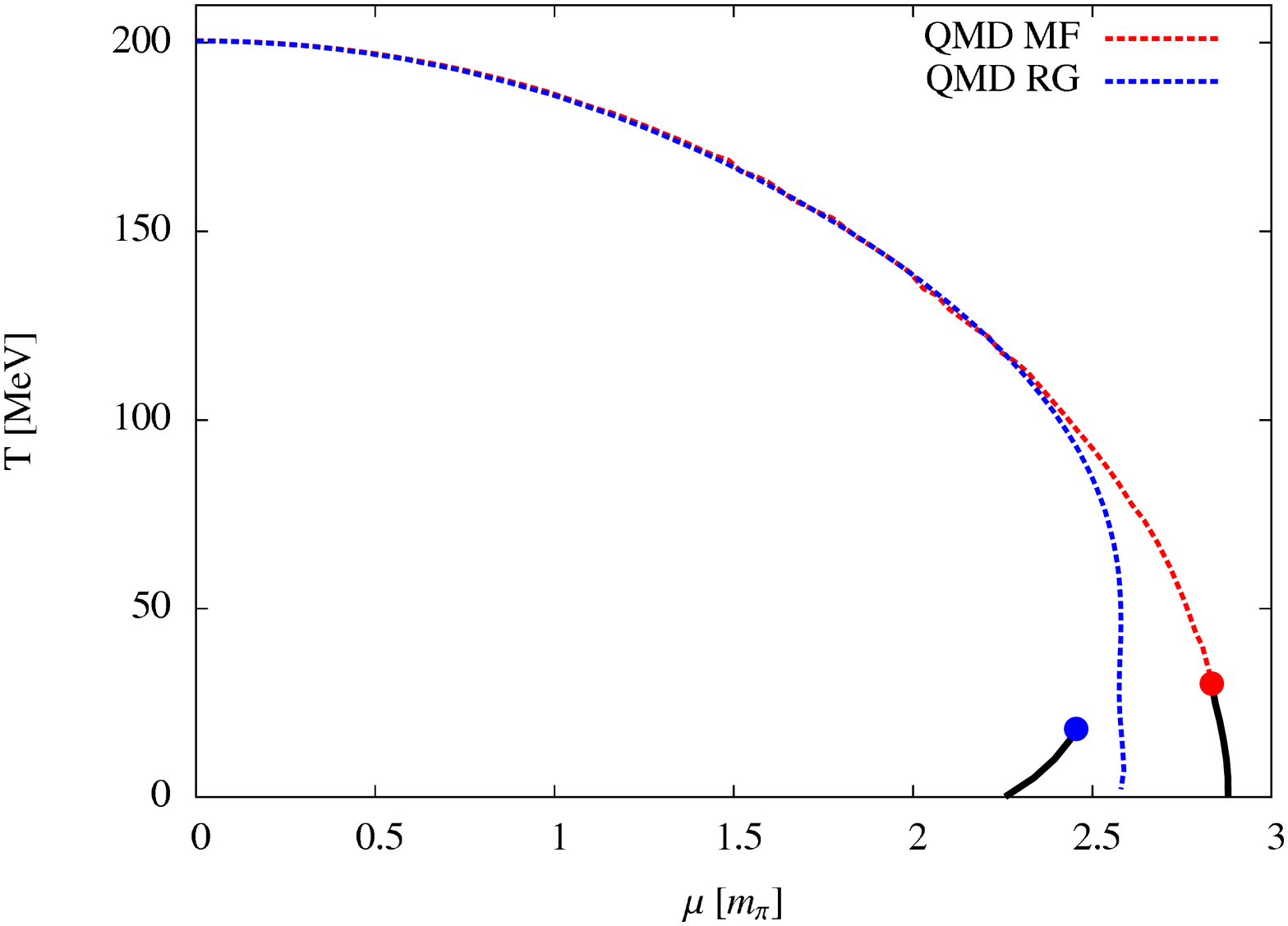}

\vspace{-.4cm}
\caption{QMD model phase diagram without diquark condensation and
  baryonic fluctuations. Shown are results from
  \cite{Strodthoff:2011tz} for the SO$(6)$
  symmetric flow (RG) and a typical mean-field diagram including
  vacuum contributions (MF).
}
\label{fig:pdd0}
\end{figure}

If we dismiss for the moment the effects of a finite
density of the bosonic baryons of \qctd, and the corresponding
fluctuations due to the collective baryonic excitations at finite density,
we may go ahead and solve the one-dimensional SO$(6)$ symmetric flow equation
in \Eq{eq:floweqdelta0} at finite quark chemical potential $\mu$ and
finite $T$. The resulting quark condensate is shown in the
three-dimensional plot in Fig.~\ref{fig:QCwoD}. The corresponding
phase diagram is shown in Fig.~\ref{fig:pdd0}.

The most important aspect of these results is that they are again
qualitatively exactly as in the quark-meson model for QCD. There is a
low-temperature $1^\mathrm{st}$-order transition line with an endpoint
at around $\mu \approx 2.5\,  m_\pi$, and a chiral crossover which is
here indicated by the dashed lines marking the half-value of the
vacuum condensate. As also known from studies of the quark-meson model
\cite{Schaefer:2004en}, the inclusion of fluctuations in the order
parameter to account for collective excitations with the FRG leads to
the capacity to describe  a low temperature phase transition to bound
quark matter. As shown in Figs.~\ref{fig:QCwoD} and \ref{fig:pdd0},
chiral symmetry is not fully restored but only jumps at the transition
and gradually decreases further into the quark-matter phase. This is
conceptually different from the restoration of chiral symmetry at the
first-order line in mean-field studies which would lead at the quark
level to the analogue of Lee-Wick matter as in the chiral Walecka
models \cite{Serot:1997xg}, but which is a mean-field artifact. 

Another nice feature of the SO$(6)$ symmetric FRG result at finite
$\mu $ is that the zero-temperature quark condensate remains precisely
constant until $\mu$ reaches the critical value for the transition to
bound quark matter as seen in Fig.~\ref{fig:QCwoD}. There is no
Silver-Blaze problem in the SO$(6)$ symmetric flow from
\Eq{eq:floweqdelta0}. The vacuum does not change at all until the
quantum phase transition to quark matter is reached.    

These results are all nice and well to illustrate general features of
quark-meson models, but they fail of course to describe an essential
part of the dynamics which are the effects of finite baryon density
and collective baryonic excitations. In \qctd this means to correctly 
describe the diquark condensation phase with superfluidity of the
bosonic baryons and the BEC-BCS crossover in the cold `{quarkionic}' 
two-color quantum gas.\footnote{The emphasis in `quarkionic' is on
  replacing the fermions in the ultracold fermionic gases by the colored quarks
which are unphysical. If it is on baryonic excitations as the physical
ones in confined quark matter, quarkyonic may be also appropriate.}    
 The advantage here as compared to the
three-color case is that we can understand these effects quite well
and study in detail what we need to do to include them. Moreover, it
is not only much more straightforward to include the baryonic degrees
of freedom in form of the diquarks of \qctd than the true fermionic
baryons in the real world with three colors, but it can also be
considered a valuable warm-up exercise for a successful quark-diquark
description of baryonic degrees of freedom in QCD.

\medskip

\subsection{Diquark condensation}

\medskip

At zero temperature, the vacuum and the mass spectrum must remain
independent of the (relativistic) chemical potential $\mu $ 
until the quantum phase transition is reached. To 
demonstrate that this phenomenological feature holds in the functional
integral representation of the grand potential has been called the 
Silver Blaze problem after the story by Arthur Conan Doyle in which
the dog did not bark when the horse named Silver Blaze was abducted
\cite{Cohen:2003kd}. This ``curious incident of the dog in the
night-time'' doing nothing led Sherlock Holmes to solve the case. 
The solution to the Silver Blaze problem for QCD with isospin chemical
potential \cite{Cohen:2003kd} equally holds for \qctd with baryon
chemical potential. However, the
relevant quantum phase transition in \qctd is not the QM model 
transition to quark matter, but that of diquark condensation at $\mu =
m_\pi/2$ with the spontaneous breaking of baryon number and baryon
superfluidity.    

Diquark condensation at $\mu = m_\pi/2 $ thus occurs way before any
quark matter or chiral transition at around $\mu \approx 2.5\,  m_\pi
$ as seen in the essentially purely mesonic model above. This is well
known from chiral effective field theory \cite{Kogut:2000ek}, lattice
simulations 
\cite{Hands:2000ei} and mean-field studies of the NJL
\cite{Ratti:2004ra} or the PNJL model \cite{Brauner:2009gu} as
mentioned at the beginning of this section. 
In order to describe it within the FRG one needs to solve the full
flow equation (\ref{eq:fullflowfinal}) with including the competing
fluctuations in the two directions of field space corresponding to the
chiral and diquark condensates, respectively, both at the same time.  
This was done numerically on a two-dimensional grid in field space in
Ref.~\cite{Strodthoff:2011tz}.

\begin{figure}[t]

\vspace*{-.2cm}

\hskip -.8cm
\includegraphics[width=1.16\linewidth]{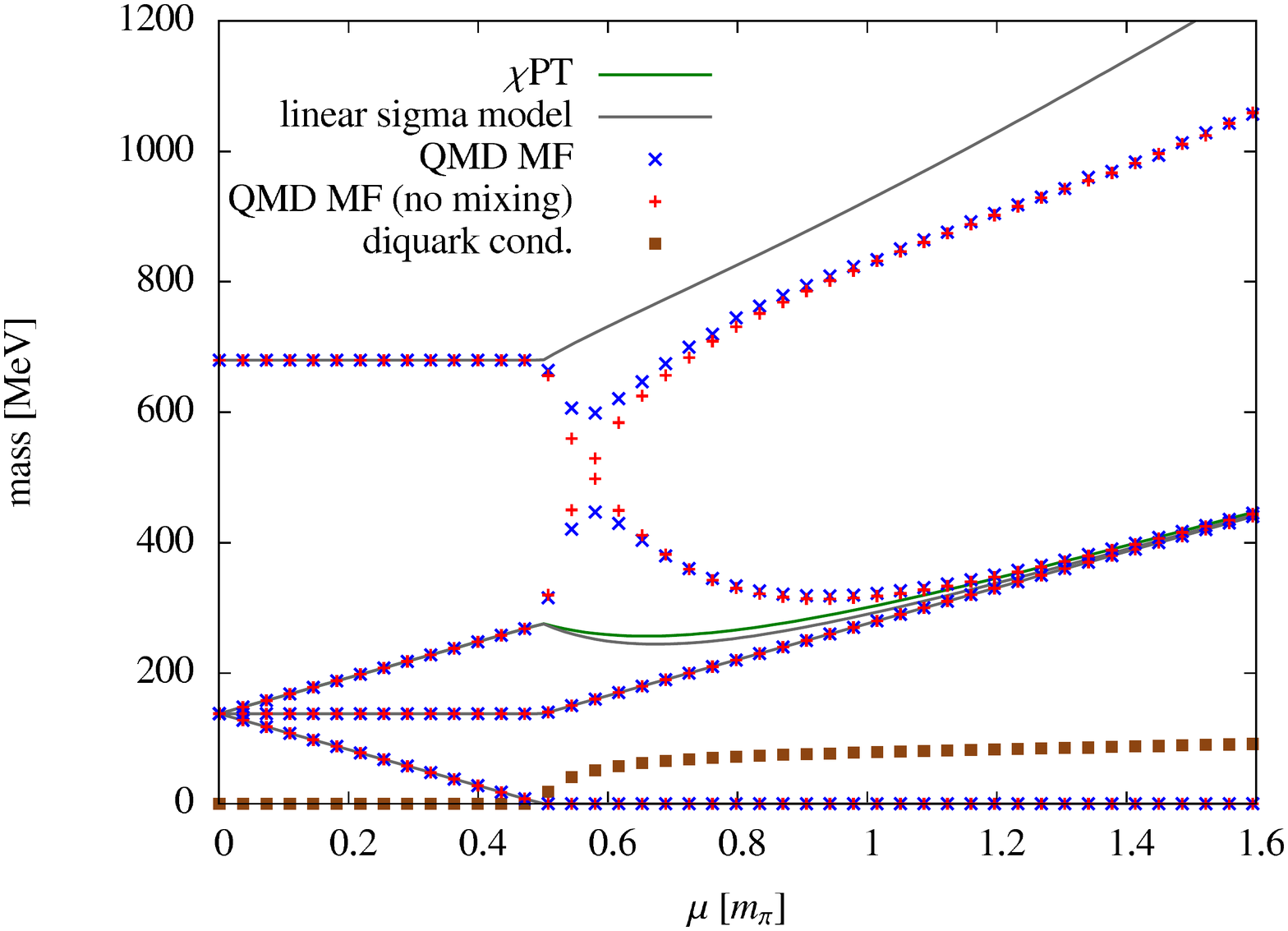}

\vspace{-.5cm}
\caption{Pole-mass spectrum at $T=0$: mean-field/RPA QMD model results
  from Ref.~\cite{Strodthoff:2011tz}.
  For comparison also shown: without the effect of diquark/sigma-meson
  mixing in the superfluid phase, linear sigma model and $\chi$PT results.  
\vspace{-.2cm} }
\label{fig:mfmass}
\end{figure}

Of course, it can also be described qualitatively already at mean-field level
in the QMD model \cite{Strodthoff:2011tz}. The resulting meson and
diquark mass spectrum is shown in Fig.~\ref{fig:mfmass}. Just as the
dichotomy between chiral symmetry breaking and the Goldstone theorem
is manifest within the mean-field/RPA framework or in the
rainbow/ladder truncation to Dyson-Schwinger equations, the Silver
Blaze problem is absent here as well. The meson masses at zero
temperature remain constant at finite $\mu$ below the transition. It
can be shown analytically that the diquark and antidiquark masses are  
given exactly by $m_{\Delta^\pm}=m_\pi \pm 2\mu$ there, and the
independence of the fermion determinant on $\mu $ for $\mu < 
m_\pi/2$ follows.

It is important to stress that at the mean-field level, this
exact feature of the theory holds only for the pole masses in the
meson and diquark propagators as obtained from the random-phase
approximation. This is true in the (P)QMD model as it is in the (P)NJL
model. The differences between the two are not significant and only
quantitative in nature in this approximation. The corresponding pole
masses are defined as the zeroes of the bosonic two-point function,
\begin{equation}
\label{eq:polecond}
\det \Gamma^{(2)}_B (p) = 0 \; , \;\;  \mbox{for} \;\; -p^2  = m_k^2
\; ,\;\;  k=1,\dots 6 \, .  
\end{equation}
Defining
\begin{equation} 
\Gamma^{(2)}_B(p)=\Gamma^{(2)}_{B,\text{tl}}(p) +\Pi_B(p) \, ,
\end{equation}
where $\Gamma^{(2)}_{B,\text{tl}}$  and $ \Pi_B $  are the tree-level
and vacuum-polarization contributions, respectively, and with the
bosonic potential $V(\vec\phi) $ from the linear-sigma model,
\Eq{eq:Vlinsig}, in the normal phase without diquark condensation the
pole-mass conditions (\ref{eq:polecond}) are then simply given by the
solutions of the following equations for $\omega$,  
\begin{equation} \label{eq:polnormal}
\begin{array}{rrl}
m_\sigma \, : & \omega^2 =& -m^2 +3\lambda\sigma^2 +
\Pi_\sigma(\omega,T) \; , \\[4pt]
m_\pi \, : & \omega^2 =& -m^2 +\lambda\sigma^2 + \Pi_\pi(\omega,T)\; , \\[4pt]
m_{\Delta^\pm} \, : & \; (\omega\pm 2\mu)^2 =& -m^2 +\lambda\sigma^2 +
\Pi_{\Delta^\pm}(\omega,T) \; . \\ 
\end{array} 
\end{equation}
Up to the chemical potential entries in $\Gamma^{(2)}_{B,\text{tl}}$
the bosonic two-point function is diagonal in the normal phase with
$d=0$, and we have introduced $\Pi_i(\omega,T)$ for the six diagonal
entries of $ \Pi_B(p) $ with $p=(-i\omega,\vec 0)$ at finite $T$.

As with chiral symmetry breaking and the occurrence of massless pions
in the chiral limit, one then readily verifies that the gap equation
for the diquark condensate along the bifurcation line $\mu_c(T)$,
which defines the onset of U$(1)_B$ breaking and diquark condensation
in the ($T,\mu$)-plane, leads to the condition, 
\begin{equation} 
\Big( - m^2 + \lambda\sigma^2  -4\mu^2 + \Pi_{\Delta^\pm}(0,T)
\Big)\Big|_{\mu=\mu_c(T)} = \, 0 \; ,
\end{equation}
because the corresponding derivative of the effective potential is
equal to $\Pi_{\Delta^+}(0,T)=\Pi_{\Delta^-}(0,T)$. This implies that a
solution with diquark-mass zero exists in \Eqs{eq:polnormal} along the
same line $\mu_c(T)$ in the ($T,\mu$)-plane.\footnote{This
  Goldstone mode persists everywhere into the diquark condensation
  phase but the simple pole-mass conditions (\ref{eq:polnormal}) are
  no-longer valid in this form there.}   

One can further verify analytically that $ \Pi_{\Delta^\pm}(\omega,T) =
\Pi_\pi(\omega,T) $ for $\mu = 0 $ as it must from SO$(5)$ symmetry, 
and, more significantly, that at zero temperature,
\begin{equation} 
 \Pi_{\Delta^\pm} (\omega,0) = \Pi_\pi(\omega\pm 2\mu,0)  \; .
\end{equation}
Therefore, the  $T=0$  solutions to \Eqs{eq:polnormal} obey
\begin{equation} 
 m_{\Delta^\pm} = m_\pi \pm 2\mu \; .  
\end{equation}

Together with the $\mu$-independence of the pion and sigma masses, 
this guarantees that there is no Silver Blaze problem at
mean-field/RPA level as seen in Fig.~\ref{fig:mfmass}, and that the
zero-temperature quantum phase transition occurs at $\mu= \mu_c(0)=
m_\pi/2$. 

In contrast, in quark-meson models one usually assigns the physical
masses of the mesons to the so-called screening masses to fix the
model parameters. These are obtained as the eigenvalues of the Hessian
of the effective potential in the leading-order derivative expansion.
They agree with the eigenvalues of 
\begin{equation} 
   \Gamma^{(2)}_B(0)=\Gamma^{(2)}_{B,\text{tl}}(0) +\Pi_B(0) \, ,
\end{equation}
and with the correct analyticity properties \cite{Das:1997gg,LeBellac}
in the continuation from discrete Matsubara to continuous real
frequencies this equivalence holds at all temperatures.   

However, it is straightforward to verify \cite{Strodthoff:2011tz} that the
screening masses have an enormous Silver Blaze problem. Their
behavior with chemical potential is utterly unphysical. That of the
sigma meson does not remain independent of $\mu$ at $T=0$ in the
normal phase as it should. The diquark and antidiquark screening
masses remain degenerate throughout the normal phase at all $T$, and
they both vanish at $\mu_c(T)$.  However,  $\mu_c(T)$ is not half
the screening mass of the pion but half its pole mass. One would have
thought that the difference is small, as we only have to extrapolate
from $p^2=0$ to $p^2=- m_\pi^2$ in the momentum dependence of the
two-point function to get from one to the other, but this is not the
case.   In the QMD model of \qctd, where the the quantum phase
transition at $\mu_c(0) = m_\pi/2$ {\em defines} the physical pion mass, one
observes that the corresponding screening mass is about 30\% too large
\cite{Strodthoff:2011tz}. And this is true both in mean-field/RPA and
with the effective potential obtained from the full flow equation
(\ref{eq:fullflowfinal}) with mesonic and baryonic fluctuations. One
thus needs to compute the pion pole-mass in those models to fix the
parameters to more realistic values. At the mean-field level an RPA
computation will do. A pole mass calculation from a flow equation for
the fully momentum dependent $\Gamma^{(2)}_{k,B}(p)$ in a truncation
consistent with that for the effective potential indeed shows that
this is much closer to $m_\pi\equiv 2\mu_c$ than the screening mass
from the effective potential in the FRG computation  as well
\cite{Strodthoff:2011tz}.

\begin{figure}[t]

\vspace*{-.2cm}

\hskip -.6cm
\includegraphics[width=1.1\linewidth]{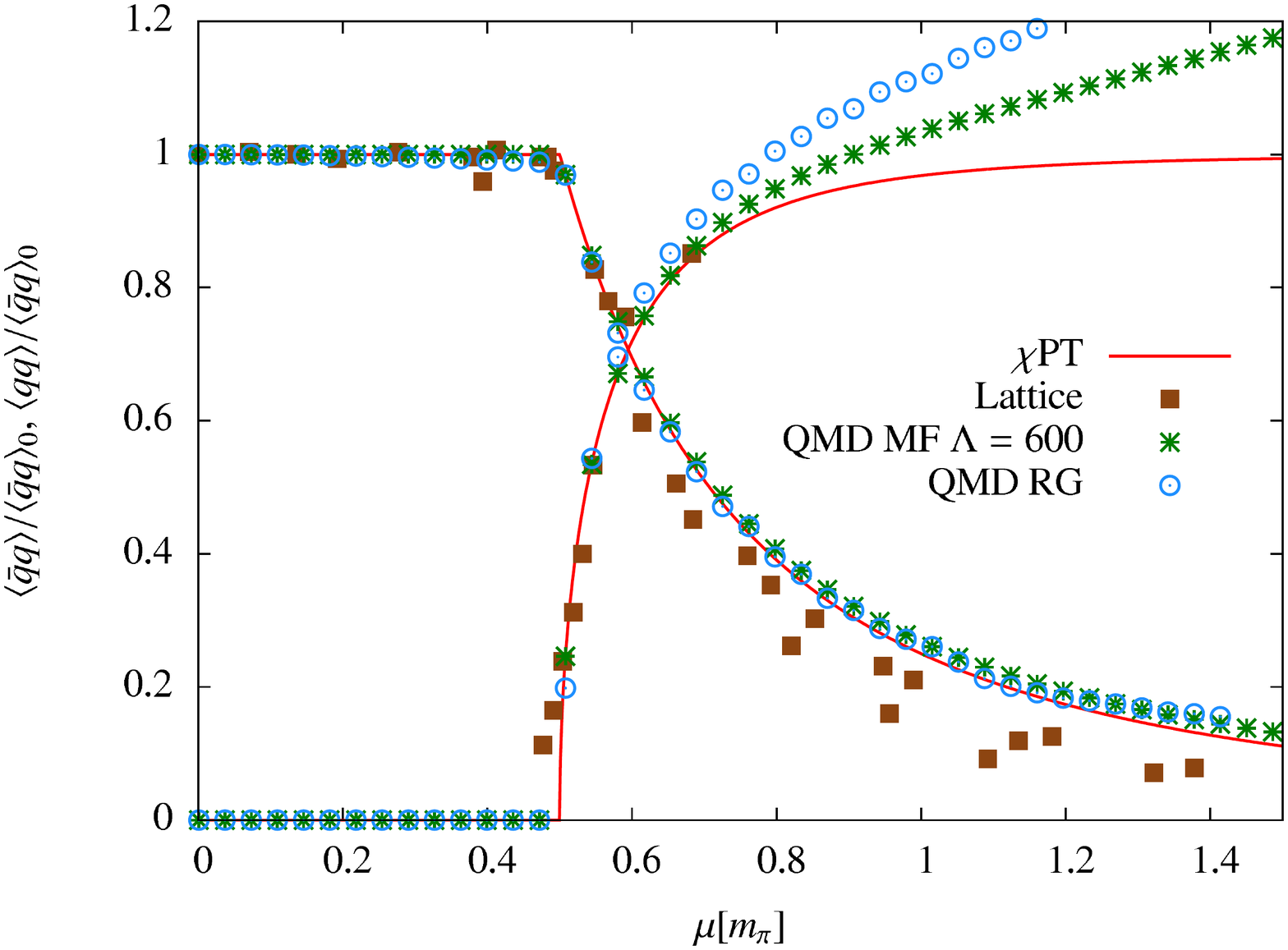}

\vspace{-.4cm}
\caption{Zero temperature condensates from full flow compared to 
  mean-field and $\chi$PT results, and lattice data (from
  \cite{Hands:2000ei}). 
\vspace{-.2cm} }
\label{fig:t0axisrg}
\end{figure}

In the diquark-condensation phase 
the sigma meson mixes with the two diquark modes, {\it
  i.e.}, the respective masses have to be determined from the zeroes
of the determinant of the corresponding $3\times 3$ submatrix in
$\Gamma^{(2)}_B$. As in the NJL model \cite{Brauner:2009gu,He:2005nk} one can
verify further exact results from the mass formulas at $T=0$. 
Also in the QMD model at mean-field/RPA level the in-medium pion pole-mass 
is equal to $m_\pi = 2\mu$ above the onset of diquark condensation 
at $2\mu = m_{\pi,0} $. Moreover, one verifies explicitly that 
one of the three modes in the diquark/sigma sector remains exactly
massless in the superfluid phase, also at finite temperature.
This is of course the Goldstone boson corresponding to the
spontaneously broken U$(1)_B$ baryon number.  
Another one becomes degenerate with the pions for large values of the
chemical potential, eventually, reflecting the restoration of chiral
symmetry.  They combine into an SO$(4)$ multiplet
as the chiral condensate vanishes for large $\mu$.  

Fig.~\ref{fig:mfmass} also shows the corresponding mass formulas from
the leading order chiral Lagrangian and the linear sigma model along with  
results of an RPA mass calculation where the mixing terms in the
sigma/diquark sector were neglected. In those results 
all off-diagonal terms in $\Gamma^{(2)}_B$ other than 
the chemical potential entries in $\Gamma^{(2)}_{B,\text{tl}}$ were
set to zero by hand to demonstrate that it is these terms which lead
to an avoided crossing in the sigma/diquark channel.  
In particular, without this mixing, the masses show more clearly that
it really is the original sigma which becomes degenerate with the pion
at high density as required by chiral symmetry. The transition marks
the BEC-BCS crossover in which the massive diquark mode changes from
its original would-be-Goldstone nature (shifted by $2\mu$) to the
heavy Higgs mode in the BCS limit at large $\mu $.

\begin{figure}[t]

\hskip -.4cm
\includegraphics[width=1.1\linewidth]{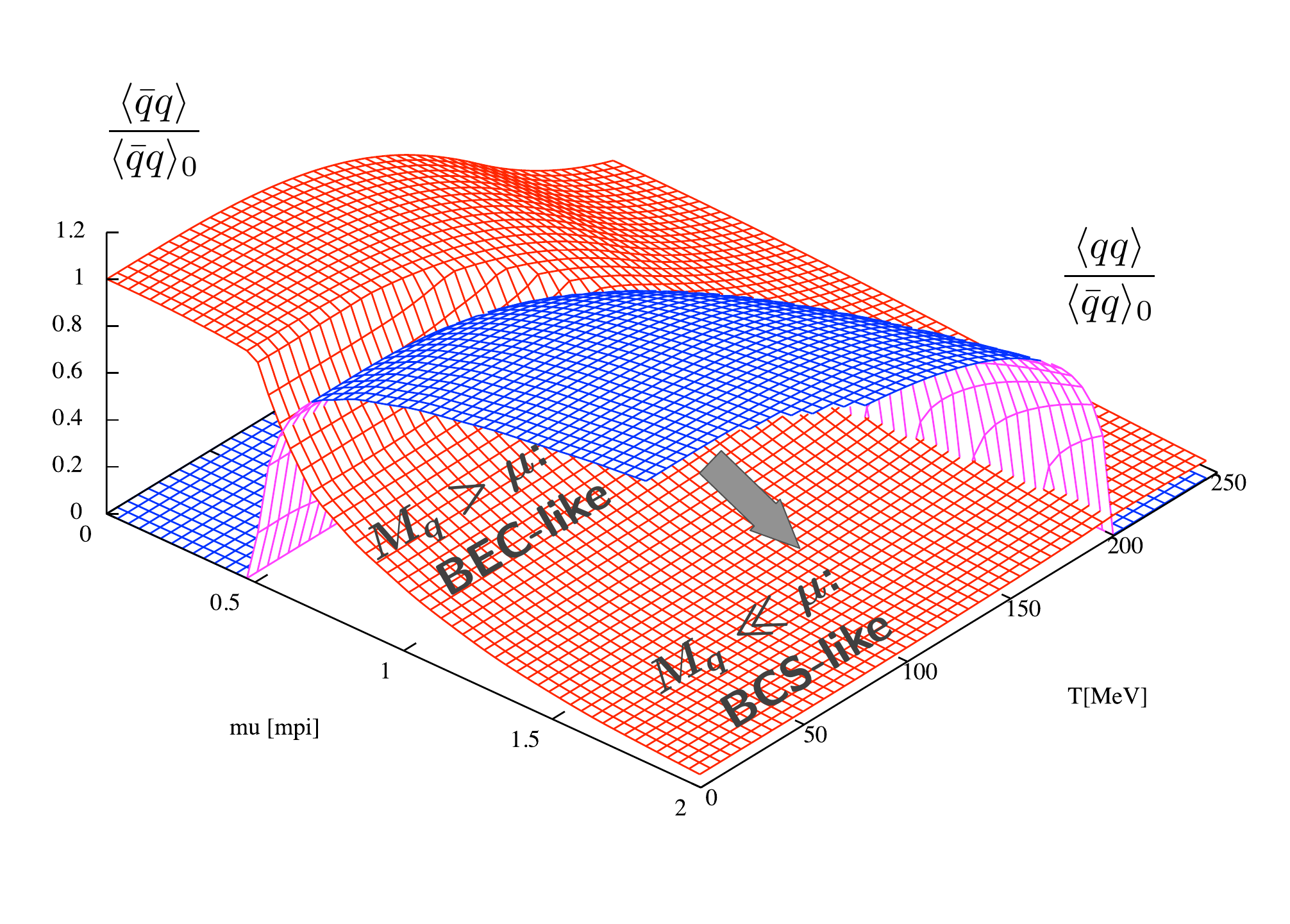}

\vspace{-.4cm}
\caption{The chiral and diquark condensates from the full 2$d$ RG flow over 
  quark chemical potential and temperature. The $\mu=0 $ plane
  corresponds to the data shown in Fig.~\ref{fig:finiteTscaling} 
  for the $m_\pi = 138$ MeV pion mass, the 2$d$ grid solution with
  explicit diquark fluctuations agrees with that of the SO$(6)$ symmetric flow
  as shown in Fig.~\ref{fig:QCwoD} there.  
\vspace{-.2cm} }
\label{fig:QaDqC}
\end{figure}

From the discussion above, it is clear that it is 
best to fix the pion mass to the zero temperature quantum phase
transition in the full FRG calculation of the effective potential from
\Eqs{eq:fullflowfinal}, (\ref{eq:bosonicflowfinal}) and
(\ref{eq:fermionicflowfinal}).   
A corresponding pole mass can be calculated from the flow of the
bosonic propagator but it is not needed once we know they must agree.
The resulting zero temperature chiral and diquark condensates are
shown in Fig.~\ref{fig:t0axisrg}. 
The linear sigma model expressions for
the $T=0$ condensates $\sigma = \langle \bar q q\rangle$ and  $\Delta
=  \langle q q\rangle$ are \cite{Andersen:2010vu}
\begin{equation}
\label{eq:nonlinsigmamodelcond}
\begin{split}
\frac{\sigma}{\sigma_0}&=\left\{
  \begin{array}{l l}
    1 & \quad \text{for $\mu<\mu_c$}\\
    \frac{1}{x^2} & \quad \text{for $\mu>\mu_c$}\\
  \end{array} \right.\\
  \frac{|\Delta|}{\sigma_0}&=\left\{
  \begin{array}{l l}
    0 & \quad \text{for $\mu<\mu_c$}\\
    \sqrt{1-\frac{1}{x^4}+2\frac{x^2-1}{y^2-1}} & \quad \text{for $\mu>\mu_c$}\\
  \end{array} \right.,
  \end{split}
\end{equation} 
where $x=2\mu/m_\pi$ and $y=m_{\sigma}/m_{\pi}$. The only difference
between these and the leading order $\chi$PT result
\cite{Kogut:2000ek}, {\it i.e.} in the non-linear sigma model, is the
$y$-dependent term in the diquark condensate which reduces to the   
$\chi$PT formula for $y \to\infty $. In this limit the vacuum
realignment from  $\langle \bar q q\rangle$-like to and  $\langle q
q\rangle$-like is described by a simple rotation with constant 
 $\langle \bar q q\rangle^2 +  \langle q q\rangle^2 $.

\begin{figure}[t]

\hskip -.5cm
\includegraphics[width=1.1\linewidth]{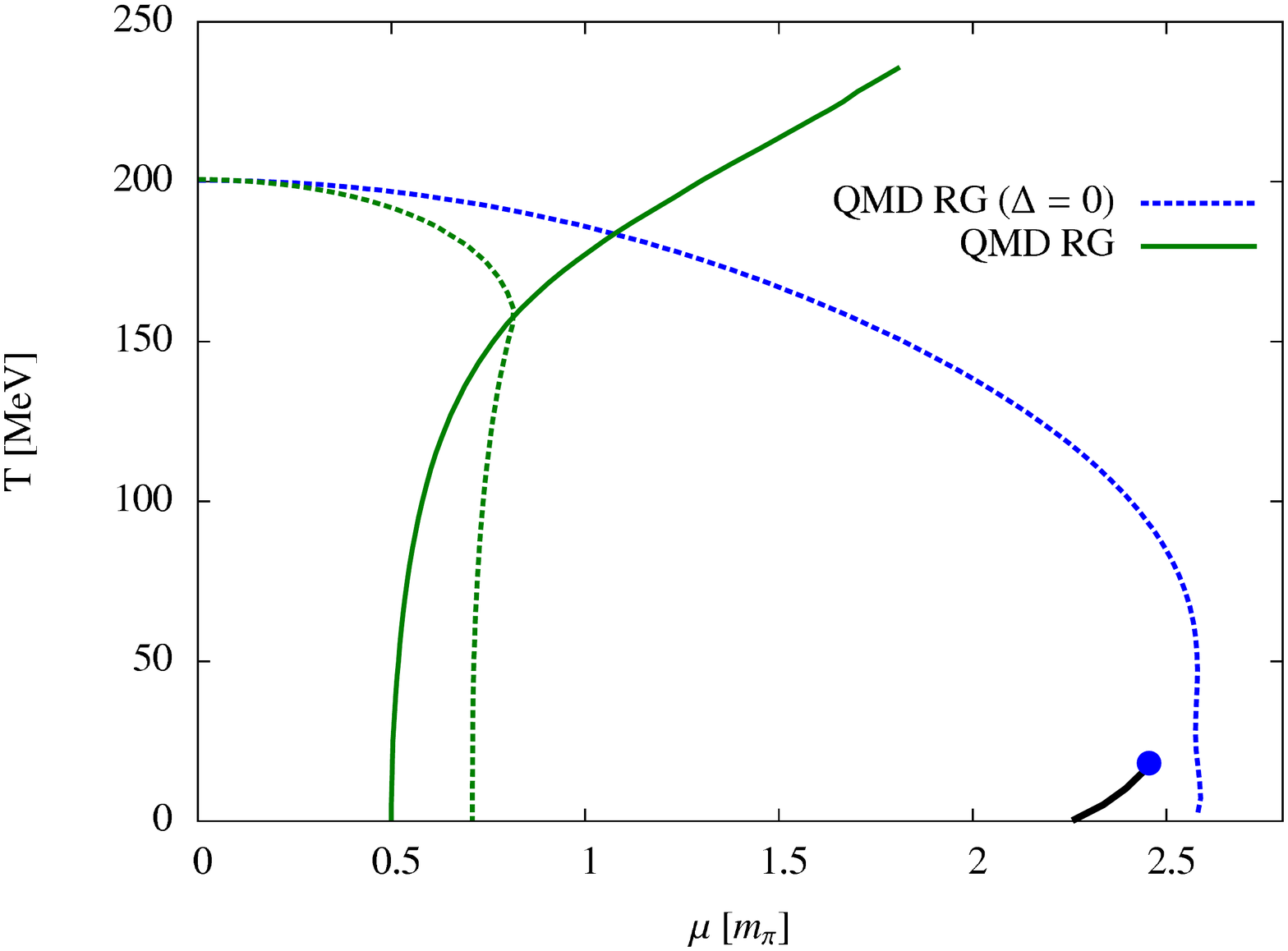}

\vspace{-.4cm}
\caption{Phase diagram from RG flow with collective baryonic
  fluctuations (and no chiral $1^\mathrm{st}$
  order transition/critical endpoint) compared to the SO$(6)$
  symmetric flow in Fig.~\ref{fig:pdd0} without diquark condensation.  
\vspace{-.2cm} 
}
\label{fig:pdrg}
\end{figure}

\medskip

\subsection{Phase diagram of the QMD model for \qctd}

\medskip

The solution to the full flow \Eq{eq:fullflowfinal}, with 
\Eqs{eq:bosonicflowfinal} and (\ref{eq:fermionicflowfinal}) 
on a two-dimensional grid in field space can be extended into the 
entire ($T,\mu$)-plane to demonstrate the effect of baryon density on
the chiral condensate in \qctd \cite{Strodthoff:2011tz}. For vanishing
gauge-field background, ${\Phi =1}$, the result is summarized in
Fig.~\ref{fig:QaDqC}.  This figure, from the full two-dimensional flow
for an effective potential with the reduced
$\mathrm{SO}(4)\times \mathrm{SO}(2)$ symmetry, should be compared to
Fig.~\ref{fig:QCwoD} from the flow for the SO$(6)$ symmetric
effective potential, to appreciate the influence of finite baryon density.

For small chemical potential in the normal phase the results agree
very well. Considering that the full solution does include independent
mesonic and baryonic fluctuations also in this regime, albeit the
latter around $\Delta = 0$, this is perhaps not quite so obvious as
one might think at first. It confirms, however, that a purely mesonic
model can produce reliable results so long as the baryon density
remains zero or sufficiently small, {\it i.e.}, the part of the phase
diagram relevant for the mesonic freeze-out. Those results are
unambiguously determined by the chiral symmetry breaking pattern, here in the
O$(6)$ universality class. Allowing additional interactions with
lower symmetry has no effect on the flow in this regime.
  
Once the quark-chemical potential approaches half the baryon mass,
corresponding to $m_B/N_c $, however, the rapidly increasing
baryon density equally rapidly suppresses the chiral condensate. With
the proper inclusion of the collective baryonic excitations, there is
no trace left of the chiral first-order transition and the critical
endpoint of the purely mesonic model. At least at zero temperature,
the baryon density is an order parameter for $N_f=N_c=2$ as well, and
it rapidly increases at the finite temperature transition line
separating the normal from the superfluid phase which should give rise
to the two-color analogue of the baryonic freeze-out.  

The corresponding phase
diagrams from the two RG solutions with and without finite baryon density
are compared in Fig.~\ref{fig:pdrg}. The dashed lines again denote the
half-value of the chiral condensate in the vacuum as an indication of
the chiral crossover line.

\begin{figure}[t]

\hskip -.4cm
\includegraphics[width=1.1\linewidth]{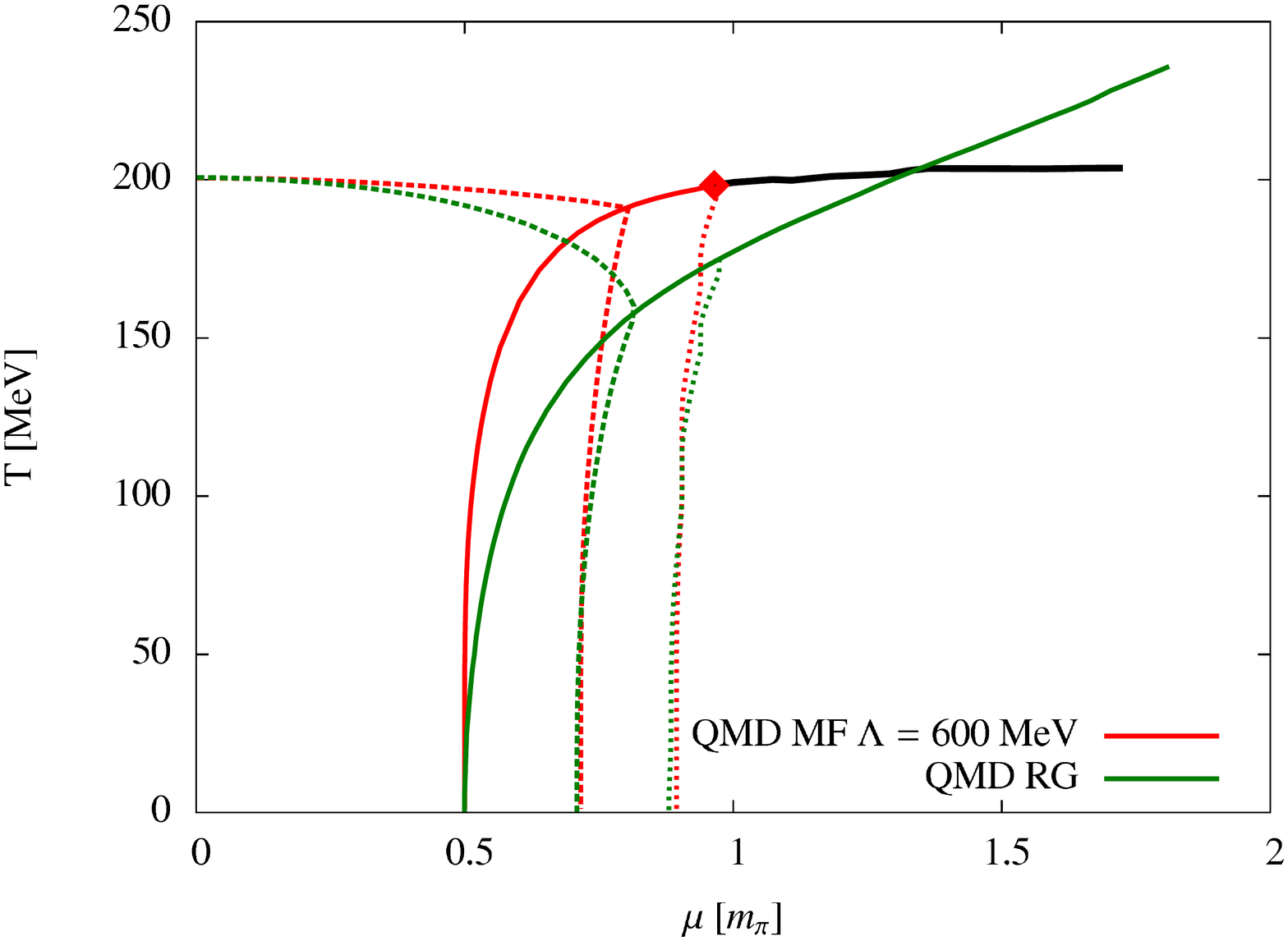}

\vspace{-.4cm}
\caption{Comparison of QMD phase diagrams from MF and RG calculations,
  including lines with $g\sigma = \mu$ in the superfluid phase to
  indicate the BEC-BCS crossover. 
\vspace{-.2cm} }
\label{fig:pdrg2}
\end{figure}

The onset of diquark condensation and superfluidity of these bosonic
baryons at low temperatures also marks the line at which the   
the residual SU$(2)_V$ flavor symmetry starts changing in nature
from the approximate Sp$(2)$ symmetry in the normal phase to
becoming the approximate  $\mathrm{SU}(2)_L\times \mathrm{SU}(2)_R$
quasi-restored chiral symmetry. Because they are both explicitly
broken and only approximate symmetries, this vacuum realignment is a
crossover. The quark mass with large chiral condensate in the 
normal phase starts out as the predominantly spontaneously generated
Dirac mass, and the bosonic baryons undergo Bose-Einstein condensation
in form of a dilute gas of strongly bound diquarks at the onset of diquark
superfluidity.  As their density increases, the underlying quark mass
rotates into a spontaneous Majorana mass leading to a BCS-like
pairing. This is the relativistic analogue in two-color QCD 
of the BEC-BCS crossover in ultracold fermionic quantum
gases. It is indicated in  Fig.~\ref{fig:pdrg2} by additional dashed
lines in the superfluid phase where the quarks' Dirac-mass equals
their chemical potential. For a more comprehensive
discussion within the NJL model, see Ref.~\cite{He:2010nb}.
   
In Figure \ref{fig:pdrg2} the phase diagram from the full RG solution
is compared to a QMD model mean-field result from
Ref.~\cite{Strodthoff:2011tz}. 
As temperature increases,
the line of the diquark-condensation
phase transition, which should be in the O$(2)$ universality class,
in the FRG solution with fluctuations differs more and more from
the mean-field result. The first-order transition line at high
temperatures in the mean-field approximation is washed out by the
fluctuations, and the associated tricritical point which was as also
predicted from next-to-leading order $\chi$PT \cite{Splittorff:2001fy}
turns out to be a mean-field artifact.

To summarize, the particular advantages of using two instead of the
usual three colors here were twofold:\\
--  to prepare for the  inclusion of diquark correlations 
and baryonic degrees of freedom in a covariant quark-diquark
description, {\it e.g.}, by a corresponding quark-meson-baryon model
for real QCD as a next step.\\    
-- to be able to test non-perturbative functional methods and models
against exact results and lattice simulations.  

The main results from the tests discussed in this section
include: the verification of O$(6)$ scaling at ${\mu =0}$; the
demonstration of the relevance of pole masses to correctly describe
the zero temperature quantum phase transition of two-color QCD in the
FRG framework, and the failure of the usual screening masses to be 
capable of that; and finally but most importantly, the non-existence
of a chiral first-order transition and critical endpoint at finite
baryon density.


\section{Universal aspects of deconfinement}

\begin{figure}[t]

\hskip .06cm
\includegraphics[width=\linewidth]{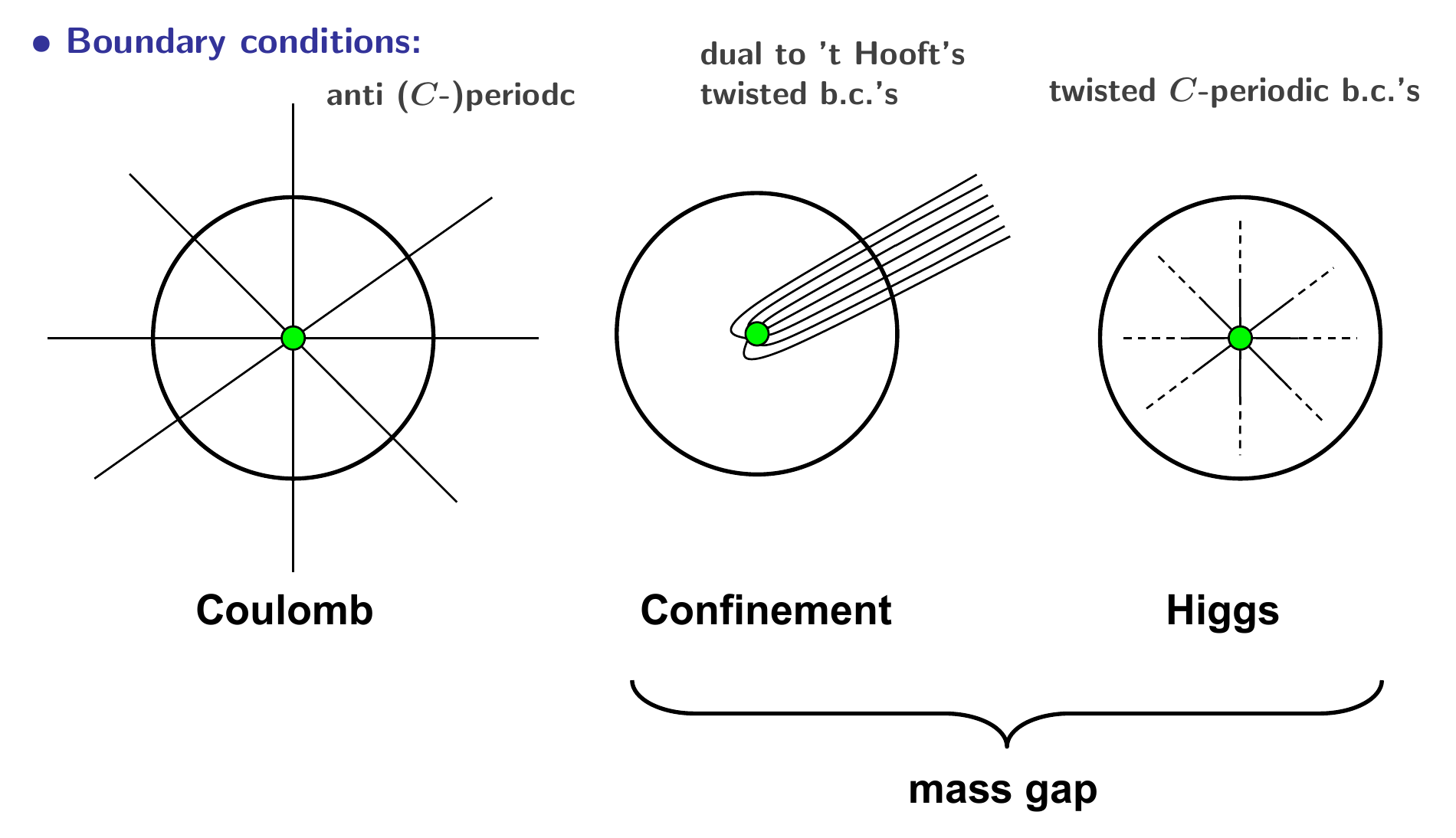}

\vspace{-.4cm}
\caption{The phases of gauge theories classified by the flux of a
  single static test charge alone in the world.   
\vspace{-.2cm} }
\label{fig:PhasesGT}
\end{figure}

The phases of gauge theories can be classified according to the
behavior of the flux of a static fundamental charge. In a
Coulomb phase, the electric flux spreads out and extends to infinity
because of Gauss' law. The Coulomb cloud of the charge is isotropic
and of long-range nature because the photon is massless. With a mass
gap, there are no such long-range Coulomb fields. Then the distinction
between Higgs and confinement phases depends on whether charge is
conserved or not. If it is, there is a kind of Gauss law and the flux
of the test charge must go somewhere. Because of the mass gap, it will
get squeezed into a string to minimize the cost of free energy. In a
Higgs phase, on the other hand, charge is not conserved, there is no
Gauss law, the condensate screens the charge and its flux lines peter
out. This rough classification is sketched in
Fig.~\ref{fig:PhasesGT}. In the language of local quantum field
theory, in terms of local field systems to measure 
the colored quantum numbers of the elementary degrees of freedom but
not necessarily with physical asymptotic states associated to these
fields (which would be called field-particle duality), this is
described by the Kugo-Ojima criterion. This criterion specifies the
same two conditions for confinement in the local field theory language:
(i) there must be a mass gap to avoid the analogue of the Coulomb cloud in
QED, but (ii) all global gauge charges must be conserved to have a Gauss
law and to avoid the Higgs mechanism.

When the charges of global gauge invariance are unbroken
one derives more generally that every gauge-invariant {\em localized} state
is a singlet under these unbroken global gauge charges. Thus, without (electric)
Higgs mechanism, QED and QCD have in common that any localized
physical state must be chargeless/colorless. The extension to {\em all}
physical states is possible only with a mass gap, however. Without
that, as in QED, non-local charged states which are gauge-invariant
can arise as limits of local ones which are not. The Hilbert space
decomposes into so-called superselection sectors of physical states
with different charges. A sector of (total) charge one differs from
the neutral sector by the presence of one charged particle, and the
energy difference between the ground states of both sectors defines
its mass. With a mass gap in QCD, on the other hand, color-electric
charge superselection sectors cannot arise: every gauge-invariant
state can be approximated by gauge-invariant localized ones (which
are colorless). One concludes that every gauge-invariant state must
also be a color singlet.  

To prepare a charged sector, one starts in a finite spatial volume
with suitable boundary conditions. The free energy difference between
the charged and the neutral sector then approaches a constant finite value,
zero, or it tends to infinity in the infinite volume limit, in
a Coulomb, a Higgs, or a confined phase, respectively.   

This is all well established and tested in pure gauge theories
with infinitely heavy test charges but without dynamical (color-)charged
particles. Beyond that, in particular at finite temperature it is
not so clear, however.
A mass gap can change gradually so the high temperature deconfined and the low
temperature confined phases can be analytically connected as in full
QCD with a crossover. But even at zero temperature where there should
be a clear distinction between differently charged superselection
sectors, there is not. The methods to fix the total color charge which will be
described below only work for the pure SU$(N)$ gauge theory without
dynamical quarks, at least when QCD is studied as an isolated theory. 
An alternative where these methods could be applied might be to go
beyond that and, in particular, include the quarks' fractional electric
charges as in the Standard Model. This possibility will be discussed
briefly in Subsection \ref{sec:fraccharge}.

In QCD alone, a non-zero total fundamental color charge cannot 
be screened completely without Higgs mechanism when dynamical quarks
are present either, but with string breaking 
it should become a boundary effect and the energy difference
to the neutral sector should reduce to the corresponding 
string-breaking scale in the thermodynamic limit. Moreover, one might
expect that the Kugo-Ojima criterion should predict symmetry breaking
and thus a phase boundary between confined and Higgs phases, for
example, but there are counter examples on the lattice for that as
well \cite{Caudy:2007sf}.       

In pure gauge theories where the methods to fix differently charged
sectors work well, we often also have a duality between magnetic and
electric sectors. For example, antiperiodic (spatial) boundary
conditions in the compact U$(1)$ lattice gauge theory, where they
are equivalent to $C$-periodic ones, can be used to prepare magnetic
sectors and compute the corresponding monopole mass
\cite{Polley:1990tf} to distinguish its Coulomb and confined phases.  
The difference in free energy of the antiperiodic
versus the periodic ensemble thereby tends to zero or a finite value 
for the confining (magnetic Higgs) or Coulomb phases, respectively. 
By duality, via the $\mathds Z$ gauge theory with integer valued link 
variables, the (electric) Higgs vs.~Coulomb phases of the
non-compact Abelian Higgs model follow an analogous pattern.

In SU$(N)$ gauge theories the analogue of antiperiodic boundary
conditions, which would not in general be gauge invariant there, are
't~Hooft's twisted boundary conditions to fix the total $Z_N$
center-flux through the planes of the spacetime torus by fixing
the total number of center vortices mod $N$ through these planes
\cite{'tHooft:1979uj}. 
Dual to those are the corresponding electric flux sectors which
describe the free energy $F_q(T,L)$ of the flux of a static
fundamental charge in a $1/T\times L^3$ box 
with boundary conditions to mimic the presence of a mirror charge in
the neighboring volume along the direction of the flux. This is a bit
more complicated than the simple anti/$C$-periodic boundary conditions
in the Abelian case but it has the analogous interpretation in terms
of mirror charges. It is reviewed in Subsec.~\ref{sec:tbc}.
Moreover, combinations of $C$-periodic and twisted
boundary conditions can be used to calculate the mass of the
't~Hooft-Polyakov monopole in SU$(N)$ with adjoint Higgs at least for
even $N$ \cite{Edwards:2009bw}.       

\medskip

\subsection{Center vortices and spin interfaces}

\medskip

The finite temperature deconfinement transition in SU$(N)$ gauge
theories in $d+1$ dimensions is very well understood in terms of the
spontaneous breakdown of their global $Z_N$ center symmetry
\cite{Greensite:2003bk}.  This symmetry is faithfully represented by
the fundamental Polyakov loops  $P(\vec x)$ which live on the $d$
spatial dimensions and describe static fundamental charges. Under the
global $Z_N$ center symmetry they transform 
like spins $s_i$ in a $d$ dimensional $q$-state Potts
model with $q = N$ and Hamiltonian \cite{Wu:1982ra},      
\begin{equation} 
\mathcal H = - J\sum_{\langle i,j\rangle} \delta_{s_i,s_j} - H\sum_i
\delta_{s_i,0} \;, \;\; s_i = 0,1,\dots q-1\;,
\label{PottsH}
\end{equation}
with nearest neighbor coupling $J$. A non-zero external field $H$,
inversely related to the quark mass $m_q$, may be included to mimic the leading
effect of heavy dynamical quarks. When $1/m_q =0$, the Polyakov loop
develops a non-zero expectation value only in the deconfined, $Z_N$-broken
phase, while the expectation value of $P(\vec x) $ vanishes in the
disordered, confined phase much like the spontaneous magnetization in
the spin model.

\begin{figure}[h]
\hskip .4cm
\includegraphics[width=0.8\linewidth]{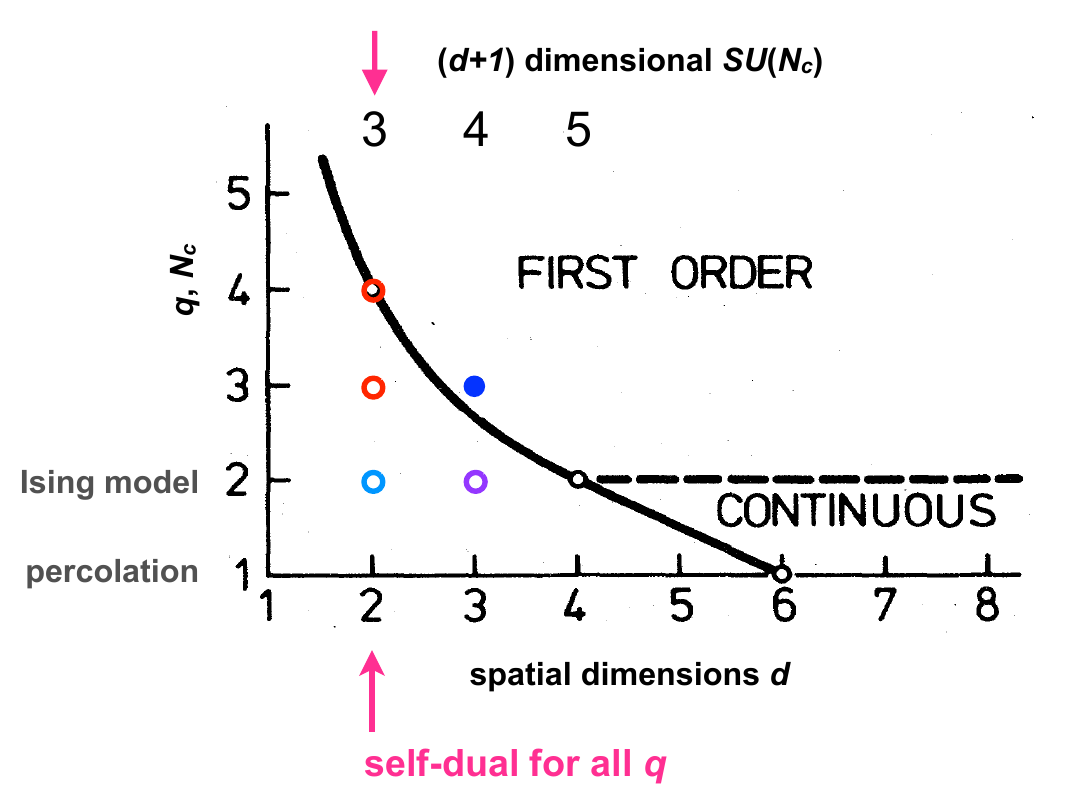}
\vspace{-.2cm}
\caption{Separation line between $1^\mathrm{st}$ and $2^\mathrm{nd}$
  order phase  transitions in the $q$-state Potts models, from
  \cite{Wu:1982ra}, and pure SU$(N)$ gauge theories.
\vspace{-.4cm} }
\label{fig:PottsSep}
\end{figure}

This is well described in terms of spacelike center
vortices which play the role of spin interfaces. They separate regions
where the Polyakov loop differs by a phase $z\in Z_N$, and their
proliferation disorders the Polyakov loop and leads to confinement.
The suppression of spatial center vortices at high temperatures 
coincides with the ordering of the Polyakov loop, and
their free energy offers an elegant order parameter for the transition 
\cite{DeForcrand:2001dp}.   

In order to be able to apply the powerful tools of universality and
scaling near a critical point, here we are particularly interested in
cases where this transition is of $2^\mathrm{nd}$ order. For pure QCD
with $N=3$ colors in $3+1$ dimensions this is of course not the
case. The transition is first order, but only just. If we either
reduce the number of colors to $N=2$ or the dimensions to $2+1$, 
or both, we obtain Yang-Mills theories with a second order deconfinement
transition within the universality class of $q$-state Potts models
with $q=2$ (Ising) in $d=3$ dimensions or with $q=2$ and $3$ in
$d=2$ \cite{Svetitsky:1982gs}. The $q=4$ Potts model in 2 dimensions
is interesting because it is known  \cite{Baxter:2000ez} to have a
$2^\mathrm{nd}$ order transition {\em and} to fall precisely on the
separatrix $q_c(d)$ with respect to the order of the transition in the
$(q,d)$-plane as shown in Fig.~\ref{fig:PottsSep}, where
only $1^\mathrm{st} $ order transitions occur for $q> q_c(d) $, {\it
  i.e.}, $q_c(2)=4$. The corresponding SU$(4)$ gauge theory in 2+1
dimensions has been studied for example in
\cite{deForcrand:2003wa,Holland:2007ar,Liddle:2008kk}. The conclusion
in \cite{Holland:2007ar} was that the transition is weakly
$1^\mathrm{st}$ order, unlike the $q=4$ Potts case. It nevertheless
seems that there is a rather wide range of intermediate length scales
where at least approximate Potts scaling can be observed
\cite{deForcrand:2003wa,Strodthoff:2010dz}. One might
then like to understand, for example, why among the wider class of
$Z_4$-symmetric Ashkin-Teller models with continuously varying
critical exponents, it is the standard $q=4$ Potts scaling that is
relevant here, and whether this can be derived from an effective
Polyakov-loop model analogous to the known cases with $N=2$ and $3$,
along the lines of Refs.~\cite{Heinzl:2005xv,Wozar:2007tz,Langelage:2010yr}.  

One general property of the 2$d$ Potts
models is that they are selfdual for all $q$. We will see
below, how this selfduality is reflected in a duality between the
spacelike center vortices and the confining electric fluxes of the
gauge theory.

In the pure SU$(N)$ gauge theories 't Hooft's
twisted boundary conditions fix the total $Z_N$-valued center flux,
{\it i.e.}, the total number modulo $N$ of center vortices in
the various planes of the Euclidean spacetime box \cite{'tHooft:1979uj}. 
At finite temperature $T$, the free energy differences of
these twisted ensembles and the periodic one define the corresponding
center-vortex free energies. If they tend to zero in the thermodynamic
limit, then these vortices condense which leads to an area law for those
Wilson loops that feel their disordering phases in the corresponding
plane of the twist.  
In a $ 1/T \times L^d$ box, we have to distinguish between temporal
and magnetic twist. The latter is defined in purely spatial planes and
corresponds to the  $Z_N$-valued magnetic flux $\vec m$ of static center
monopoles, with the direction of $\vec m $ perpendicular to the
spatial plane of the twist. Their free energy tends to zero with
$L\to\infty$ as $\exp\{-\sigma_s(T) L^2 \}$ at all $T$
\cite{vonSmekal:2002gg}, corresponding to the area law for spatial
Wilson loops with spatial string tension $\sigma_s(T)$.   

\begin{figure}[t]

\vspace*{.1cm}
\centerline{\includegraphics[width=\linewidth]{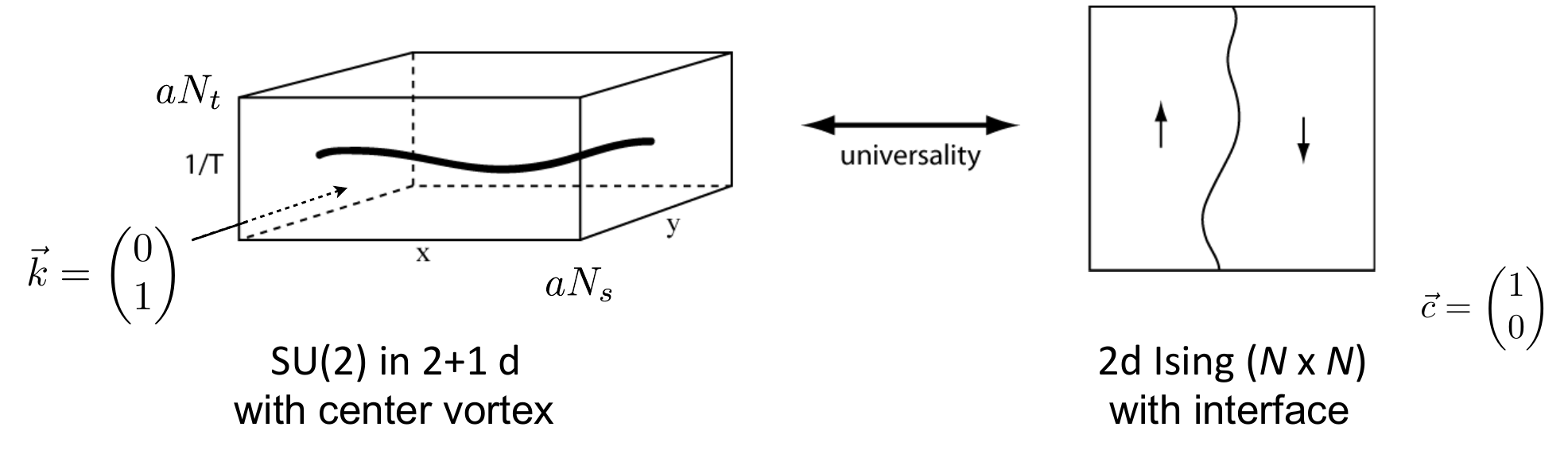}}
\vspace{-.2cm}
\caption{Spatial center vortices share their universal behavior with
  interfaces in a spin model. Polyakov loops on either side of the
  vortex differ by the corresponding center phase. A thin
  vortex in SU$(2)$ thus acts like an interface with antiferromagnetic
  bonds in the Ising model.  \vspace*{-.6cm}} 
\label{fig:VortexInterface}
\end{figure}

Relevant for the deconfinement transition are only the
temporal twists. These are characterized by $d$-dimensional vectors
$\vec k$ of integers $\bmod\ N$, {\it i.e.}, with components 
$k_i \in Z_N$ representing the twist in the temporal plane of orientation
$(0,i)$, with total center flux $\exp( 2\pi i\,
k_i/N)$ through that plane. See Fig.~\ref{fig:VortexInterface} for an
illustration of such a vortex ensemble in SU$(2)$.  We denote the 
partition functions of the various ensembles with temporally twisted boundary
conditions by $Z_k(\vec k)$. The corresponding center-vortex free
energies per $T$ are then given by  
\begin{align} 
F_k(\vec k) \equiv -\ln R_k(\vec k) \equiv&
 - \ln\big(Z_k(\vec k)/Z_k(0)\big) \; ,\\
  & \quad k_i = 0,1,\dots N-1\; , \notag
\end{align}
where $Z_k(0)$ stands for the periodic ensemble. 

Intuitively, center vortices can lower their free energy by spreading out.
As temperature is increased, however, the temporal ones can no-longer
spread arbitrarily but get squeezed more and more until the phase
transition is reached above which they are completely suppressed. In
the vicinity of a second order deconfinement transition these vortex
free energies show  the universal behavior of interfaces in the
respective $d$-dimensional Potts model. Interfaces in the spin models
are typically introduced as frustrations along which the coupling of
adjacent spins favors cyclically shifted spin states rather than
parallel ones for the usual ferromagnetic couplings $J>0$. They form
$d-1$ dimensional surfaces dual to links at which the 
$\delta_{s_i,s_j} $ in Eq.~(\ref{PottsH}) are replaced by 
$\delta_{s_i,s_j+ m \bmod q} $ and are conveniently studied by
introducing analogous cyclically shifted boundary conditions,
\begin{equation}
s_{\vec x + \vec e_i L} = s_{\vec x} + c_i \bmod q \;
,\;\mbox{with}\;\;  c_i =  0,\, 1,\,  \dots q-1 \; .
\end{equation}
Here, the interface free energies per temperature,
\begin{equation}
F_I(\vec c) \equiv -\ln R_q(\vec c) \equiv - \ln\big(Z_q(\vec
c)/Z_q(0)\big) \; , 
\end{equation} 
are obtained from ratios $R_q(\vec c)$ of Potts model partition
functions $Z_q(\vec c)$ with cyclically shifted boundary conditions
labeled by $\vec c$ over the periodic one, $Z_q(0)$.
Interfaces are suppressed in the low temperature ordered phase, and
these ratios  $R_q(\vec c)$  tend to zero in the thermodynamic limit.
Interfaces in the spin model below $T_c$ thus behave as the temporal center
vortices in the ordered, deconfined phase of SU$(N)$ above
$T_c$. Complementary to that, in the disordered(confined) phase
above(below) $T_c$, it is the interface(vortex) {\em free energies} that
tend to zero such that the ratios $R_q$, $R_k $ approach 1  for all
boundary conditions. Only at the critical temperature  $T=T_c $ do
these ratios converge to non-trivial and universal values.

In two dimensions these universal numbers,  
\begin{equation}
0< R_{q,c}^{(m,n)} < 1 \; , \;\;  R_{q,c}^{(m,n)} \equiv R_q(\vec
c)\big|_{T_c}\; ,   
\end{equation}
for all cyclic boundary
conditions $\vec c = (m,n)$ and all Potts models with $2^\mathrm{nd}$ 
order transitions, {\it i.e.}, for $q=2$, $3$ and $4$,
follow from the exact expressions of the corresponding partition
functions at criticality obtained in Ref.~\cite{Park:1988}. In 
terms of Jacobi elliptic theta functions,
\begin{equation}
\vartheta_3(z) \, = \, \sum_{n=-\infty}^\infty e^{-\pi n^2 z} \; ,
\end{equation}
and with the definitions (for square lattices),
\begin{align} 
  T_q(m,n) \, &\equiv \, \frac{\sqrt{x_q}}{\mathcal N} \, 
           \vartheta_3(x_q m^2) \, \vartheta_3(x_q n^2) \; , \;\;
           \mbox{where}  \\
  x_q \, &=\,  1-\frac{\arccos(\sqrt{q}/2)}{\pi}  \,=\, \Bigg\{ \frac{3}{4},\,
  \frac{5}{6},\, 1\Bigg\} \; , \notag \\
& \hskip -.4cm \mbox{for} \;\; q = \{2,\,3,\, 4\}\; ,
\;\mbox{and}  
\;\; \mathcal N = \Gamma[1/4]^2/(2 \pi^{3/2})\; , \notag
\end{align} 
the partition functions of the $q$-state Potts models with different
cyclic boundary conditions at criticality can
be expressed \cite{Park:1988}, for the Ising model with
$q=2$ as, 
\begin{align}
Z_{2,c}^{(0,0)} &= 4 T_2(4,4) - T_2(1,1) \; , \label{eq:crit2}\\ 
Z_{2,c}^{(1,0)} &= Z_{2,c}^{(0,1)} = 
        Z_{2,c}^{(0,0)} + 2 T_2(2,1) - 4 T_2(4,2) \; ,\notag \\ 
Z_{2,c}^{(1,1)} &= Z_{2,c}^{(0,0)} - 2 Z_{2,c}^{(1,0)} \; , \notag
\end{align}
for the $q=3$ state Potts model as,
\begin{align}
Z_{3,c}^{(0,0)} &= 6 T_3(6,6) - 3 T_3(3,3) + 2 T_3(2,2) - T_3(1,1)  \;
, \notag\\ 
Z_{3,c}^{(1,0)} &=   Z_{3,c}^{(0,1)} =  Z_{3,c}^{(2,0)} =
Z_{3,c}^{(0,2)}  \label{eq:crit3} \\ 
               &= Z_{3,c}^{(0,0)} + 3 T_3(3,1) - 6 T_3(6,2) \; ,\notag \\ 
Z_{3,c}^{(1,1)} &=  Z_{3,c}^{(1,2)} = Z_{3,c}^{(2,1)} = Z_{3,c}^{(2,2)}  \notag \\ 
&= Z_{3,c}^{(0,0)}/2 -  Z_{3,c}^{(1,0)} \; ,   \notag
\end{align}
and the distinct ones for $q=4$ as
\begin{align}
Z_{4,c}^{(0,0)} &= 6 T_4(2,2) - T_4(1,1) \; , \label{eq:crit4} \\ 
Z_{4,c}^{(1,0)} &= 2 T_4(2,2) - T_4(1,1) + 4 T_4(4,1) - 4 T_4(4,2) \; ,\notag \\ 
Z_{4,c}^{(1,1)} &= 4 T_4(2,2) + T_4(1,1) - 4 T_4(2,1) \; ,\notag \\ 
Z_{4,c}^{(2,0)} &= - 2  T_4(2,2) - T_4(1,1) + 4 T_4(2,1) \; ,\notag \\ 
Z_{4,c}^{(2,1)} &= Z_{4,c}^{(2,0)}
                    - 4 T_4(4,1) + 4 T_4(4,2) \; ,\notag \\ 
Z_{4,c}^{(2,2)} &= 6 T_4(2,2) + 3 T_4(1,1) - 8 T_4(2,1)\; . \notag
\end{align}
Generally, the partition functions obey $Z_{q}^{(n,m)} =
Z_{q}^{(m,n)}$ and  $Z_{q}^{(q-n,m)} = Z_{q}^{(n,m)}$.
The ratios $R_{q,c}^{(m,n)} = Z_{q,c}^{(m,n)}/Z_{q,c}^{(0,0)} $
readily follow.

These universal ratios can be used to determine the critical couplings
$\beta_c$ of the transition in the 2+1 dimensional gauge theories with
high precision \cite{Edwards:2009qw}, by requiring that $\beta =
\beta_c$ at $R_k = R_{q,c}$ with $q=N$ and the corresponding boundary
conditions. For SU$(2)$ in $2+1$ dimensions, for example, 
where,
\vspace{-.2cm}
\begin{equation}
   R_{2,c}^{(1,0)} = \frac{1}{2^{3/4}+1}\; , \;\; \mbox{and} \;\; 
R_{2,c}^{(1,1)} = \frac{2^{3/4}-1}{2^{3/4}+1}\; ,
\end{equation}
from the 2$d$ square Ising model, this was used in
\cite{Edwards:2009qw} to determine the 
critical couplings of the deconfinement transition 
for lattices with up to $N_t = 9$ links in the Euclidean time
direction with a precision  typically two orders of magnitude better
than previous literature values where they were available
\cite{Teper:1993gp,Engels:1996dz,Liddle:2008kk}.  
In particular, the high precision allows to reliably determine the
subleading $1/N_t$-corrections to the linearly increasing behavior of
$\beta_c$ with $N_t$ near the continuum limit in $2+1$ dimensions,
\begin{align}
\label{eq:betavsNt}
\beta_c(N_t) = 
1.5028(21)\, N_{t}\, +& \,  0.705(21)\\ 
& \hskip .2cm -0.718(49)\, \frac{1}{N_{t}}  + \cdots \; . \notag   
\end{align}  
The slope of the leading term determines the critical temperature
$T_c$ in units of the dimensionful continuum coupling $g_3^2 $ of the
(2+1)$d$ theory, yielding $T_c/g_3^2 = 0.3757(5)$. By standard
arguments, from this result one can also read off the temperature
dependence of the coupling at a given $N_t$ \cite{Edwards:2009qw},
\begin{align} 
\beta(t) &= \beta_c +  4N_t \, \frac{T_c}{g_3^2} \, t -
\frac{0.270(2)}{N_t} \, \frac{g_3^2}{T_c} \, \frac{t}{1+t} \, + \,
\mathcal O(1/N_t^2) \; , \notag\\
 &  \hskip 4cm t = T/T_c - 1 \; . \label{tbeta}
\end{align}
For each fixed $N_t$, we thus have precise control of the temperature
$T=1/(aN_t)$, where $a$ is the lattice spacing, by varying the lattice
coupling $\beta$. The physical length of the spatial volume then
follows from  
\begin{equation} 
L = a N_s = \frac{N_s}{N_t T} = \frac{N_s}{N_t T_c}\, \frac{1}{1+t(\beta)} \;.
\label{eq:Lfss}
\end{equation}

\medskip

\subsection{Twisted boundary conditions,
                       electric and\\ magnetic center flux}
\label{sec:tbc}

\medskip

\begin{figure}[t]

\vspace*{.1cm}

\hskip .1cm
\centerline{\includegraphics[width=\linewidth]{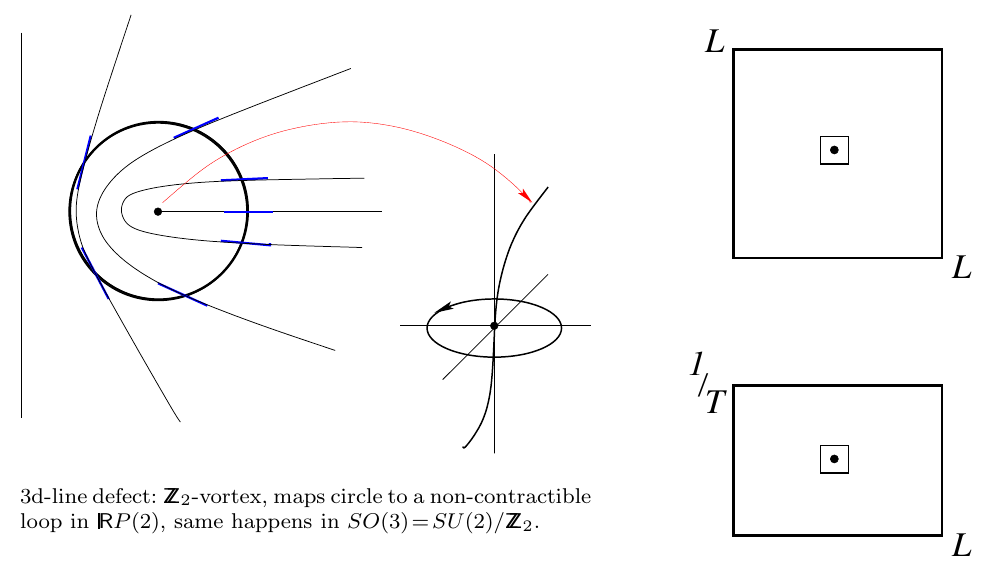}}

\vspace{-.2cm}
\caption{$Z_2$-vortex, {\it e.g.}, as in nematic liquid crystals
  or pure SU$(2)$ gauge theories (left). Center vortices in the $d+1$
  dimensional gauge theory lower their free energy by spreading out:    
  magnetic ones through the $d(d-1)/2$ spatial $\vec m$-planes spread
  at all $T$, while those through the $d$ temporal $\vec k$-planes get
  squeezed with temperature (right).    
\vspace*{-.6cm}} 
\label{fig:Z2Vortex}
\end{figure}

As mentioned above, the different choices of twisted b.c.'s in pure
SU$(N)$ gauge theories at finite temperature 
are labeled by two $Z_N$-valued vectors, $\vec m$ and $\vec k$, and fix 
the total numbers of vortices mod $N$ through the orthogonal planes of
the Euclidean $1/T\times L^d$ spacetime box as sketched in
Fig.~\ref{fig:Z2Vortex}. There are $d(d-1)/2$ purely spatial planes and the
$d(d-1)/2$-dimensional vector $\vec m$ denotes the total 
conserved, $Z_N$-valued and gauge-invariant magnetic flux
through those, as generated by a static center monopole. The ones
through the $d$ temporal planes labeled by the $d$-dimensional vector
$\vec k $ are the universal partners of the Potts interfaces.

In order to understand the inequivalent choices for imposing boundary
conditions on the gauge fields $A$, which are blind to the center $Z_N$ of
SU$(N)$, one first chooses $A(x)$ to be periodic with the lengths $L_\mu$ 
of the system in each direction $\hat\mu$ up to gauge transformation
$\Omega_\mu(x) \in \mathrm{SU}(N)$, which is physically equivalent to
periodic boundary conditions,
\begin{equation} 
 A_\nu(x\!+\!L_\mu \hat\mu) =  \, A^{\Omega_\mu}_\nu(x) 
          \equiv \Omega_\mu(x) \, \big(A_\nu(x) - {i} \, \partial_\nu \big)\,  
        \Omega_\mu^\dagger(x)   
     \; .       \label{tbcs}
\end{equation} 
In the $1/T\times L^d$ box at finite temperature,
we use
\begin{equation}
 L_0 = \beta = 1/T\; , \; \mbox{and}\;\; L_i = L\, , \; i= 1,\dots d\;. 
\end{equation}
Then, compatibility of two successive translations in a ($\mu,\nu$)-plane
requires that (no summation of indices) 
\begin{align}
\label{coc}
&\Omega_\mu(x\! +\! L_\nu \hat\nu) \Omega_\nu(x) = Z_{\mu\nu} \,
 \Omega_\nu(x\! +\!  L_\mu \hat\mu) \Omega_\mu(x)
\\
   &\mbox{with}  \;\; Z_{\mu\nu} = e^{2\pi i \, n_{\mu\nu}/N} \;, 
   \mbox{ and }  n_{\mu\nu} = - n_{\nu\mu}  \in  Z_N  \; .  
\nonumber
\end{align}
The total number modulo $N$ of center vortices in a ($\mu,\nu$)-plane
is specified in each sector by the corresponding component
of the twist tensor $n_{\mu\nu}$. In $d=3$ spatial dimensions magnetic
center flux $\vec m$ through the box is given by $n_{ij} =
\epsilon_{ijk} m_k $, in $d=2$ by $n_{ij} = \epsilon_{ij} \, m
$. The time components $ n_{0i} \equiv  k_i$ define the temporal twist
$\vec k \in Z_N^d$.

With these inequivalent choices of boundary conditions, 
the finite-volume theory in $3+1$ dimensions decomposes into
sectors of fractional Chern-Simons number ($\nu + \vec k\cdot\vec m
/N$) \cite{vanBaal:1982ag}  
with states labelled by $|\vec k , \vec m , \nu \rangle $, where the
integer $\nu $ is the usual instanton winding number.
These sectors are connected by homotopically non-trivial
gauge transformations $\Omega [\vec k,\nu]$, 
\begin{equation}
\label{hom-nt}
      \Omega [\vec k',\nu'] \, |\vec k , \vec m , \nu \rangle \, =\,  
          |\vec k\! +\!\vec k', \vec m , \nu\! +\!\nu' \rangle \, .
\end{equation}
Note that such a gauge transformation must be multivalued in SU$(N)$ 
and hence singular to change the twist. 

A Fourier transform of the twist sectors $Z_k(\vec k, \vec m, \nu)$, which
generalizes that for $\theta$-vacua as Bloch waves from $\nu$-vacua in
two ways,  by replacing $\nu \to (\nu + \vec k\cdot\vec m /N)$ 
for fractional winding numbers, and with an additional
$Z_N^3$-Fourier transform w.r.t.~the temporal twist $\vec k$, yields, 
\begin{equation}
 Z_e(\vec e,\vec m,\theta )\,=\, 
\frac{1}{N^3} \, \sum_{\vec k, \,\nu} \,  e^{i\omega(\vec k, \nu)} \,
 Z_k(\vec k,  \vec m, \nu)  \, .  \label{ZnFT}
\end{equation}    
Up to a geometric phase,
\begin{equation} 
 \omega(\vec k,\nu) =
2\pi \vec e\cdot\vec k/N  + \theta (\nu \!+\! \vec k\cdot\vec m /N) \;
,
\end{equation}  
the states in the new sectors are then invariant under the non-trivial
$\Omega[\vec k,\nu]$ also, 
\begin{equation}
\label{hom-nt-inv}
      \Omega [\vec k,\nu] \, |\vec e , \vec m , \theta \rangle \, =\,  
         \exp\{ -i \omega(\vec k,\nu) \} \, 
  |\vec e , \vec m , \theta \rangle \, .
\end{equation}
Their partition functions $Z_e$ are classified, in addition to their magnetic
flux $\vec m $ and vacuum angle $\theta $, by their $Z_N$-valued  
{\em gauge-invariant electric flux} in the $\vec e$-direction
\cite{'tHooft:1979uj,'tHooft:1998}. 
Here, we do not consider finite $\theta$ and omit the argument
$\theta$ in the partition functions henceforth. 
Magnetic flux is irrelevant for the deconfinement transition also, 
because for all combinations of magnetic flux $\vec m$ with either
electric flux $\vec e$ or temporal twist $\vec k$
\cite{vonSmekal:2002gg},  
\begin{equation} 
Z_e(\vec e,\vec m)  \stackrel{L\to\infty}{\longrightarrow}  Z_e(\vec e,0)
\, , \;\mbox{and}   \;\;    Z_k(\vec k,\vec m) 
              \stackrel{L\to\infty}{\longrightarrow} Z_k(\vec k,0)\;.
\end{equation}
There is no analogue of magnetic center flux in the $3d$ spin
systems, and center monopoles always condense because they are screened
by the spatial string tension which is non-zero on either side of the
deconfinement transition, likewise. Different 
combinations of temporal $\vec k$ and spatial $\vec m$-twists,
with different fractional topological charge, can however
be used to measure the topological susceptibility without cooling
\cite{vonSmekal:2002uz}.  

The essential structure here is, however, that the temporally twisted
vortex ensembles $Z_k(\vec k) $ in the $d+1$ dimensional pure SU$(N)$
gauge theory are related to those with fixed units of electric flux 
$Z_e(\vec e)$ (and a perhaps more intuitive physical interpretation)
by a $d$-dimensional discrete $Z_N$-Fourier transform
\cite{'tHooft:1979uj}, 
\begin{equation}
  Z_e(\vec e ) \,= \, \frac{1}{N^d}  \, \sum_{\vec k\in Z_N^d}
  \, e^{2\pi i \, \vec e\cdot\vec k/N}\, Z_k(\vec k) \; , \label{eq:ZeZk}
\end{equation}  
no matter what the magnetic flux is, which is why the 
argument $\vec m$ is dropped here and in the following also.

The role of the electric flux ensembles is best understood in terms of the 
translationally invariant flux between a fundamental color charge at
some point $\vec x$ in the finite volume and its mirror charge at
$\vec x + \vec e L$ in a neighboring volume in the direction of the
flux $\vec e$ \cite{deForcrand:2001nd},     
\begin{equation}
\frac{Z_e(\vec e)}{Z_e(0)} \, = \, \frac{1}{N}\, \Big\langle
\mbox{tr}\Big( P(\vec x) P^\dagger (\vec 
x+ \vec e L) \Big) \Big\rangle_{\mbox{\scriptsize no-flux}}\, , \label{mirror}
 \end{equation}
where the $P(\vec x)$'s are untraced fundamental Polyakov loops, including
any potentially non-trivial transition function $\Omega_0(x)$  in the time
direction. With time-ordering from right to left, 
\begin{equation}
\label{eq:polynp} 
P(\vec x) = \Omega_0^\dagger(0,\vec x) \, T\Big( \exp\Big\{i\int_0^\beta
    A_0(t,\vec x) \, dt\, \Big\}\Big)  \; . 
\end{equation}
The gauge field at $t=\beta$ differs from that at $t=0$ by the gauge
transformation $\Omega_0(0,\vec x)$ and we must undo this to make $\tr\,
P(\vec x) $ gauge invariant also under non-periodic gauge
transformations $g(x)$ which change the transition functions as
\begin{equation}
\Omega_\mu'(x) = g(x+L_\mu \hat\mu) \,\Omega_\mu(x)\, g^\dagger(x) \;
. \label{eq:gttr} 
\end{equation} 
With this, the Polyakov line in \Eq{eq:polynp} transforms as 
\begin{equation} 
 P'(\vec x) =  g(0,\vec x)\, P(\vec x) g^\dagger(0,\vec x) \; ,
\end{equation}
and its trace is invariant under such gauge transformations
even when $g(\beta,\vec x) \not= g(0,\vec x) $.

In order to derive \Eq{mirror} we first decompose the total electric 
flux vector $\vec e \equiv e_i \vec e_i$, with $ e_i = 1,\,\dots N-1$,
and perform successive translations by $L $ in the direction of the
individual unit vectors $\vec e_i$. For each step we use the boundary
conditions (\ref{tbcs}) for the gauge fields $A_\mu$ and the cocycle
condition (\ref{coc}) for the transition functions $\Omega_\mu $, to show
that in a fixed twist sector with boundary conditions $\vec k$, and
$Z_{i0} = e^{- 2\pi i \, k_i/N}$,
\begin{align}
P(\vec x + L\vec e_i)  &= \, \Omega_0^\dagger(0,\vec x+L\vec e_i) \, 
\Omega_i(\beta,\vec x) \, \tilde P(\vec x) \, \Omega_i^\dagger(0,\vec
x) \notag  \\  
&= Z_{i0}\,  \Omega_i(0,\vec x)\, \Omega_0^\dagger(0,\vec x) \, 
 \tilde P(\vec x)\, \Omega_i^\dagger(0,\vec x) \notag  \\  
&= Z_{i0}\,   \Omega_i(0,\vec x)\,  P(\vec x) \, \Omega_i^\dagger(0,\vec x) 
\label{eq:polytrans}
\end{align}
where we have used  $ \tilde P(\vec x) = T\big( \exp i\int_0^\beta
    A_0(t,\vec x) \, dt\, \big) $ to denote the normal Polyakov
line as suitable for periodic boundary conditions. If we repeat this
step for an arbitrary translation by $L\vec e \,\in L\, Z_N^d $, we pick up
a product of center elements for the twists along
the way,
\begin{equation}
P(\vec x + L\vec e) =\, e^{-2\pi i \, \vec e\cdot\vec k/N} \,
\Omega(0,\vec x)\,  P(\vec x) \, \Omega^\dagger(0,\vec x) \; , 
\label{eq:polytrfull}
\end{equation}
and $\Omega(0,\vec x)$ stands for the product of spatial transition
functions when going from $\vec x $ to $\vec x + L\vec e$. 
Changing the order of two spatial steps in different directions, $L\vec e_i
\leftrightarrow L\vec e_j$ in the path from $\vec x $ to $\vec x +
L\vec e$,  multiplies  $\Omega $ by the corresponding spatial twist $Z_{ij}$.
Because $Z_{ij}^* Z_{ij} = 1$, their order does therefore 
not matter in \Eq{eq:polytrfull}. We have,
\begin{equation}
 P(\vec x) \, \Omega^\dagger(0,\vec x) \, P^\dagger(\vec x + L\vec
 e)\, \Omega(0,\vec x)\,  =\, e^{2\pi i \, \vec e\cdot\vec k/N} \, \mathds 1
 \; , 
\label{eq:polyunity}
\end{equation}
in an ensemble with fixed temporal twist $\vec k$. When all boundary
conditions are spatially periodic, $\Omega(0,\vec x) = \mathds 1$. 
We can then insert the unity in \Eq{eq:polyunity} into the partition
functions with fixed boundary conditions and sum over all temporally
twisted sectors  to establish \Eq{mirror} from \Eq{eq:ZeZk}. 
The subscript `no-flux' in \Eq{mirror} indicates that the
expectation value is meant to be taken in the enlarged ensemble
$Z_e(0) \equiv \sum_{\vec k}  Z_k(\vec k) /N^d $ corresponding to this
sum. If there is no magnetic flux, $\vec m=0$, we can always apply
time-independent (non-periodic) gauge transformations with 
\begin{equation} 
g(\vec x + L\vec e_i) = g(\vec x) \, \Omega^\dagger_i(0,\vec x) \;\;
\Rightarrow \; \Omega_i'(0,\vec x) = \mathds 1\; .
\end{equation}
We can thus assume without loss that $\Omega(0,\vec x) = \mathds 1$,
as in \Eq{mirror} which then remains manifestly invariant under
spatially periodic gauge transformations in this form. When $\vec m
\not =0$ we can still use \Eq{mirror} for a single direction, but for
diagonal electric fluxes in a plane with magnetic twist
we must include the combination of spatial transition
functions $\Omega(0,\vec x)$ in the expectation value on the right as
in \Eq{eq:polyunity} which is then, however, invariant under non-periodic
gauge transformations as well.\footnote{This slight subtlety was
  overlooked in \cite{deForcrand:2001nd}. To verify the invariance of
  \Eq{eq:polyunity} under general, non-periodic gauge transformations is a
  straightforward exercise in manipulating transition functions 
  using their cocycle condition (\ref{coc}) and transformation law
  (\ref{eq:gttr}).}   

\Eq{mirror} emphasizes the physical interpretation of the electric
flux sectors, as providing the necessary mirror charges for the sectors
of SU$(N)$ with non-vanishing total color charge, dual to those with
fixed boundary conditions. Unlike the 
standard product  $ \mbox{tr}\big( \tilde P(\vec x)
 P^\dagger (\vec  x+ R \vec e_i) \big) $ of static charges
at distance $R$ in a periodic ensemble, which is not gauge invariant,
the electric fluxes (\ref{mirror}) determine the gauge-invariant
color-singlet free energy or potential between static charges at
distance $L|\vec e|$. 
They contain no ultraviolet divergent perimeter terms and no
short-distance Coulomb contributions either. 

\medskip

\subsection{Electric fluxes and selfduality in $2+1$ dimensions}

\medskip

Independent of universality and scaling, exact maps between the spin
systems and their dual theories, in terms of disorder variables on the
dual lattice, are provided by Kramers-Wannier duality \cite{Savit:1979ny}. 
In 2 dimensions, the dual of a $Z_N$ spin model is a $Z_N$ spin
model again. In particular, the 2$d$ $q$-state Potts models
(\ref{PottsH}) are selfdual for all $q$. In $3$ dimensions, the spin
models are dual to $Z_N$ gauge theories. The best known example is the
$3d$ Ising model whose dual partner is the $Z_2$ gauge theory, and this
system is relevant for the dual universal behavior of center vortex and
electric flux ensembles in the $3+1$ dimensional SU$(2)$ gauge
theory \cite{deForcrand:2001nd,vonSmekal:2002ps}. In 4 dimensions, the
dual of a $Z_N$ gauge model is a $Z_N$ gauge model again \cite{Savit:1979ny}. 

One particular aspect of the SU$(N)$ gauge theories in $2+1$
dimensions with $N=2$, $3$ and $4$, that has been overlooked until
recently, follows from the selfduality of the corresponding 2$d$
$q$-state Potts models and universality. The center vortex and
electric flux ensembles of the  $2+1$ dimensional gauge theories are     
mirror images of one another within the universal scaling
window around criticality
\cite{vonSmekal:2010xz,vonSmekal:2010du,Strodthoff:2010dz}. 
This is obvious from the fact that, as we will see, the exact
duality transformation of the 2$d$ $q$-state Potts models in a finite
volume is precisely of the structure of 't Hooft's relation
between the temporal center vortex and electric flux ensembles
in \Eq{eq:ZeZk} discussed above.

The duality transformation and the selfduality of the  
$q$-state Potts models in 2 dimensions has long been known for
infinite systems. A particularly simple proof 
based on the random bond-cluster representation was given in
\cite{Wu:1982ra}. Far less is known in a 
finite volume with translationally invariant boundary conditions,
however, where ensembles with different boundary conditions in general
mix under duality transformations. One case where this is known is the
3$d$ Ising/$Z_2$-gauge theory system
\cite{Gruber:1977,Caselle:2001im,Caselle:2010}. For the Potts model, using the
random cluster methods developed in \cite{Park:1988}, we were
able to obtain the following exact 
duality transformations for all $q$ on a finite 2$d$ torus as discrete
2$d$ Fourier transforms over all ensembles $Z_q^{(m,n)}$ with
cyclically shifted boundary conditions $\vec c = (m,n)$
\cite{vonSmekal:2010xz},   
\begin{align} 
Z_q^{(-s,r)}(\widetilde K) &= \Bigg( \frac{e^{\widetilde K}
  -1}{e^K-1} \Bigg)^{N_\mathrm{sites}} \frac{1}{q} \sum_{m,\, n}
\, e^{2\pi i \, (rm+sn)/q } \; Z_q^{(m,n)}(K) \; , \notag \\
& \hskip 1cm m, n, r,s =  0, 1, \dots q-1\; , \label{Duality} 
\end{align}
where $N_\mathrm{sites}$ is the total number of sites of the  2$d$ 
lattice ({\it i.e.}, $N_\mathrm{sites} = N^2$ on an $N\times N$ square
lattice), $K=J/T$ is the coupling per temperature, 
and $\widetilde K$ its dual obtained from 
\begin{equation}
\big(e^{\widetilde K}-1 \big)\big(e^K
-1\big )\, =\,  q \; , \label{eq:dualK}
\end{equation}
as usual, with temperature mirrored around criticality at
\begin{equation}
K= \widetilde K = K_c = \ln(1+\sqrt q) \; .  
\end{equation}
For $q=2$, and with $K \to K/2$, conventionally, this duality relation reduces
to an analogous known result for the Ising model \cite{Muenster:2001}. With
$q=3$ it agrees with a result 
obtained for the planar or vector Potts model \cite{Bugrii:1997ry} 
which is equivalent to the standard one in that case. 

Before we turn to the sketch of the proof, note that the structure of 
the finite-volume duality transformation in the Potts models
(\ref{Duality}) is precisely the same as that in the relation between
the temporal center vortex and electric flux ensembles in \Eq{eq:ZeZk}.  
The temporal center vortices are the universal partners of the
interfaces in the spin model, {\it c.f.}, Fig.~\ref{fig:VortexInterface}, 
and the electric fluxes are their duals, with $d=2$ again universally
related to ensembles with interfaces, but at the dual temperature,
swapped around criticality. In $d=3$ spatial dimensions the pattern is
the same \cite{vonSmekal:2002ps}, the center vortex free energies in
SU$(2)$ share their universal behavior with those of the interfaces in
the $3d$ Ising model, and electric fluxes correspond to ensembles of
the dual $3d$ $Z_2$ gauge theory with anti-periodic boundary
conditions. But they are not exact mirror images of one another in
$d=3$.     

The general formula (\ref{Duality}), for all $q$-state Potts models on
a 2$d$ torus, was first given to my knowledge in \cite{vonSmekal:2010xz}.

I have not published its proof nor have I seen it in the
literature before. So I will sketch it here, in the rest of this 
subsection which will not be needed later on and which the reader may
thus choose to skip ad libitum. 

The proof uses mostly standard techniques
\cite{Baxter:2007,Wu:1982ra}, and it is based on the setup and results
of Ref.~\cite{Park:1988}.   

We start from the Hamiltonian (\ref{PottsH}) without external field
($H=0$) but with cyclic interfaces such that  
\begin{equation} 
-\beta \mathcal H^{(m,n)} = K \sum_{\langle i,j\rangle}
\delta(s_i,G^{(m,n)}_{ij}s_j) \; ,
\end{equation}
where the notation is such that we have introduced matrices
$G^{(m,n)}$ on the links connecting nearest neighbors $\langle
i,j\rangle$, and $\delta(s,s') = \delta_{s,s'} = 1$ when $s=s'$ and
zero otherwise. For the cyclic interfaces we introduce straight 
lines (seams) $L_x$ and $L_y$ in the $x$ and $y$ directions on the
dual lattice and use 
\begin{equation}
 G_{ij}^{(m,n)} \,=\, \left\{ \begin{array}{ll}  
             S^m\; , & \langle i,j\rangle \in L_y^*\; ,\\ 
             S^n\; , & \langle i,j\rangle \in L_x^* \; ,\\ 
             \mathds 1 \; , & \mbox{otherwise} \; , \\
   \end{array}
 \right. \label{eq:seams}
\end{equation} 
where $S$ is a $q\times q$ matrix for an elementary cyclic shift,
\begin{equation}
S =  \begin{pmatrix} 
  0 & 0 & 0 & \cdots & 1 \\
  1 & 0 & 0 & \cdots & 0 \\
  0 & 1 & 0 & \cdots & 0 \\
  \cdot & \cdot & \cdot &  \cdot & \cdot \\
  0 & 0 & \cdots & 1 & 0 
\end{pmatrix}
\end{equation}
such that along the $x$-direction the spins get shifted $m$ times at
the links dual to ({\it i.e.}, crossing) $L_y$
and $n$ times in the $y$-direction at those dual to $L_x$. The
cyclically shifted boundary conditions implemented in this way are
special cases of the general permutations of spin states across the
boundary considered in 
Ref.~\cite{Park:1988}. We will use the random-cluster representation
of the Potts models on a finite torus from there. With 
\begin{equation} 
 v =  e^K - 1
\end{equation}
the random-cluster representation of the partition function with
$(m,n)$-boundary conditions then follows from   
\begin{align} 
Z_q^{(m,n)} &= \, \sum_{\{s\}} \exp\Big\{ 
K \sum_{\langle ij\rangle} \delta(s_i,G^{(m,n)}_{ij}s_j)
\Big\} \label{eq:cluster} \\ 
            &= \, \sum_{\{s\}} \prod_{\langle ij\rangle} \Big( 1 + v\,
\delta(s_i,G^{(m,n)}_{ij}s_j)  \Big) = \sum_\mathrm{graphs}  v^{N_b} q^{N_c}  \; .
\notag
\end{align}
This is the standard random-cluster representation of the infinite
Potts models \cite{Fortuin:1971dw,Baxter:2007,Wu:1982ra}. It is valid
also on the $2d$ torus in this form as long as we only consider the
cyclic shifts of the spin states across the seams in \Eq{eq:seams} as
will be shown below. The product in the second line is multiplied out
and the spin sum represented as a sum 
over all graphs where a link with aligned spins,
$\delta(s_i,G^{(m,n)}_{ij}s_j) = 1$, is said to be occupied
while those with $\delta(s_i,G^{(m,n)}_{ij}s_j) = 0 $ are empty. The
possible configurations are then identified with graphs consisting of
clusters connected by occupied links, whereby each disconnected  site
counts as its own cluster. So all sites are part of a cluster. 
Each occupied link (or bond) in a graph contributes a factor $v$, and
their total number in a given graph is called $N_b$. Each of the $N_c$
clusters in a graph can be flipped independently as a whole and thus
gives a combinatoric weight $q$ to the particular graph in the
partition sum.  

On the $2d$ torus one needs to classify the clusters by the possible 
winding numbers $(\omega_1,\omega_2)$ of the non-selfintersecting
closed loops around the torus that one can draw on them:

-- If one cluster
extends across both boundaries in such a way that independent $(0,1)$
and $(1,0)$ loops are possible, the graph is said to have a torus
cluster (TC) of which it can have at most one. 

-- A cylinder cluster (CC) extends across the boundaries in such a way
that only non-trivial loops with winding  numbers
$(\omega_1,\omega_2)$ are possible, where $\omega_1$  and 
$\omega_2$ are coprime. We call the number of 
cylinder clusters in a graph $N_{\mathrm{CC}}$, and they must all 
have the same pair of winding numbers. 

-- And finally, island-type clusters are the  
ones on which only $(0,0)$ loops are possible. A graph with only
island-type clusters is called a torus lake (TL) because the empty
links form stacks which extend across the boundaries in both
directions such that independent $(0,1)$ and $(1,0)$ loops are possible
on the dual lattice, perpendicular to these stacks.  

Generally, the empty links on the original lattice $\mathcal L$ are
dual to links which can be connected to form a cluster on the dual
lattice $\mathcal L^*$. Therefore, a lake on the original lattice is a
cluster on the dual lattice \cite{Park:1988}.

With this classification, and more general seams corresponding to
arbitrary and independent permutations $G_x$ and $G_y$ of the spin
states across each boundary in $x$ and $y$ directions the random
cluster formulation of the Potts models on the torus as derived in
\cite{Park:1988} becomes,

\vspace*{-.4cm}

\begin{widetext}
\vspace*{-.2cm}
\begin{align}
Z_q^{G_x,G_y} &=  \sum_{\{\mathrm{CC}\}} \,  v^{N_b} q^{N_c} \, 
\Bigg(\frac{C(G_x,G_y;\omega_1,\omega_2) }{q}\Bigg)^{N_{\mathrm{CC}}} 
+ \sum_{\{\mathrm{TC}\}}\,  v^{N_b} q^{N_c-1} \, T(G_x,G_y)  +
\sum_{\{\mathrm{TL}\}} \, 
v^{N_b} q^{N_c} \; . 
\label{eq:PottsTorusG} 
\end{align}
Here this simplifies to,
\begin{align}
Z_q^{(m,n)} &=  \sum_{\{\mathrm{CC}\}} \,  v^{N_b} q^{N_c} \, 
\delta_q(m\omega_1+n\omega_2) \,
+\, \delta_{m,0} \delta_{n,0} \sum_{\{\mathrm{TC}\}} \, v^{N_b} q^{N_c}   +
\sum_{\{\mathrm{TL}\}}  \, v^{N_b} q^{N_c} \; . \label{eq:PottsTorusC}
\end{align} 
\end{widetext}


\noindent
The first sum is over all graphs with cylinder clusters, the second
over all graphs with one torus cluster and the third over all graphs
with only island-type clusters in a torus lake. See
Ref.~\cite{Park:1988} for more details.  

In the second line (\ref{eq:PottsTorusC}) we have used that $G_x =
S^m$ and $G_y = S^n$ here, and that the weight of an
$(\omega_1,\omega_2)$ cylinder cluster 
from \cite{Park:1988} then reduces to 
\begin{equation} 
    C(G_x,G_y;\omega_1,\omega_2)  =  \tr\Big(G^{\omega_1}_x
    G_y^{\omega_2} \Big) 
           =  \tr\Big(S^{m\omega_1} S^{m\omega_2} \Big) \; , 
\end{equation}
whereby, with the Kronecker delta modulo $q$, 
\begin{align}
        \tr\Big(S^{m\omega_1} S^{m\omega_2}\Big) &= q \,
        \delta_q(m\omega_1+n\omega_2)  \\ 
&= \Bigg\{\begin{array}{ll}
 q\; , & m\omega_1+n\omega_2 = 0 \bmod q \; , \\[4pt]
 0\; , & \mbox{otherwise}\, .
  \end{array} \notag
\end{align}
Moreover, unlike the more general case with arbitrary 
permutations $G_x$, $G_y$ in \cite{Park:1988}, for the cyclic
ones in \Eq{eq:seams} considered here, a torus cluster is only
possible when $m=n=0$, 
{\it i.e.}, with periodic boundary conditions. But then it can have $q$
different states just as any other island-type cluster, therefore
\begin{equation} 
T(m,n) \equiv T(S^m, S^n) = \Bigg\{\begin{array}{ll}
 q\; , & m = n = 0 \; , \\[4pt]
 0\; , & \mbox{otherwise}\, .
  \end{array} 
\end{equation}
This means that when we restrict to cyclically shifted boundary
conditions, all clusters that are possible for such  
given $(m,n)$ boundary conditions have the same weight, $ v^{N_b}
q^{N_c} $, and this establishes that the infinite volume form of the
cluster representation in \Eq{eq:cluster} remains formally valid. One
only needs to figure out the allowed cylinder and torus cluster and
torus lake graphs as in \Eq{eq:PottsTorusC} which is the starting
point for the duality transformation in the finite volume.

The first task in obtaining the dual cluster representation is 
to re-express the number of occupied links $N_b$ and the number of
clusters $N_c$ in \Eq{eq:PottsTorusC} in terms of the corresponding
numbers on the dual lattice $\mathcal L^*$.

To achieve this, recall that for an arbitrary $2d$ lattice, the Euler
formula relates the number of vertices $V$, edges $E$, faces $F$,
and open ends $N_\mathrm{open} $ to
the genus $g$ of the surface on which it is drawn 
via the Euler characteristic, 
\begin{equation}
\chi = V - E + F = 2 - 2g - N_\mathrm{open} \; .
\end{equation}
The Euler characteristic of the torus with $g=1$ and without open
ends,  $N_\mathrm{open} =0 $, is $\chi = 0$. Moreover, $V= N_s$ is the
number of sites of $\mathcal L$, and $E$ is the total number of links,
that is the sum of occupied ones
$N_b$ and empty ones which is identical to the number $N_b^*$ of 
 occupied ones on the dual lattice $\mathcal L^*$. Thus $E = N_b +
 N_b^*$. Finally, the number of faces is the number of sites on
 $\mathcal L^*$, {\it i.e.}, $F=N_s^*$. Therefore
\begin{equation}
 N_b = N_s + N_s^* - N_b^* \; . \label{eq:nolink}
\end{equation}
It also follows from the Euler relation on the torus with $\chi = 0$
that the total number of sites $N_s$ of the lattice $\mathcal L$ is
given by the number of clusters $N_c$ in a graph (recall that every
site is part of a cluster) plus the number of occupied links $N_b$
minus the number of independent non-homotopic (and
non-selfintersecting) loops $N_l $ on the clusters of the
graph. Therefore, the number of clusters $N_c$ in a graph is expressed as 
\begin{equation} 
N_c = N_s - N_b + N_l \; . \label{eq:Euler}
\end{equation}
All TC graphs, with one torus cluster on the original lattice, are
graphs with one torus lake on the dual lattice and vice versa, 
{\it i.e.}, $\{\mathrm{TC}\} =  \{\mathrm{TL}^*\} $ 
and  $\{\mathrm{TL}\} =  \{\mathrm{TC}^*\} $. 

On a TL graph, where all clusters are islands whose coastlines have
$(0,0)$ winding numbers, one can walk around every pond which is a
lake with $(0,0)$ coastline inside a cluster. Therefore $N_l =
N_\mathrm{ponds} $. On a TC graph, one can walk around all ponds
individually,  but walking around all of them is homotopic to a
point. In addition there are the $(0,1)$ and $(1,0)$ loops, thus $N_l
= N_\mathrm{ponds} + 1$. Together with the fact that the islands of a
TL graph are the ponds of its dual graph TC$^*$ and vice versa,
this implies for  
\begin{align} 
 \{\mathrm{TL}\}  =  \{\mathrm{TC}^*\} \; : 
      \quad & N_l = N_\mathrm{ponds} = N_c^* - 1 \; , \label{TL-TC} \\  
 \{\mathrm{TC}\}  =  \{\mathrm{TL}^*\} \; : 
      \quad & N_l = N_\mathrm{ponds} +1  = N_c^* + 1 \; . \notag
\end{align}
The ponds become islands on the dual graph. So $N_\mathrm{ponds}$ is
equal to the number of dual clusters $N_c^*$ on TL$^*$ graphs, while
$N_c^*$  is the number of islands (and thus $N_\mathrm{ponds}$) 
 plus one on TC$^*$ graphs.

If we insert the identities (\ref{TL-TC}) together with the
Euler relation (\ref{eq:nolink}) into  \Eq{eq:Euler}, this leads to 
\begin{align} 
 \{\mathrm{TL}\}  =  \{\mathrm{TC}^*\} \; : 
      \quad & N_c = -N_s^* + N_b^* + N_c^* - 1 \; , \label{eq:NcTL} \\  
 \{\mathrm{TC}\}  =  \{\mathrm{TL}^*\} \; : 
      \quad & N_c = -N_s^* + N_b^* + N_c^* + 1 \; . \label{eq:NcTC} 
\end{align}
CC graphs are selfdual. They contain an equal number of cylinder clusters
and cylinder lakes, and all of them have the same winding numbers
$(\omega_1,\omega_2)$. The dual graph is the one with clusters and
lakes swapped, and it thus has the same number of cylinder clusters,
$N_\mathrm{CC} = N_\mathrm{CC}^*$, 
with the same set of winding numbers. On a CC graph one can walk
non-homotopically around every pond and along the
$(\omega_1,\omega_2)$ loop of every cylinder cluster, therefore,

\newpage

\vspace*{-1cm}   
\begin{align}   
 \{\mathrm{CC}\} &=  \{\mathrm{CC}^*\} \; :   \notag \\
    & N_l =  N_\mathrm{ponds}  + N_\mathrm{CC} =  N_\mathrm{islands}^* 
 + N_\mathrm{CC}^*  = N_c^* \; , \notag \\  
    & N_c = -N_s^* + N_b^* + N_c^* \; . \label{eq:NcCC}
\end{align}
Now we use the relations in \Eqs{eq:nolink}, (\ref{eq:NcTL}),
(\ref{eq:NcTC}),  and (\ref{eq:NcCC}),  to re-express $N_b$ and $ N_c$ in
\Eq{eq:PottsTorusC} in terms of the corresponding numbers  $N_b^*$ and $ N_c^*$ 
on the dual lattice $\mathcal L^*$. Labeling the graphs in the sums by
their respective dual cluster configurations, this yields for the
partition function with $(m,n)$ boundary conditions,

\vspace*{-.4cm}

\begin{widetext}
\begin{align}
&Z_q^{(m,n)} =  v^{N_s+N_s^*} q^{-{N_s^*}} \, \Bigg\{ 
\sum_{\{\mathrm{CC}^*\}} \,  \Bigg(\frac{q}{v}\Bigg)^{N_b^*}  q^{N_c^*} \, 
\delta_q(m\omega_1+n\omega_2) \,
+\, \delta_{m,0} \delta_{n,0} \sum_{\{\mathrm{TL}^*\}} \, 
\Bigg(\frac{q}{v}\Bigg)^{N_b^*} q^{N_c^*+1}   +
\sum_{\{\mathrm{TC^*}\}}  
\, \Bigg(\frac{q}{v}\Bigg)^{N_b^*} q^{N_c^*-1} \, \Bigg\}  \;
, \label{eq:dualP1}
\end{align}
and its discrete Fourier transform over  $m, n = 0,\, \dots q-1$ becomes,
\begin{align}
&\sum_{m,\, n} \, e^{2\pi i \, (rm+sn)/q } \;  Z_q^{(m,n)} = 
 v^{N_s+N_s^*} q^{1-{N_s^*}} \, \Bigg\{ 
\sum_{m,\, n} \, e^{2\pi i \, (rm+sn)/q } \,
\sum_{\{\mathrm{CC}^*\}} \,  \Bigg(\frac{q}{v}\Bigg)^{N_b^*}  q^{N_c^*-1} \,
\delta_q(m\omega_1+n\omega_2) \label{eq:dualP2} \\
&\hskip 8.2cm +\, \sum_{\{\mathrm{TL}^*\}} \,
\Bigg(\frac{q}{v}\Bigg)^{N_b^*} q^{N_c^*}   +  q^2\, \delta_{r,0}
\delta_{s,0} \sum_{\{\mathrm{TC^*}\}}  \,
\Bigg(\frac{q}{v}\Bigg)^{N_b^*} q^{N_c^*-2} \, \Bigg\} \; . \notag 
\end{align} 
\end{widetext}


\noindent
We have factorized another $q$ in \Eq{eq:dualP2} to get the right power
$N_c^*$ in the middle term for which the sums over $m$ and $n$
collapse to the single $m=n=0$ term. The additional factor $q^2$ in front of
the last term in \Eq{eq:dualP2} arises because the TL graphs are
independent of the boundary conditions $m$ and $n$ and therefore their
sums yield $q \delta_{r,0}$ and $q \delta_{s,0}$, respectively. In the
first term, the graph sum extends over all kinds of graphs with
cylinder clusters for all pairs of coprime windings $\omega_1$ and
$\omega_2 $, and the periodic Kronecker delta, 
\begin{equation} 
 \delta_q(m\omega_1+n\omega_2) = \frac{1}{q} \sum_{t=0}^{q-1} 
 e^{-2\pi i\, (m\omega_1+n\omega_2)\,t/q}
\end{equation}
selects the right ones for each pair of boundary conditions
$(m,n)$. The duality transformation is thus completed  upon realizing
that for any pair of coprime $(\omega_1,\omega_2)$, we can write,

\begin{align} 
&\sum_{m,\, n} \, e^{2\pi i \, (rm+sn)/q }\; \sum_{t=0}^{q-1} 
 e^{-2\pi i\, (m\omega_1+n\omega_2)\,t/q} = \\
 &\hskip .2cm = \sum_{t=0}^{q-1} \, q\delta_q(\omega_1 t - r) \,
 q\delta_q(\omega_2 t - s) = q^2 \delta_q(\omega_2 r -\omega_1 s) \;
 .\notag  
\end{align} 
The last step might not be immediately obvious. It is verified by
showing the following:\footnote{I thank Nils Strodthoff for verifying
  that these two conditions are indeed satisfied.}  

(a) There is at most one $t$ in $\{0,\dots
q-1\}$ which solves the two conditions $\omega_1 t = r \bmod q$ and 
$\omega_2 t = s \bmod q $ at the same time. If it does, then the
condition $\omega_2 r - \omega_1 s = 0 \bmod q$ is also satisfied. 

(b) If there is no solution $t$ to the first two conditions, then the
new condition $\omega_2 r - \omega_1 s = 0 \bmod q$ does not have one
either. 

Therefore, we have shown that the terms in brackets on the right in
\Eq{eq:dualP2} agree with the random cluster representation
(\ref{eq:PottsTorusC}) of a Potts model partition function 
$Z_q^{(-s,r)} $ with boundary conditions $(-s,r)$, and with the
replacement $ v\to q/v$  which is equivalent to replacing the coupling
per temperature $K$ by its dual $\widetilde K$ according to their
duality relation in \Eq{eq:dualK}. Analogously rewriting the prefactor on the
right in \Eq{eq:dualP2} with \Eq{eq:dualK},
\begin{equation} 
 v^{N_s^*} = (e^K-1)^{N_s^*} = \frac{q^{N_s^*}}{(e^{\widetilde
     K}-1)^{N_s^*}} \;  ,
\end{equation} 
and rearranging \Eq{eq:dualP2} to solve for the terms in brackets on
the right, equalling $Z_q^{(-s,r)}(\widetilde K) $, then finally yields the
selfduality relation (\ref{Duality}) of the $2d$ $q$-state Potts models
(\ref{PottsH}) for all $q$. The result here is actually a bit more general than
that given in (\ref{Duality}) where we have used
$N_\mathrm{sites}\equiv N_s = N_s^*$ for a $2d$ square lattice.   
More generally, the duality relation valid for {\em any} lattice
$\mathcal L$ on a $2d$ torus without open ends, and symmetric in
$\mathcal L $ and $\mathcal L^*$, here follows from \Eq{eq:dualP2} as
\begin{align} 
\big(e^{\widetilde K} -1\big)^{-N_s^*} \; Z_q^{(-s,r)}(\widetilde K)
&= \label{Duality2} \\
& \hskip -1.2cm  \big(e^K-1 \big)^{-N_s} \;  \frac{1}{q} \sum_{m,\, n}
\, e^{2\pi i \, (rm+sn)/q } \; Z_q^{(m,n)}(K) \; . \notag 
\end{align}
At criticality, with $K=\widetilde K$, the prefactors cancel and one
verifies that the values of the partition functions for the various
cyclic boundary conditions in \Eqs{eq:crit2},
(\ref{eq:crit3}), and (\ref{eq:crit4}) reproduce themselves under the
discrete Fourier transform (\ref{Duality2}) for $q=2$,
$3$ and $4$ as they must. The finite-volume duality relation is valid,
however, for all $q$, including $q>4 $ with first order transition.   

In the 2+1 dimensional gauge theories, temperature is the same on both sides of
the $Z_N$-Fourier transform in \Eq{eq:ZeZk}. As a consequence of the
selfduality of the corresponding spin models, however, the free
energies of spatial center vortices and those of the confining
electric fluxes are mirror images of one another within the universal
scaling window around the 
second order phase transition.

\medskip

\subsection{Universality and finite-size scaling}

\medskip

The ratios of partition functions with $(\vec k,\vec m)$-twisted and periodic 
boundary conditions in the SU$(N)$  gauge theories are obtained from
multiplying coclosed stacks $\Omega^*(\mu,\nu) $ of plaquettes
$ U\!_{\dAlember} $ by the center elements such that one plaquette in
every plane of orientation $(\mu,\nu)$ is multiplied  by the
$Z_{\mu\nu} = e^{2\pi i\, n_{\mu\nu}/N}$ corresponding to the twist
tensor $n_{\mu\nu} $ ({\it c.f.}, \Eqs{coc}),   
\begin{align}
\label{eq:twistlat}
\frac{Z_k(\vec k,\vec m)}{Z_k(0,0)} =& \frac{\int
  \prod\! dU \, 
\exp\big\{\! -\! \beta\! \sum\hskip -3pt_{\dAlember } \frac{1}{N} \,
  \mathrm{Re}\big(\, Z_{\dAlember}  \, \tr\, U\!_{\dAlember} \big) \big\} }
     {\int \prod \!dU \,  
 \exp\big\{\! -\!\beta \!\sum\hskip -3pt_{\dAlember } \frac{1}{N}
         \, \mathrm{Re}( \,\tr \, U\!_{\dAlember} )\big\} }
   \; , \notag \\[4pt]
&\hskip .6cm   Z_{\dAlember}   = \Bigg\{ \begin{array}{rl} 
 Z_{\mu\nu}  \, ,    & \dAlemb \in \Omega^*(\mu,\nu)\, ,\\[4pt]
 1 \, ,  & \mbox{otherwise} \, .   
\end{array} 
\end{align}
These coclosed stacks of plaquettes $\Omega^*(\mu,\nu)$ are dual to
non-homotopic closed surfaces or lines on the dual lattices in $3+1$
or $2+1$ dimensions, respectively. 

If the total number of plaquettes in all stacks $\Omega^*(\mu,\nu)$ 
is $N\!_{\dAlember} $, the ratio of partition functions in
(\ref{eq:twistlat}) can be converted into a product of
$N\!_{\dAlember} $ ratios, each of which represents the expectation
value of the $n^\mathrm{th}$ twisted plaquette $ Z_{\dAlember}^{(n)}
U\!_{\dAlember}^{\, (n)} $ in an ensemble $Z_k^{(n-1)} $ with $n-1$
already twisted ones, where $n$ runs lexicographically through the
plaquettes of one $\Omega^*(\mu,\nu)$ stack after another, to implement 
twisted b.c.'s  in all $(\mu,\nu)$-planes according to the non-zero
components of the twist tensor $n_{\mu\nu}$,
\begin{align}
\frac{Z_k(\vec k,\vec m)}{Z_k(0,0)} &=  \prod_{n=1}^{N\!_{\dAlember}} 
\frac{Z_k^{(n)}}{Z_k^{(n-1)}} = \prod_{n=1}^{N\!_{\dAlember}} 
\Big\langle e^{\frac{\beta}{N}  \mathrm{Re}\left(
    (1-Z_{\dAlember}^{(n)}) \, \tr\,
  U\!_{\dAlember}^{\, (n)} \right) } \Big\rangle_{(n-1)}   \, , \notag \\[2pt] 
&\quad Z_k^{(0)} \equiv Z_k(0,0) \;
,\;\;  Z_k^{(N\!_{\dAlember})} \equiv Z_k(\vec k,\vec m) \; .
\end{align}
In this way, one ratio of partition functions requires
$N\!_{\dAlember} $ independent Monte-Carlo simulations
\cite{deForcrand:2000fi,deForcrand:2001nd}.  
  
For the temporal $\vec k$-twists one introduces coclosed stacks
$\Omega^*$ aligned with the time direction between one pair of
adjacent time-slices, where they are dual to spatial lines in $2+1$ and
surfaces in $3+1$ dimensions, just as the respective spin interfaces
in $2$ and $3$ dimensions as illustrated in
Fig.~\ref{fig:VortexInterface} above. 

Once the reduced temperature $t(\beta) $ is determined in terms of the
lattice coupling $\beta$, as from \Eq{tbeta} for $2+1$ dimensional
SU$(2)$, and with the spatial lattice size  $L $ from \Eq{eq:Lfss}, a
finite-size-scaling analysis can be performed. In the vicinity of the
$2^\mathrm{nd}$ order phase transition generalized couplings such as
the vortex-ensemble ratios $R_k(\vec k)$, for 
sufficiently large $L$, only depend on the ratio of $L$ and the
large correlation length $\xi $ which in the infinite volume diverges
as
\begin{equation} 
\xi = f^\pm |t|^{-\nu} + \, \cdots  \; , \;\; t \to 0^\pm \; .
\end{equation}
with the correlation-length critical exponent $\nu $, where for the
2$d$ Potts models $\nu = 4/3$ for $q=1$ (percolation), $\nu = 1$ for
$q=2$ (Ising model), $\nu = 5/6$ for $q=3$ and $\nu = 2/3$ for $q=4$
\cite{Baxter:2007} . Here we are particularly interested in the  
ratios $R_q^{(m,n)}$ which, up to finite size corrections $\sim
N^{-\omega}$ on an $N\times N$ square lattice, depend on $x= N^{1/\nu} t$ in a
universal way. The corresponding universal scaling functions $f_I(x)$ for the
interface free energies per temperature, suppressing the indices
$(m,n)$ for the boundary conditions, are given by 
\begin{align}
F_{I}(N,K) &= \, f_{I}(N^{1/\nu} t) \, +\,  c_{I} \, N^{-\omega} +\,
\cdots \; , \;\; \mbox{with} \notag\\
  f_I(0) &= -\ln R_{c} \; , \;\; \mbox{and} \label{eq:int-fss}\\  
f_{I}(x)  &= \sigma_0 \, (-x)^\mu +\, \cdots  \; , \;\; x \to -\infty \; , \notag
\end{align}
where $K = K_c/(t+1)$, $K_c= \ln(1+\sqrt{q})$, and $R_c$ stands for
the universal ratio $R_{q,c}^{(m,n)} $ with the particular combination of
boundary conditions that is being used in the $q=2$, $3$ and $4$ state
Potts models as per \Eqs{eq:crit2}, (\ref{eq:crit3}) or (\ref{eq:crit4}).

\begin{figure}[t]

\vspace*{.1cm}

\hskip .1cm
\centerline{\includegraphics[width=0.9\linewidth]{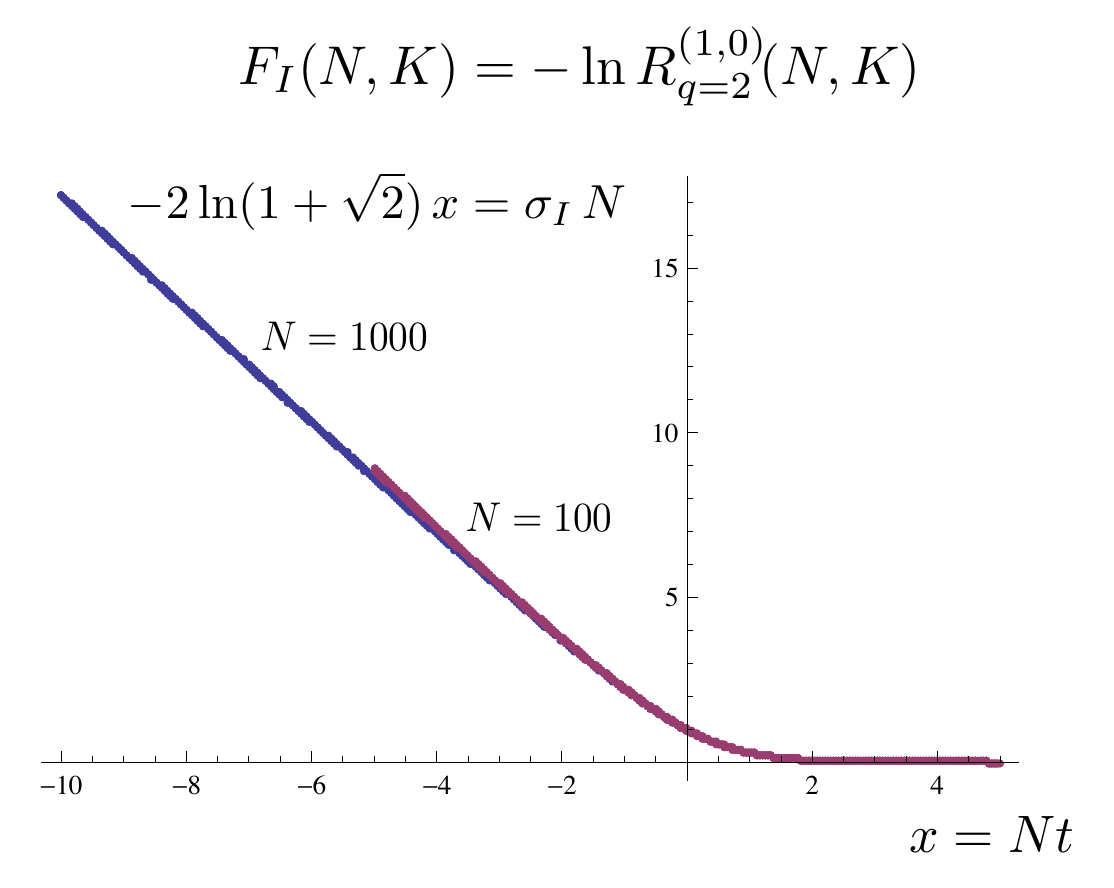}}

\vspace{-.2cm}
\caption{The universal scaling function $f_I(x)$ calculated for a $(1,0)$
  interface corresponding to one antiperiodic direction 
  from the ratio exact finite volume partition functions
  $Z_{q=2}^{(1,0)}/Z_{q=2}^{(0,0)}$ of the $N\times N$
  square Ising model in Ref.~\cite{Wu:1999}.
\vspace*{-.2cm}} 
\label{fig:IsingInterfaceFSS}
\end{figure}

For the $2d$ square Ising model, with $\nu =1$, the finite volume 
partition functions $Z_{q=2}^{(m,n)}(N,K)$ have been obtained exactly
for all combinations of periodic and anti-periodic boundary conditions
in Ref.~\cite{Wu:1999}.   We can use these results for the
ratios $R_{q=2}^{(1,0)}$ and $R_{q=2}^{(1,1)}$ to obtain the
corresponding universal scaling functions $f_I(x)$ from \Eq{eq:int-fss}.
In Fig.~\ref{fig:IsingInterfaceFSS}, we plot
$F_{I}(N,K)$ for the $(1,0)$ interface, for example, over $x=Nt$ for
$N=100$ and $N=1000$. We have tested that the difference between
$F_I(N,K)$ and $f_I(x)$ vanishes as $N^{-\omega}$ with $\omega = 2$
\cite{Edwards:2009qw}. The corresponding small residual finite-volume
effects are observed, {\it e.g.}, in the deviation of the $N=100$
result around $x=-5$ from that for $N=1000$. To check which lattice
size is needed for any desired accuracy, one can calculate the
deviations of $F_I(N,K)$ from the asymptotic slope $\sigma_0 $ for
$x\to -\infty$ which is also known analytically.

Generally, the interface tensions (per temperature) in the $d$
dimensional spin models are defined in the ordered phase as %
\begin{equation} 
\label{eq:int-def}
\sigma_I = \lim_{N\to\infty} N^{-(d-1)}  F_I(N,K)\; , \;\; K > K_c \; .
\end{equation}   
Comparison of this definition with \Eqs{eq:int-fss} thus entails that,
if $\sigma_I $ is finite in this limit, the interface tensions near
criticality must behave as
\begin{equation} 
\label{eq:int-defcr}
\sigma_I = \sigma_0 (-t)^\mu + \, \cdots \; ,\; \mbox{with} \;\; \mu =
(d-1)\nu \; ,\;\; t \to 0^- \; .
\end{equation}   
The exponent of the interface tension  $\mu $ is tied to the
correlation length exponent $\nu$  by one of the so-called
hyperscaling relations.  
In particular, for the $2d$ square Ising model with one antiperiodic
direction, with $\mu=\nu=1$ and thus $\sigma_I = -\sigma_0 t \, +
\, \cdots $, the asymptotic slope $\sigma_0 $ follows from Onsager's  
famous result \cite{Onsager:1943jn}, valid for all $0\le T\le T_c$ in
the ordered phase with $N\to \infty$, for the tension of a straight 
$(1,0)$ interface in the thermodynamic limit, 
\begin{equation} 
\label{eq:ifOnsager}
\sigma_I^{(1,0)}  = 2 K + \ln\tanh K  \; .
\end{equation}
When the spin-coupling per temperature $K=J/T$ is expanded about
criticality at $K_c = \ln(1+\sqrt{2})/2$ this leads to,\footnote{We
  use the typical Ising model conventions for $J$ such that $K\to K/2$
  here as compared to the general Potts model formulae above.} 
\begin{equation} 
 \sigma_0^{(1,0)} = 2\ln(1+\sqrt 2) \; ,
\end{equation}
and this agrees with the asymptotic slope of $f_I(x)$ for $x\to
-\infty$ in Fig.~\ref{fig:IsingInterfaceFSS} as obtained from the
exact finite volume partition functions   $Z_{q=2}^{(1,0)}$ and
$Z_{q=2}^{(0,0)}$ of Ref.~\cite{Wu:1999}.

\begin{figure}[t]

\vspace*{.1cm}

\hskip .1cm
\centerline{\includegraphics[width=0.9\linewidth]{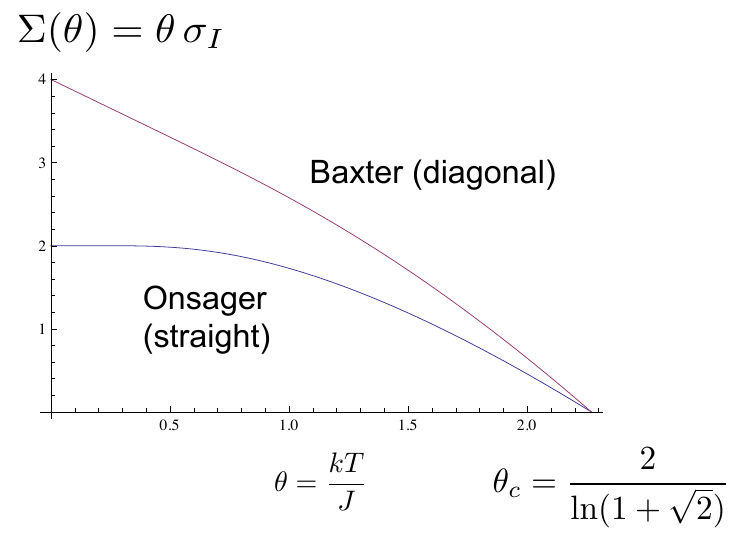}}

\vspace{-.2cm}
\caption{The exact interface tensions per coupling, $\Sigma(\theta) =
  \theta\sigma_I $ over temperature $\theta =T/J$, for the straight
  $(1,0)$ interface from Onsager, \Eq{eq:ifOnsager}, and for the diagonal
  $(1,1)$ interface from Baxter, \Eq{eq:ifBaxter}. 
\vspace*{-.6cm}} 
\label{fig:IsingInterface}
\end{figure}

Many alternative and simplified derivations of Onsager's result
(\ref{eq:ifOnsager}) were given since then and can be found in text
books \cite{Baxter:2007,Itzykson:1989}. A particularly interesting
one for us is given in Baxter's book \cite{Baxter:2007}, as it
provides the tension of a diagonal $(1,1)$ interface. In our notations
his result can be written as
\begin{equation} 
\label{eq:ifBaxter}
\sigma_I^{(1,1)}  = 2 \ln\sinh(2K)  \; .
\end{equation}
As the free energies so far, these are all interface tensions per
temperature. The proper free energy of an interface is given by
$- T\ln R_q =  T \sigma_I N $. The corresponding proper interface
tensions,
\begin{equation} 
\Sigma(\theta) \equiv \, \theta \,\sigma_I \; ,\;\; \theta = T/J = K^{-1}\; ,
\end{equation}
per coupling $J$ over temperature, for $\theta = 0$ to $\theta_c =
 K_c^{-1}$, following from \Eqs{eq:ifOnsager} and 
(\ref{eq:ifBaxter}) are compared in Fig.~\ref{fig:IsingInterface}. At
$T=0$, every antiferromagnetic bond-coupling costs $2J$ in energy. 
The diagonal or zigzag interface is twice as long as the straight one.
It needs twice as many antiferromagnetic couplings, hence, in units of
$J$, $\Sigma(0) = 2$ for the straight and $4$ for the diagonal interface.  
In the isotropic limit, when the correlation length is much larger
than the lattice spacing, on the other hand, the underlying lattice
does not matter anymore and the diagonal interface through a large
finite volume is only a factor of $\sqrt{2}$ longer than the straight
one, {\it i.e.}, from \Eq{eq:ifBaxter} around $K=K_c$,
\begin{equation} 
 \sigma_0^{(1,1)} = 2\sqrt{2}  \ln(1+\sqrt 2) = \sqrt 2\, \sigma_0^{(1,0)} 
 \; .
\end{equation}
The same square-root ratios near $T_c$ are observed for the $3d$ Ising
interfaces \cite{Pepe:2001cx},
\begin{equation} 
\label{eq:stringform}
 \sigma_0^{(1,0,0)} \, : \, \sigma_0^{(1,1,0)} \, : \,
 \sigma_0^{(1,1,1)} \, \sim \, 1\, : \, \sqrt{2} \, : \, \sqrt{3} \; . 
\end{equation}
The spin-interface tension $\sigma_I$ corresponds to the dual string tension
$\tilde\sigma $ for spatial center vortices in $2+1$ and spatial 't
Hooft loops in $3+1$ dimensions for which a dual area law holds in the
high-temperature $Z_N$-broken phase. The same square root
ratios (\ref{eq:stringform}) are  observed for the dual string tension 
in SU$(2)$ \cite{deForcrand:2001nd,vonSmekal:2002ps}.  From the
duality \Eq{eq:ZeZk}, on can show that they then must also hold for the
electric fluxes in the confined $Z_N$-disordered phase below $T_c$,
and they are interpreted as the smoking gun of string formation: The
free energy of orthogonal fluxes is minimized when their length is,
{\it i.e.}, when diagonal strings through the volume form.

At least for the $2+1$ dimensional SU$(2)$ gauge theory string formation just
below the deconfinement transition temperature is thus proven from
universality and the observed square-root ratios of the exact Ising
model interface tensions in the isotropic limit.  

Unlike the spin models, however, whose interface tensions 
at $T=0$ follow ratios $1\,:\, 2 \,:\, 3$ as for isotropic
flux, the electric fluxes in the SU$(N)$ gauge theories are expected
to also show the square-root ratios for string formation at zero
temperature. The numerical evidence from early Monte-Carlo results
on the lattice is affirmative that this is indeed the case
\cite{Hasenfratz:1989tp}.   

The (2+1)$d$ SU$(2)$ vortex-ensemble ratios $R_k(\vec k)$ were
calculated on $N_t\times N_s^2$ lattices for  $N_t = 2$ to $10$, each
with spatial sizes up to $N_s=96$  and in a suitable window of lattice
couplings $\beta$ around $\beta_c$ to test the finite-size scaling (FSS) in
Refs.~\cite{Edwards:2009qw,vonSmekal:2010xz,vonSmekal:2010du}.  
For each fixed $N_t$ one generally observes 
very good scaling of the available data for all $N_s$ when plotted
over the FSS variable $x = L^{1/\nu} t$.  With $\nu=1$, the physical
spatial length $L $ from
\Eq{eq:Lfss} in units of $T_c^{-1} $, and the reduced temperature 
$t(\beta)$ from the lattice coupling $\beta $ via \Eq{tbeta}, we
define,
\begin{equation}
x \, \equiv L T_c t \,  = \frac{N_s}{N_t} \, \frac{t(\beta)}{1+t(\beta)} 
\end{equation}
For the resulting vortex free energies $F_k$, for $\vec k = (1,0) $
and $(1,1)$,  we then obtain accurate one-parameter fits via 
\begin{equation}
F_k(x) = f_I(-\lambda x) \; ,
\end{equation}
to the exact universal scaling functions $f_I(x)$ for $(1,0)$
and $(1,1)$ b.c.'s computed from the results of
Ref.~\cite{Wu:1999} as in
Fig.~\ref{fig:IsingInterfaceFSS}. This determines the single
non-universal parameter $\lambda \equiv \lambda(N_t)$ 
which is dimensionless and relates 
the SU$(2)$ FSS variable to the Ising one, 
\begin{equation}
 x_\mathrm{Ising} \,=\, -\lambda(N_t) \, x_{\mathrm{SU}(2)}
\end{equation}
whereby the minus sign reflects the interchange of high and low
temperature phases between the two.

\begin{figure}[t]

\vspace*{.1cm}

\hskip .16cm
\includegraphics[width=0.96\linewidth,trim=0.8cm 0 0
0,clip=true]{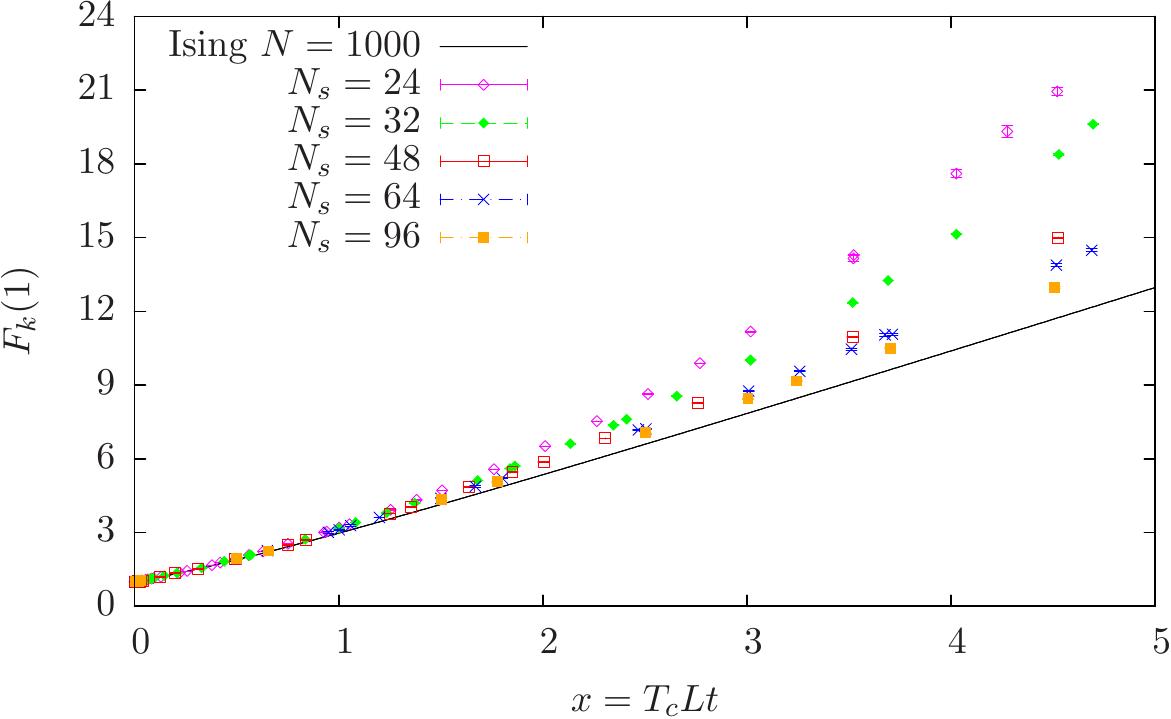} 

\vspace*{-2.3cm}
\leftline{\small\hskip 5cm $\nwarrow $}
\vspace{-.16cm}
\leftline{\scriptsize \hskip 5.3cm $f_I(-\lambda x)\sim \rho x$}

\vspace{1.4cm}
\caption{Vortex free energies $F_k$ in SU$(2)$ above $T_c$ for $N_t=4$.}
\label{fig:fss-nt4-free-energy}
\end{figure}

\begin{figure*}[t]

\vspace*{.1cm}

\hskip .2cm\includegraphics[width=0.98\linewidth,trim=.8cm 0 0
0,clip=true]{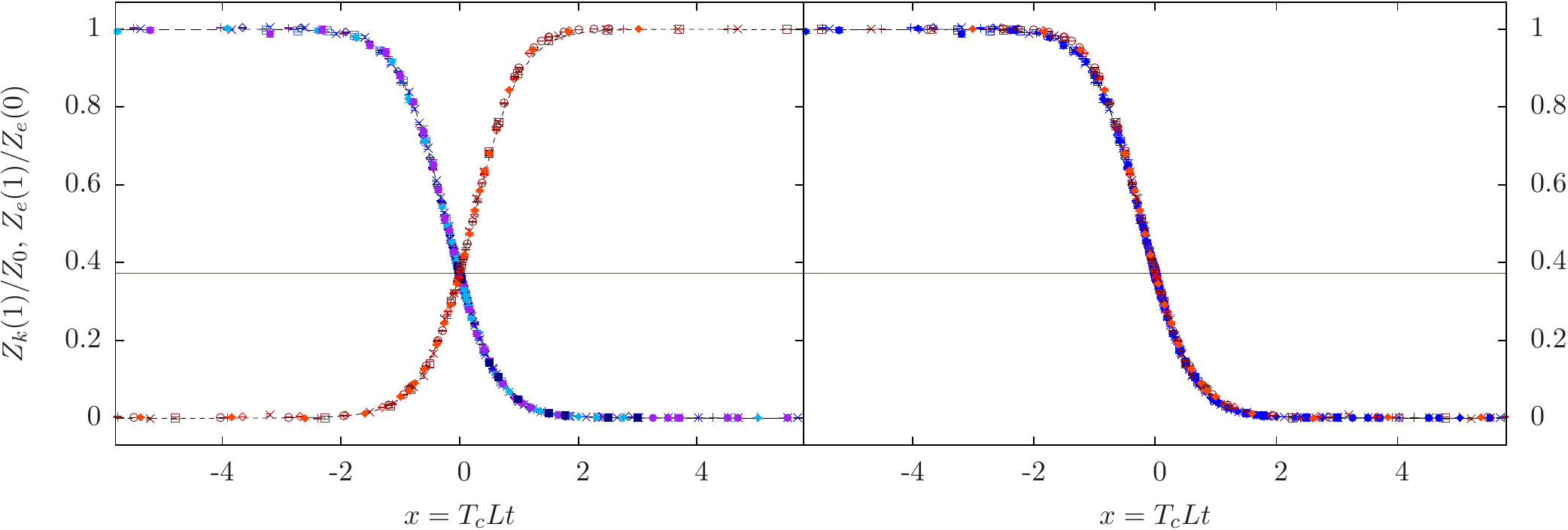} 

\vspace*{-5.3cm}
\leftline{\footnotesize\hskip 1.1cm $R_k(x) $ \hskip 5.5cm
  $R_e(x) $ \hskip .4cm $R_k(x),\, R_e(-x) $}

\vspace{4.6cm}
\caption{The ratio  $R_k(\vec k)$ of the partition function with twist
  $\vec k = (1,0)$ over the periodic ensemble in the $2+1$ dimensional
  SU$(2)$ gauge theory compared to $R_e(\vec e)$ for one unit of
  electric flux  $\vec e =   (1,0)$ relative to the no-flux ensemble (left), 
  and to its mirror image (right). The data here is obtained for $N_t
  = 4$ and spatial volumes up to $N_s=96$ \cite{vonSmekal:2010xz}. The dashed lines represent
  the universal scaling function from the $2d$ Ising
  model, $\exp\{-f_I(-\lambda x)\} $ for  $R_k$ and
  $\exp\{-f_I(\lambda x)\} $ for  $R_e$ from selfduality. \vspace{-.2cm}} 
\label{fssoffsetduality-multiplot}
\end{figure*}

How hard it is to precisely determine the dual string tension $\tilde
\sigma $ directly from the data for the vortex free energy is shown in
Fig.~\ref{fig:fss-nt4-free-energy}. We need 
to get to asymptotically large $x$, {\it c.f.}, \Eqs{eq:int-fss}, but
that requires larger and larger $L = a N_s$ in order to stay within the
universal scaling window for sufficiently small $t$. Note that $x=4$
here roughly amounts to $x\approx -6$ in
Fig.~\ref{fig:IsingInterfaceFSS} where $N_s = 100$ is not large enough
to suppress finite-size corrections even in the Ising model. The
computational costs in the $d+1$ dimensional gauge theory roughly
increase as $N_s^{d+2(d-1)}$ so it becomes rather expensive to beat the 
FSS corrections to the dual string tension by brute force, and this was
observed even more so in $d=3$ \cite{deForcrand:2001nd}.

Here, our one-parameter fits to the vortex free energies as per
$ F_k(x) = f_I(-\lambda x) $ to the exact universal scaling functions 
help tremendously because we can obtain accurate data at
comparatively low computational costs for these fits from small $x$
values, {\it i.e.}, small volumes. Once $\lambda $ is fixed, however,
we can compute the dual string tension and, as we will see,
also the string tension for electric flux from these.

This is because with the hyperscaling relations $\mu = (d-1) \nu$, 
the product of interface tension and correlation length,
$\sigma_I\xi^{d-1}$ is independent of $t$ near criticality.  In fact,
this product is another example of a universal amplitude
ratio \cite{Pelissetto:2000ek}. To be precise, if we use the so-called
exponential correlation length in the disordered phase of the spin
model above $T_c$, $\xi_\mathrm{gap}^+ $, 
then the universal constant is given by  
\begin{align} 
\label{eq:xiGap}
&\sigma_I (\xi_\mathrm{gap}^+)^{d-1}  = {R_\sigma^+}_\mathrm{gap} \;
,\;\;\mbox{where}  \\[4pt]
&{R_\sigma^+}_\mathrm{gap} = 
\Bigg\{ \begin{array}{l} 
 1\; , \; \mbox{for}\;\; q = 2,\,3,\,4 \;, \; \mbox{in} \;\; d=2 \;, \\[4pt]
 0.40(1) \; , \; \mbox{for}\;\; q = 2\; ,
\; \mbox{in} \;\; d=3 \; .
\end{array}
\notag
\end{align}
The $d=2$ results are again exact
\cite{Wu:1982ra,Baxter:2007,Itzykson:1989},\footnote{The $q=2$ result is found
in any of the textbooks, for $q=3$ it follows from the hard hexagon
model \cite{Baxter:2007}, and for $q=4$ probably from the Baxter-Wu
(triangular three-spin) model \cite{Baxter:1973zz,Baxter:2007}, but I
am unsure about the status of a mathematical proof of the latter.}  
 while the $3d$ Ising ratio is determined numerically, {\it e.g.}, see
 the review in Ref.~\cite{Pelissetto:2000ek}.

The exponential correlation length corresponding  to $
\xi^+_\mathrm{gap} $ on the gauge theory side is that of the Polyakov
loops in the confined phase below $T_c$, which is given by the string
tension $\sigma$ and temperature as
\begin{equation} 
 \frac{1}{\xi^{-}} = \frac{\sigma}{T} \; .
\end{equation}
Therefore, we have a universal relation between
the string tension per $T$ below $T_c$, playing the role of
${\xi^+_\mathrm{gap}}^{-1} $ in \Eq{eq:xiGap}, 
and the dual string tension $\tilde\sigma$ above $T_c$, as the
interface tension $\sigma_I $ in \Eq{eq:xiGap},
\begin{align}   
 \tilde\sigma \,&=\, {R_\sigma^+}_\mathrm{gap}\,  (\sigma/T)^{d-1},  \;\;
 \mbox{around criticality, or} \notag\\ 
 \sigma \,&= \rho T_c^2 (-t)^\nu +\cdots  \, , \; T < T_c\; ,
 \;\;\mbox{and} \\
 \tilde\sigma \,&= {R_\sigma^+}_\mathrm{gap} (\rho
 T_c)^{d-1} t^{(d-1)\nu} + \cdots \, , \; T > T_c \; ,\notag
\end{align}
where we have introduced an unknown non-universal constant $\rho$ which
drops out from the universal ratio, and we have replaced one factor $ T $
by $T_c$ in the string tension in second line as we work at leading
order in the reduced temperature $t$ here. Instead of determining 
the constant $\rho$ from the vortex free energies in the
high temperature phase at asymptotically large $x = (T_cL)^{1/\nu}  \,
t $, here we can use the exact universal scaling function $f_I(x)$ 
 from the Ising model: We know the asymptotic slope of
$f_I(x)$ which is $2\ln(1+\sqrt{2})$, therefore, from 
\begin{equation}
F_k(x) = f_I(-\lambda x) \; \Rightarrow\;\; \rho = 2\lambda
\ln(1+\sqrt{2}) \; .
\end{equation}
Once we extract $\lambda$ from our one-parameter fits
at small $x$ where we have very accurate data, this then equally
accurately determines the string tension and its dual around $T_c$.
  Moreover, we can determine this single non-universal parameter
  $\lambda $ for $N_t = 4,\, 5 \dots 10$ which allows a polynomial fit, 
\begin{equation} 
 \lambda(N_t) = \lambda_\infty + b/N_t + c/N_t^2 \; , 
\end{equation}
with the extrapolated result $\lambda_\infty = 1.354(25)$
\cite{vonSmekal:2010xz}.
This then determines the leading behavior of the {\em continuum} 
string tension and its dual around the phase transition,
\begin{align} 
 \sigma &= \lambda_\infty T_c^2 2\ln(1+\sqrt 2)\, |t| + \cdots , \;\;
 t\to 0^-\;, \;\; \mbox{and} \notag \\
 \tilde\sigma &= \lambda_\infty T_c 2\ln(1+\sqrt 2)\, t + \cdots ,
 \;\;  t\to 0^+\; .  
\end{align}

\begin{figure}[t]

\vspace*{.1cm}

\hskip .12cm
\includegraphics[width=0.96\linewidth]{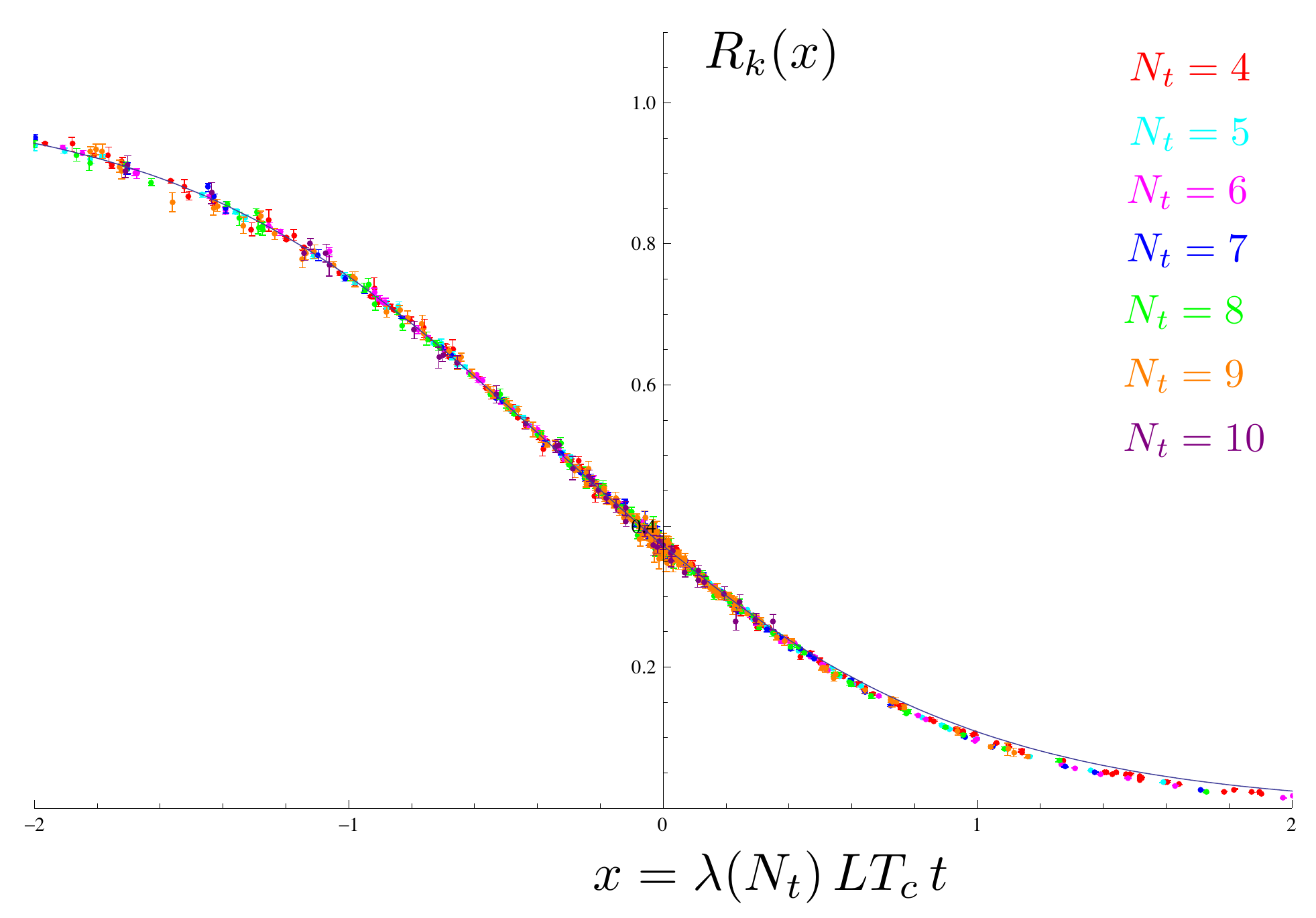}

\vspace{-.1cm}
\caption{SU$(2)$ center-vortex ensembles with renormalized FSS
  variable $\lambda x$ for $N_t$ up to $10$, and $N_s$ 
  up to 96 as in Fig.~\ref{fssoffsetduality-multiplot}, from     
  \cite{vonSmekal:2010du}.}   
\label{fig:Ntscale}
\end{figure}

Once we have fixed the single non-universal parameter $\lambda$ we can plot
the calculated center vortex free energies $F_k $ or the corresponding ratios
$R_k(\vec k)$ in comparison with $\exp\{-f_I(-\lambda x)\} $ as in
Fig.~\ref{fssoffsetduality-multiplot} for $\vec k = (1,0)$. The data
for all different spatial lattice sizes collapse beautifully onto the
universal curve, the finite-size scaling description works rather well.  
Together with the corresponding  $\vec k = (1,1)$ result for two orthogonal
center vortices we can perform the $2d$ $Z_2$-Fourier transform
(\ref{eq:ZeZk}) to obtain the ratios of electric flux $\vec e$ over
the no-flux ensemble in \Eq{mirror},
\begin{equation}
R_e(\vec e) \equiv \frac{Z_e(\vec e)}{Z_e(0)} =  \frac{1}{N}\, \Big\langle
\mbox{tr}\Big( P(\vec x) P^\dagger (\vec 
x+ \vec e L) \Big) \Big\rangle_{\mbox{\scriptsize no-flux}}\, . 
\end{equation}
These are equally well described by $\exp\{-f_I(\lambda x)\} $ without
refitting the non-universal  parameter $\lambda$, they are mirror
images of the $R_k(\vec k)$ in a surprisingly large scaling window
around criticality as seen in the right panel of  \Eq{mirror}, where
we swapped $x \to -x$ for the electric fluxes to demonstrate
that. This is the manifestation of the selfduality of the
corresponding $2d$ spin model.

Moreover, if we rescale the FSS variables by the parameters
$\lambda(N_t)$ determined from the one-parameter fits for the 
individual $N_t= 4\, ,\dots 10$ data sets, the scaling of the data over
all spatial lattice sizes $N_s$ can be extended to include different
$N_t$ lattices. With this renormalization, one thus essentially
obtains good continuum results already with the coarsest $N_t=4$
lattice here, as seen in the zoom-in plot around criticality 
in Fig.~\ref{fig:Ntscale}. The rescaled $x$-range there, roughly 
corresponds to $x = [-1.33,1.33] $  in 
Fig.~\ref{fssoffsetduality-multiplot}, and the dashed line is obtained
from the universal scaling function as $\exp\{-f_I(x)\} $ and thus now
independent of $N_t$. The slight deviations of the vortex free
energy from this curve near $x=1$ seen here, are predominantly due to        
the leading finite-size corrections $\sim L^{-\omega} $, analogous to
what is being observed in the ordered phase of the Ising model
in Fig.~\ref{fig:IsingInterfaceFSS} at negative $x$.

\begin{figure}[t]

\vspace*{.1cm}

\hskip .12cm
\includegraphics[width=0.96\linewidth]{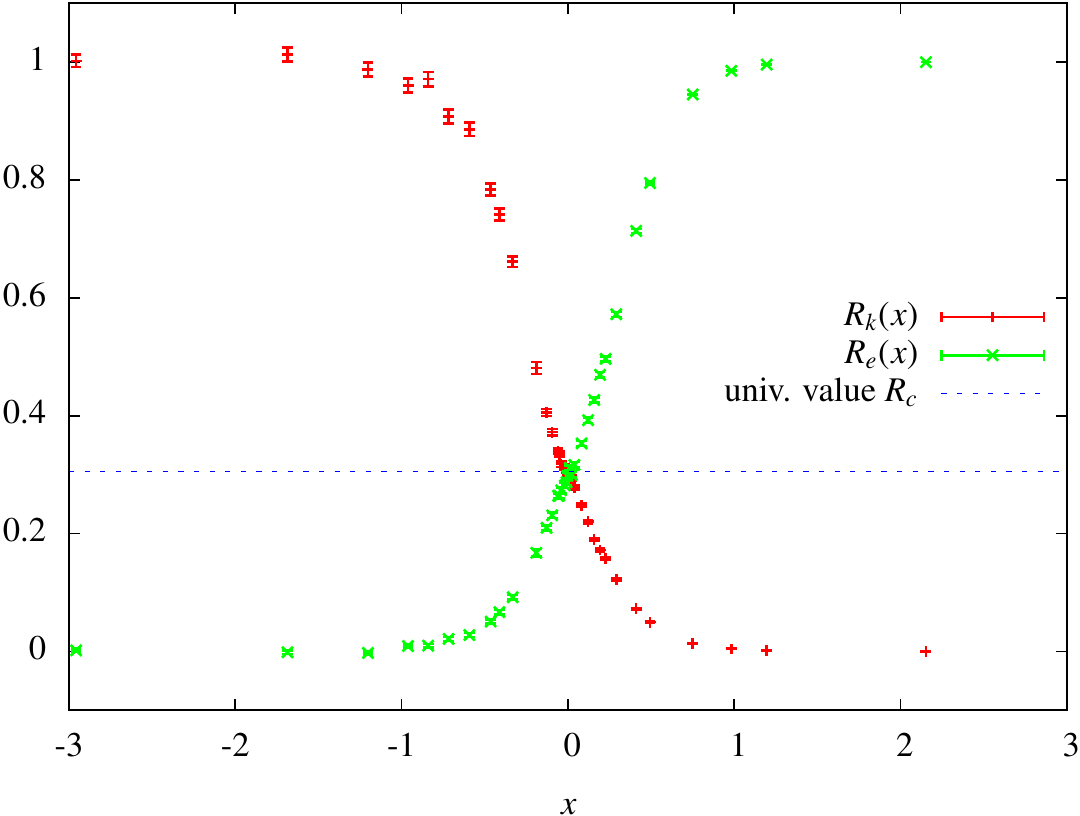} 

\vspace*{-.43cm}
\leftline{\footnotesize\hskip 4.2cm $ = \pm LT_c |t|^\nu $}

\vspace{-.1cm}
\caption{Selfduality in $2+1$ dimensional SU$(3)$ from
  \cite{Strodthoff:2010dz}.}   
\label{fig:su3selfduality1}
\end{figure}

The selfduality of the underlying Potts model is also observed in the $2d$
SU$(3)$ gauge theory \cite{Strodthoff:2010dz}, as shown in
Fig.~\ref{fig:su3selfduality1}. The dashed line in this figure marks the
universal ratio $R_{3,c}^{(1,0)} $ from \Eqs{eq:crit3}. The two
independent ones that one can form on a symmetric 
lattice for $q=3$ evaluate to 
\begin{align} 
R_{3,c}^{(1,0)}&=R_{3,c}^{(2,0)} = 0.30499982\dots \; , \;\; \mbox{and} \\
R_{3,c}^{(1,1)}&=R_{3,c}^{(2,2)} = R_{3,c}^{(1,2)} =  0.19500018\dots
\; . \notag 
\end{align}
Even though the data in Fig.~\ref{fig:su3selfduality1} was obtained on
a quite small $N_t =2$ and $N_s=24$ lattice, the intersection point of
the vortex and electric flux ensembles $R_k$ and  $R_e$ is almost
right on the universal line. To exploit selfduality, where possible,
turns out to be the fastest converging method to determine critical
couplings by far.

\begin{figure}[t]

\vspace*{.1cm}

\hskip .12cm
\includegraphics[width=0.96\linewidth]{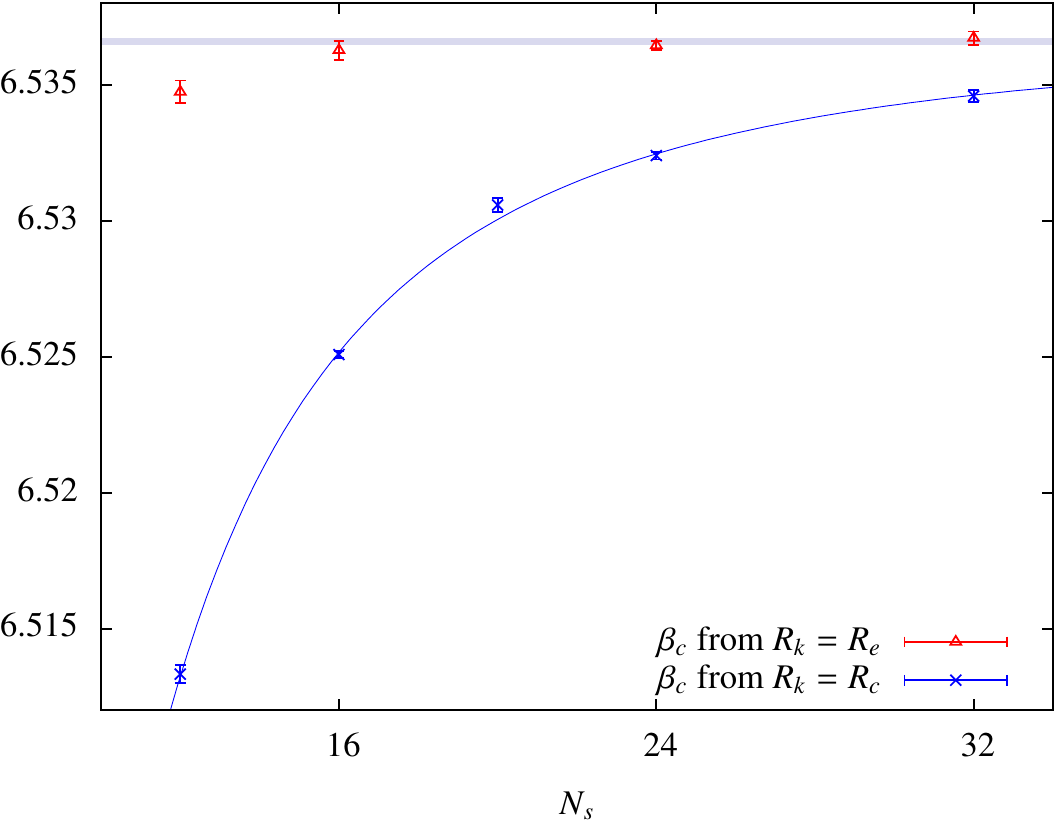} 

\vspace{-.1cm}
\caption{Critical couplings for (2+1)$d$ $SU(2)$ ($N_t= 4$) from
  selfduality compared to those of Ref.~\cite{Edwards:2009qw}, with
  $N_s$ up to 96 and the infinite-volume extrapolated result
  $\beta_c=6.53661(13) $ shown as the narrow grey band here, from 
  \cite{vonSmekal:2010du}.}   
\label{fig:su2crit}
\end{figure}

There is a long history of methods to extract critical couplings or
temperatures from simulations in finite volumes, going back to using
pairwise intersections of Binder cummulants on successively larger
lattices \cite{Binder:1981sa}. Hasenbusch later demonstrated that the ratios of
partition functions with different boundary conditions 
could be used in the same way to obtain a much more rapid convergence
with very good estimates already from rather small lattices
\cite{Hasenbusch:1993wn}.  At criticality, these ratios tend to
the universal values $0<R_{c}<1 $ in the thermodynamic limit. In
\cite{Edwards:2009qw} it was therefore shown how to obtain critical
couplings for gauge theories from intersecting the ratios $R_k$ of finite
volume partition functions with these universal fixed points, once 
their values are known. For (2+1)$d$ SU(2) this led to an even much 
faster convergence than their pairwise intersections. At the time we
thought this is the best method, but now look at the comparison of the
results from \cite{Edwards:2009qw} with the intersection points from
selfduality in Fig.~\ref{fig:su2crit}, again for SU$(2)$ in $2+1$ dimensions as
a benchmark. Already with $N_s =16$, and without any extrapolation,
the result from selfduality is within the errors of the
extrapolated best infinite volume result,  $\beta_c=6.53661(13) $ from
Ref.~\cite{Edwards:2009qw}, as indicated by the narrow grey band in
Fig.~\ref{fig:su2crit}.  

The reason for this impressive result is that the leading finite-size
corrections do not change the position of the intersection point in
$\beta$. They only move it upwards along a straight vertical line with
increasing $L$ in plots such as the one of
Fig.~\ref{fig:su3selfduality1}.

To see this explicitly, first use a finite-size scaling ansatz for the
vortex ensemble ratios $R_k$ as functions of the lattice coupling $\beta$
around criticality of the form   
\begin{equation}
R_k(\beta)=R_c+ b \left(\beta -\beta_{c} \right)N_s^{1/\nu}+c
N_s^{-\omega}+\cdots, \label{fssans}  
\end{equation} 
Defining pseudocritical couplings $\beta_c(N_s,N_t)$ in a finite
volume by requiring that $R_k(\beta) = R_c$ for the method of
\cite{Edwards:2009qw} leads to, 
\begin{equation}
\beta_c(N_s,N_t)=\beta_{c}(N_t)-(c/b)\, N_s^{-(\omega+1/\nu)} + \cdots .
\label{eq:pseudocrit}
\end{equation}
These extrapolate to $\beta_{c}(N_t)$ from large spatial lattice sizes
$N_s$ at fixed numbers of time slices $N_t$. As a byproduct this
method gives numerical estimates of the correction to scaling
exponent $\omega $. 

With selfduality, however, one must then have $R_e(\beta ) = R_k(\beta)$
for like values of electric flux $\vec e$ and and temporal twist $\vec
k$ at $\beta=\beta_c$. In fact, it is straightforward to verify 
 that then,
\begin{equation}
R_e(\beta)= R_c -  b \left(\beta -\beta_{c} \right)N_s^{1/\nu}+c
N_s^{-\omega}+\cdots,  
\end{equation} 
with the same coefficients $b$ and $c$ as in~(\ref{fssans}). 
 Therefore, the leading finite-size corrections to $\beta_{c}$ when defined
by $R_e = R_k$ cancel. At criticality, 
\begin{equation}
R_e(\beta_c) = R_k(\beta_c)
= R_c + c N_s^{-\omega} + \cdots \; ,
\end{equation}
so the leading corrections only
move the intersection point of $R_e(\beta) $ and $R_k(\beta)$
vertically without shifting the so defined critical coupling. 

The results for SU$(3)$ from the  $R_e = R_k$ intersection
points are compared to the extrapolated pseudocritical couplings  
from intersecting $R_k(\beta)$  with its universal critical value 
$R_c$ via \Eq{eq:pseudocrit} in Tab.~\ref{tab1}. 

In 2+1 dimensions the critical couplings grow linearly with $N_t$ to
leading order at large $N_t$, {\it c.f.} \Eq{eq:betavsNt}, and the
critical temperature in units of the dimensionful continuum coupling
$g_3$ is given by the slope, 
\begin{equation}
\beta_c(N_t)/(2N_c)  =  (T_c/g^2_3) \, N_t + \cdots \; . 
\end{equation}
From the values for $\beta_c$ with $N_t=4$, $6$ and $8$ in
Tab.~\ref{tab1},  one obtains for SU$(3)$ \cite{Strodthoff:2010dz},
\begin{equation}
\frac{T_c}{g_3^2}=0.5475(3)
\end{equation}
corresponding to ${T_c}/{\sqrt{\sigma}}=0.9938(9)$ with a zero
temperature string tension $\sqrt{\sigma}/g_3^2 = 0.5509(4) $ from a
weighted average of the four values in \cite{Bringoltz:2006zg}. This
is consistent with  $T_c/{\sqrt{\sigma}}=0.9994(40)$ from \cite{Liddle:2008kk}. 
 
Moreover, because the spatial center vortex free energies $F_k$ 
for sufficiently large $L$ depend only on $L^{1/\nu} t$, and $t
\propto (\beta -\beta_c)$ at leading order in $N_t$, when
expanding 
\begin{equation} 
F_k(\beta ) = -\ln R_c + d(N_s) (\beta-\beta_c)+ \cdots \;,
\end{equation}
one expects the slope at $\beta_c$ to behave as 
\begin{equation}
d(N_s)\sim N_s^{1/\nu} \; .
\end{equation}
Our current best estimates from fitting these slopes are:
$\nu = 0.99(3)$ for SU$(2)$ where one expects $\nu =1$ for the $2d$
Ising model, $\nu=0.818(24)$ for SU(3) where 
$\nu=5/6\approx 0.833$ for the 2$d$ 3-state Potts model, and 
$\nu = 0.673(9)$ for SU$(4)$ as compared to $\nu = 2/3$ for the $q=4$
Potts model. An update on our results including more on SU$(4)$ 
will be published elsewhere.

\begin{table}[t]

\vspace*{.2cm}

\centering
    \begin{tabular}{|c|c|c|c|}
\hline  $N_t$&$\beta_c$($R_k = R_e$)  & $\beta_c$ ($R_k = R_c$) &Lit.\\ 
\hline  2& 8.15309(11)& 8.15297(57) & 8.1489(31)$^\dagger$ \\ 
\hline  4& 14.7262(9)& 14.7194(45)  & 14.717(17)$^\dagger$ \\
\hline  6& 21.357(25)& - & 21.34(4) $^\ddagger$\\
\hline  8& 27.84(12)& - &  -\\ 
\hline 
\end{tabular}
\caption{SU$(3)$ critical couplings from selfduality (weighted means),
  and intersection with the universal value (extrapolated) from
  \cite{Strodthoff:2010dz},
  previous literature values from $^\dagger$\cite{Liddle:2008kk},
  $^\ddagger$\cite{Engels:1996dz}.} 
\label{tab1}

\vspace*{-.2cm} 
\end{table}

For SU$(4)$, the $Z_4$ center symmetry  alone does not uniquely specify
the effective spin model to describe the dynamics of Polyakov
loops. SU$(4)$ is a rank-three group and has three fundamental
representations, $4$, $\bar 4$ and $6$. So even the simplest effective
Polyakov-loop model will consist of two distinct real terms, with
nearest neighbor couplings between loops in $4$/$\bar 4$
representations and between loops in the $6$ representation, {\it
  c.f.} Ref.~\cite{Wozar:2006fi}. Depending on the relative weight
between the two, the corresponding spin model could be any of the
$Z_4$-symmetric Ashkin-Teller models with three energy levels per link and
continuously varying critical exponents between the $q=4$ Potts model
class with $\nu=2/3$ and $\beta = 1/12$, and that of the planar or
vector Potts model, or simply the clock model, which corresponds to
two non-interacting Ising models in this case with $\nu =1 $ and
$\beta = 1/8$.  In between these limits the model corresponds to two
interacting Ising models with spins $s_i$ and $\sigma_i$ and it thus
has two order parameters, the usual magnetic ones $\langle s_i \rangle
$ or $\langle \sigma_i \rangle $, which are the same in the isotropic
model, with critical exponent $\beta_m = (2-y)/(24-16y) $, and an
electric one $\langle s_i\sigma_i \rangle $ with exponent $\beta_e =
(12-8y)^{-1}$, with $y=0$ for the standard Potts model ($\beta_e =
\beta_m$), and $y=1 $  for the non-interacting case ($\beta_e =
2\beta_m$) of the clock model \cite{Baxter:2007}.  

The present conclusion from the studies in
Refs.~\cite{deForcrand:2003wa,Holland:2007ar,Liddle:2008kk} 
is that the deconfinement transition in 
(2+1)$d$ SU(4) gauge theory is weakly $1^\mathrm{st}$ order.
We do observe, however, at least approximately at the length scales
corresponding to our spatial lattice volumes, a universal scaling
which seems closest to the standard $q=4$ Potts case
\cite{Strodthoff:2010dz}. The critical couplings from the  universal
amplitude ratios in \Eqs{eq:crit4} are fully 
consistent with those determined independent of Potts scaling and
universality, {\it e.g.}, for $N_t = 4$ we find $\beta_{c}=26.294(2)$
from \Eqs{eq:crit4} as compared to $\beta_{c}=26.283(9) $ independent
of that. The correlation length exponent is also consistent with $2/3$
and the data scales nicely for all volumes up to $N_s=80$. This maybe
still too small to see the $1^\mathrm{st}$ order nature, and the
conclusions may well depend on $N_t$ as well. But even if the
transition is weakly $1^\mathrm{st}$ order in the infinite volume ($N_s
\to\infty $) and continuum limits ($N_t \to\infty $), it seems
legitimate to ask, why it is the standard Potts model
universality that is observed before those limits are reached,
and why not any of the other Ashkin-Teller models, and can we derive
the correct effective Polyakov-loop model to demonstrate that? 

\medskip

\subsection{Fractional electric charge and quark confinement}
\label{sec:fraccharge}

\medskip

Center symmetry is explicitly broken and the finite temperature
transition of QCD becomes a smooth crossover when dynamical quarks in
the fundamental representation are included. 
This crossover at zero chemical potential is very well studied on the 
lattice \cite{Borsanyi:2010bp,Bazavov:2010sb}. In presence of fields
such as the quarks in QCD which faithfully represent the center of the gauge
group, there are no twisted boundary conditions on the torus and
therefore no center vortex ensembles to study. The detailed
understanding of the deconfinement phase transition in the pure gauge
theory might appear to be a rather academic exercise then.  

This neglects the electric charge of quarks, however, which is commonly
expected to only require small perturbative corrections. On the other
hand, the inclusion of the quarks' fractional electric charge is well
known to lead to a global $Z_6$ symmetry of the full fermion and Higgs
sector of the Standard Model. This $Z_6$ symmetry combines the centers
of the color and electroweak gauge groups, see also
\cite{Bakker:2005ph}, and \cite{Baez:2009dj} for a review.
Since quarks carry fractional electric charges $Q=\frac{2}{3}e$ or
$-\frac{1}{3}e$, their color and electromagnetic phases in the pair of
combined transformations,  
\begin{equation}
(e^{i2\pi/3},e^{i2\pi Q/e}),\;(e^{-i2\pi/3},e^{-i2\pi Q/e})\in
\mathrm{SU}(3)\times\mathrm{U}(1)_\mathrm{em} \, , \notag
\end{equation}
cancel precisely. They act trivially on all other
particles in the Standard Model, which are blind to the center of
SU$(3)$ and carry integer electric charge. Electric charge $Q$ is related to
hypercharge $Y$ and the third component of weak isospin $t_3$ by
$Q/e=t_3+Y/2$, with $e^{i2\pi t_3} \equiv -1 \in$ SU$(2)$ and $Y$
quantized in units of $1/3$. The symmetry is therefore generated by  
\begin{equation}
(e^{i2\pi /3},-1,e^{i\pi Y})\in \mathrm{SU}(3) \times \mathrm{SU}(2)\times
\mathrm{U}(1)_Y \, , 
\end{equation}
which gives six elements including the identity.

A global symmetry brings with it the possibility of a phase transition
characterized by spontaneous symmetry breaking. In this case, one
expects the transition to be driven by topological defects (vortices)
that carry both color and electromagnetic flux, that is, center
vortices with an additional electromagnetic Dirac string
\cite{Creutz:2004xk,vonSmekal:2005mq}. Could the physical realization
of this symmetry have non-trivial implications for confinement and the
phase structure of QCD as part of the Standard Model then? We have
started to address this question recently in a toy model with
half-integer electrically charged quarks in a two-color QCD world with
electromagnetism \cite{Edwards:2010ew,Edwards:2012tr}. 

The starting point is \qctd plus electromagnetism with 2 flavors
of Wilson fermions in $3 +1$ dimensions. By including `up' and `down'
quarks with fractional charges $\pm e/2$ relative to the
U$(1)_\mathrm{em}$ gauge action, we obtain a model with a global $Z_{2}$
symmetry. The lattice action is  
\begin{equation}
S = -\sum_{\dAlember}\left( \frac{\beta_\mathrm{col}}{2} \, \tr\,
  U\!_{\dAlember}   
  +\beta _\mathrm{em} \cos \theta_{\dAlember} \right) +  S_{f,W}, 
\end{equation}
where $S_{f,W}$ is the usual Wilson fermion action with the
distinction that parallel transporters for quarks are  products of 
SU$(2)$ color matrices and U$(1)_\mathrm{em}$ phases, 
\begin{equation}
\label{eqn:transporters}
 U_\mu(x) e^{i\theta_{\mu}(x)/2},\;\; U_\mu
 (x)\in\mathrm{SU}(2)\,,\;\theta_{\mu}(x)\in (-2\pi,2\pi] \; . 
\end{equation}
The SU$(2)$ plaquettes $U\!_{\dAlember} $ and U$(1)_\mathrm{em}$
plaquette angles $\theta_{\dAlember}  $ are formed from $U_\mu$ and
$\theta_\mu$ in the usual way. In this model, 'fractional charge'
means that the parallel transporters for quarks contain \emph{half}
the U$(1)_\mathrm{em}$ angle relative to the $\theta_\mu$'s that appear
in the plaquette angle $\theta_{\dAlember} $. That is, an
$e^{i\theta_\mu/2}=-1$ electromagnetic link for quarks appears as 
an $e^{i\theta_\mu}=+1$ link in the U$(1)_\mathrm{em}$  gauge action. An
important point is that the 'volume' of the compact U$(1)_\mathrm{em}$ is
determined by the quarks, which carry the smallest quantum of electric
charge. The range of  $\theta_{\mu}(x)$ is chosen such that we
integrate over all possible electromagnetic transporters for the
quarks, amounting to a double counting in the U$(1)_\mathrm{em}$ gauge action
and for all integer charged particles. This is consistent with the
premise that our compact U$(1)_\mathrm{em}$ is the result of symmetry
breaking in a SU$(3) \rightarrow \mathrm{SU}(2) \times
\mathrm{U}(1)_\mathrm{em}$ unified theory, {\it e.g.}, see
Ref.~\cite{Preskill:1984gd}.   

It is clear from Eq. \eqref{eqn:transporters} that a color
center element $-\mathds 1 \in$ SU$(2)$ combined  with an electromagnetic phase
$e^{i\theta_\mu/2}=-1$ act as the identity on the quarks. The model
therefore retains a color and electromagnetic combined  $Z_2$ center
symmetry, despite the introduction of dynamical quarks. This is the
analog of the hidden $Z_6$ symmetry of the Standard Model.

\begin{figure}[t]

\vspace*{.1cm}

\includegraphics[width=\linewidth]{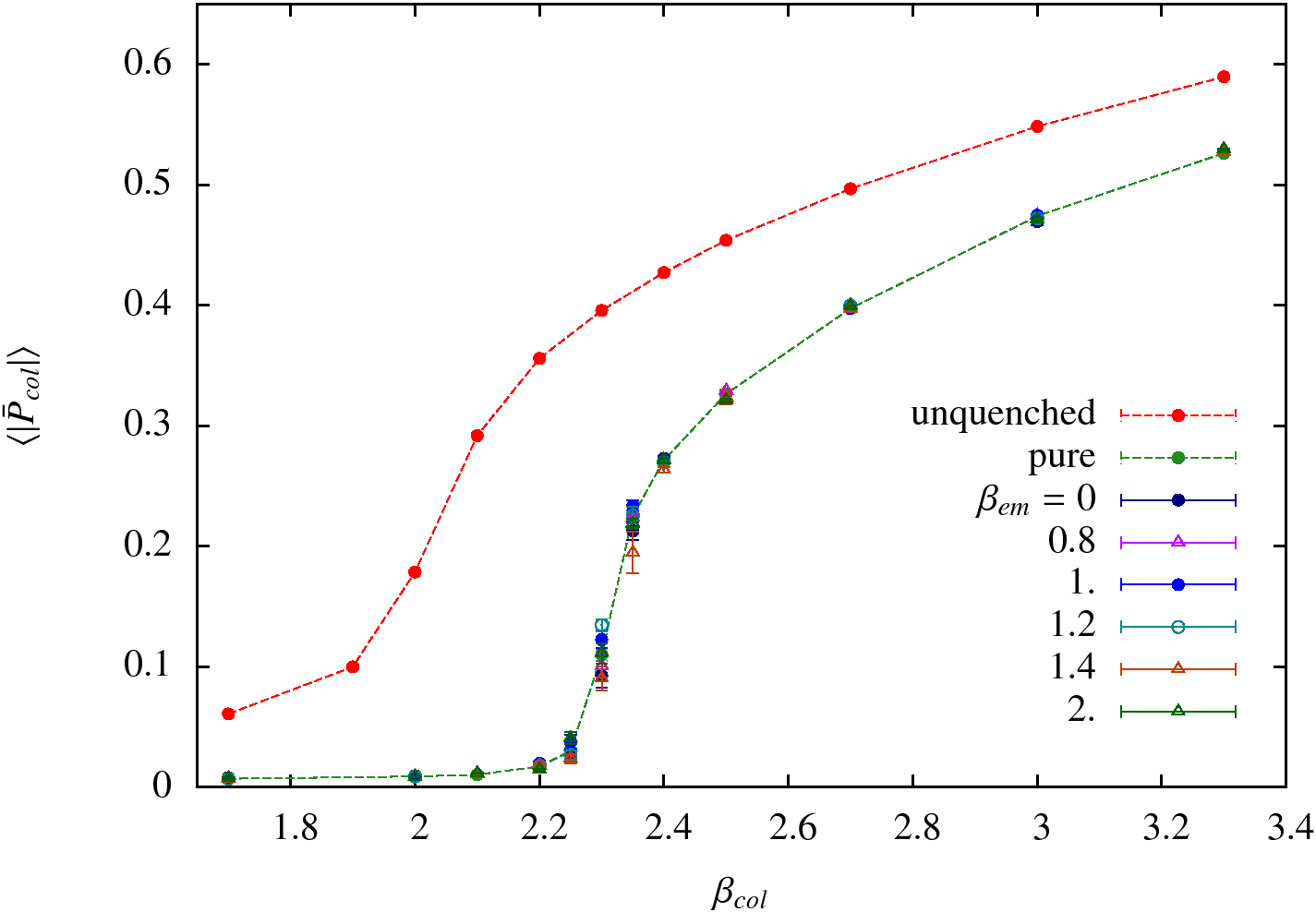} 

\vspace{-.1cm}
\caption{The SU$(2)$ Polyakov loop in the pure gauge theory (green), in
  full two-color QCD with $\kappa = 0.15 $ (red), and in the SU$(2) \times
  \mathrm{U}(1)_\mathrm{em}$ toy model with electromagnetic couplings 
   varying from total U$(1)_\mathrm{em}$ disorder at
   $\beta_\mathrm{em} =0$ to deep into the Coulomb phase at
   $\beta_\mathrm{em} = 2$ on a $16^3\times 4$ lattice with Wilson
   quarks.}    
\label{fig:plot164k015hotbe2}
\end{figure}

The results in \cite{Edwards:2010ew,Edwards:2012tr} indicate that the
usual ordering of color links by the dynamical quarks is negated by
the inclusion of fractional electric charge. This is understood by
analogy with the spin systems. In QCD alone, terms in a loop 
expansion of the fermion determinant that wind around the temporal
direction favor the center sector in which the traced Polyakov loop
$P_\mathrm{col}= 1 $. They break center symmetry and lead to an
ordering in much the same way as an external magnetic field $H$ does
for the Potts model spins in \Eq{PottsH}. The
inclusion of fractional electric charge in our model bestows quark
loops with an additional U$(1)_\mathrm{em}$ phase which may undo this
effect. Consider, for example, the hopping expansion of the Wilson
fermion action for $N_t=4$ time slices to leading order in the hopping
parameter, $\mathcal O(\kappa^{4})$, 
\begin{align}
\label{eqn:hopping}
S_{f,\text{eff}} &= -16\kappa^4 \, \Bigg( \sum_{\dAlember}
 \,  \cos{\frac{\theta_{\dAlember}}{2}} \cdot\tr \,U\!_{\dAlember} \\ 
 &  \hskip .6cm
 +8 \, \sum_{\vec{x}} 
  \cos\big(P_{\theta/2}(\vec x)\big) \cdot \mathrm{Re}\,
  P_\mathrm{col}(\vec x) \, \Bigg) 
  +\dots \; . \notag 
\end{align}
If the U$(1)_\mathrm{em}$ Polyakov loop angle for quarks
$P_{\theta/2}(\vec{x})=\sum_{t=0}^{N_t-1}\theta_t(t,\vec{x})/2$ is
disordered, then the $P_\mathrm{col}=1$ sector is no longer favored
and the SU$(2)$ center symmetry is \emph{dynamically} restored. 
The effect is analogous to placing a Potts model in a fluctuating
magnetic field, or to the Peccei-Quinn mechanism \cite{Peccei:1977hh},
in which the coupling of the CP violating term in QCD to an axion
field allows for the dynamical restoration of CP symmetry. 

Since the parallel transporters for quarks possess a $Z_{2}$ degree of
freedom that the U$(1)_\mathrm{em}$ gauge action is blind to, this is
possible even in the Coulomb phase for integer electric charges. 
Indeed, the SU$(2)$ Polyakov loop in the toy model with half-integer
charged dynamical Wilson quarks is indistinguishable from the quenched
result at values of the hopping parameter, or the quark mass, that
would otherwise cause a significant amount of ordering in standard
2-color QCD. As shown in Figure~\ref{fig:plot164k015hotbe2}, this is
equally true when the U$(1)_\mathrm{em}$ links are totally disordered
in the confined phase $\beta_\mathrm{em} \lesssim 1.01$, and in the
Coulomb phase at $\beta_\mathrm{em} \gtrsim 1.01$ of the
U$(1)_\mathrm{em}$. Since the U$(1)_\mathrm{em}$ gauge action is blind
to the $Z_{2}\subset \mathrm{U}(1)_\mathrm{em}$ disorder as seen by the quarks, 
{\it i.e.}, $e^{i\theta_\mu /2}=\pm  1$, it is unable to remove it.

The quenched result in Fig.~\ref{fig:plot164k015hotbe2} reflects the
second order nature of the phase transition of pure SU$(2)$ in the
universality class of the $3d$ Ising/$Z_2$ gauge system with typical
finite-size corrections. It clearly turns into a crossover in full
\qctd at $\kappa = 0.15$ as in the Ising model with a
symmetry-breaking external field. When the half-integer
electromagnetic charges are included, the SU$(2)$ Polyakov loop falls
straight back onto the quenched curve indicating that there might be
true disorder in the vacuum when QCD is embedded into the Standard Model.    

This is not the full story, however. A comparison of the $\pi$ and
$\rho$ meson masses in the toy model and in \qctd with equal
hopping parameter $\kappa $ and SU$(2)$ gauge coupling
$\beta_\mathrm{col} $ as in \cite{Skullerud:2003yc} reveals that the
mass scales have changed dramatically in the toy model
\cite{Edwards:2012tr}. One plausible interpretation is that there is
strong $Z_2$ disorder even for $\beta_\mathrm{em}\to \infty $ which
adds to the strength of the confining potential. This effect might be 
unphysical. One needs to go back and verify scaling of the masses (or
correlation lengths) in physical units as the continuum limit is approached
which might not even exist without unification in the toy model.  

Meanwhile, we have explored the phase diagram 
at the leading order $\mathcal{O}(\kappa^4)$ in the hopping expansion
for $N_t=4$, with the effective action  in \Eq{eqn:hopping},
which allows us to cheaply check our intuition about the resulting
couplings of SU$(2)$ and U(1)$_\mathrm{em}$ links, without having to worry
about the chiral limit for $\kappa$. We thus take the leading-order
hopping expansion at face value, as a model beyond its range of
applicability as an approximation to the full toy model. This
effective model then shares its qualitative features with a
fractionally charged fundamental Higgs model \cite{Greensite:2012}.
 
\begin{figure}[t]

\vspace*{.1cm}

\includegraphics[width=0.96\linewidth]{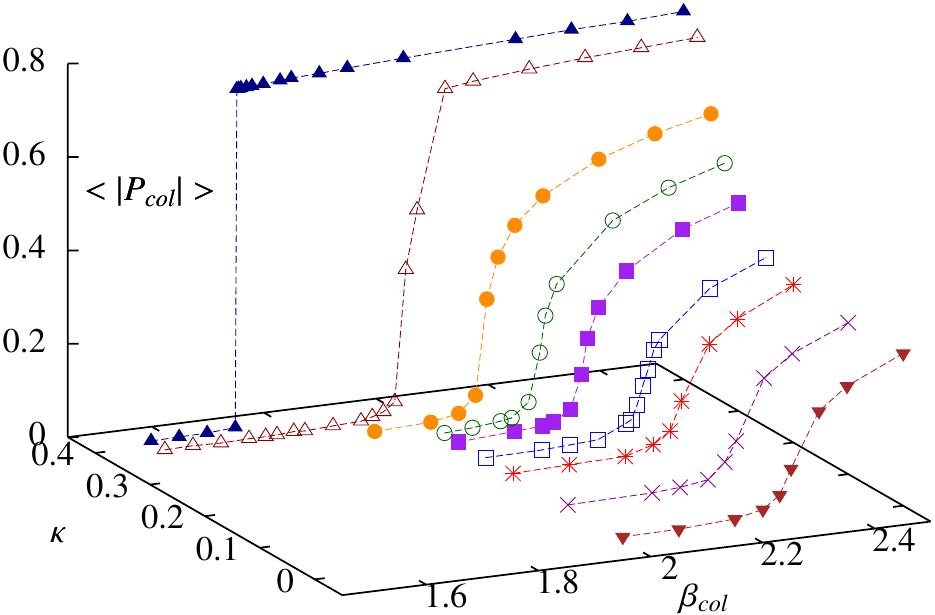} 

\vspace{-.1cm}
\caption{The SU$(2)$ Polyakov loop on the $16^3\times 4$ lattice with
  the leading order interactions from \Eq{eqn:hopping} in the 
  $\beta_\mathrm{em} \to\infty $ limit from Ref.~\cite{Edwards:2012tr}.}    
\label{fig:plot3d164-withquenched}
\end{figure}

Fig.~\ref{fig:plot3d164-withquenched} shows the results for the
SU$(2)$ Polyakov loop in the effective SU$(2)\times\mathrm{U}(1)$
model with the leading order interactions in \Eq{eqn:hopping}, where
the U$(1)_\mathrm{em}$ phases  $e^{i\theta_\mu(\vec{x}) /2}$ are
restricted to $\pm 1$, which amounts to $\beta_{em}\to\infty$ in the
U$(1)_\mathrm{em}$ gauge action. As $\kappa$ increases, the
disorder-order deconfinement transition of the SU$(2)$ 
Polyakov loop moves to smaller values of $\beta_\mathrm{col}$ and
sharpens  dramatically. Note that in the 
combined limit $\kappa,\; \beta_{em} \rightarrow \infty$, the
plaquette-plaquette coupling in \Eq{eqn:hopping} forces the
SU$(2)$ plaquettes to take the values $\pm 1$. For very large $\kappa$
the transition line should therefore terminate with the first order
bulk transition of the $4d$ $Z_2$ gauge theory at $\beta_\mathrm{col} \sim
0.44$ \cite{Wegner:1984qt}. As the SU$(2)$ plaquette is driven to
unity for large $\beta_\mathrm{col}$, the U$(1)_\mathrm{em}$
plaquettes corresponding to quark loops, $\cos(\theta_{\dAlember}/{2})$, 
receive an effective coupling that suppresses $Z_2$ disorder. The
relevant U$(1)_\mathrm{em}$ Polyakov loop $\cos{P_{\theta/2}}$ 
remains disordered at small values of $\kappa$, but transitions to
unity for large $\beta_\mathrm{col}$ and $\kappa$ (not shown). Due to
the $Z_2$ disorder transition, the model might have three distinct
phases, the confined phase for small $\kappa $ and  small
$\beta_\mathrm{col} $, the deconfined one for small $\kappa $ but
$\beta_{c} < \beta_\mathrm{col} $ and a Higgs phase in the corner
where both couplings  $\kappa $ and $\beta_\mathrm{col} $ are large.  
We currently investigate the same effect of $Z_2$ disorder from 
half-integer electric charge in the phase diagram of fundamental
SU$(2)$ Higgs models \cite{Greensite:2012}.

Given the fractional electric charge of quarks and the existence of a
global symmetry that relates the centers of the color and electroweak
gauge groups, it may be misleading to study non-perturbative phenomena
such as confinement and perhaps also dynamical mass generation in QCD
alone. In the toy two-color model, the coupling of Wilson quarks 
with half-integer electric charge relative to a compact U$(1)_\mathrm{em}$
has a dramatic effect on the color sector. This is not due to
electromagnetic fluctuations which should have small effects deep in
the Coulomb phase for large $\beta_\mathrm{em}$, but it can be due to a
frozen $Z_2$ disorder that the low energy theory inherits from the
grand unified theory. Such frozen disorder could have an effect akin
to a random external field in Potts models  \cite{Eichhorn:1996} which
are known to have phase transitions where they would not otherwise
have them, with constant external fields.  

Here the SU$(2)$ Polyakov loop shows
an order-disorder transition that is consistent with the spontaneous
center symmetry breaking transition of the pure gauge theory at values
of the hopping parameter where center symmetry is clearly
\emph{explicitly} broken in standard two-color QCD. This happens despite
the fact that static quarks are now described by the combined SU$(2)$
and half-integer U$(1)_\mathrm{em}$ loops which continue to show a
crossover behavior. The scales of both transitions are still widely
separated and this problem needs to be solved in such models. 

Combined twisted boundary conditions are nevertheless possible with
more than one gauge group and the corresponding center vortex
ensembles might become relevant for a deconfinement phase transition
in the standard model if its hidden center-like symmetry is 
relevant to the real world. Whether this is the case or not seems hard
to decide without grand unification. Meanwhile one might want to find
out whether SU$(3)\times Z_3$ models of the kind of our two-color toy
model in the $\beta_\mathrm{em} \to \infty$ limit are at all suited as
effective low energy theories for hadron physics. If we did have
absolute confinement in the real world, separated by a true phase
transition from the quark-gluon plasma, this would obviously make a
conceptually and phenomenologically important difference as compared
to our present understanding that quarks are only exponentially
suppressed in a hadronic phase that is analytically connected to the
high-temperature plasma. 

\section{Concluding remarks}

Many interesting facets of strongly interacting matter can be
studied in QCD-like theories. Changing the number of quark flavors,
their masses or the number of colors often leads to simplifications or
idealizations which allow to understand qualitative features and
mechanisms from such powerful concepts as universality, scaling
and finite-size scaling. They allow to relate phenomena
from diverse areas of physics by often surprising analogies.
The liquid-gas transition of nuclear matter, Heisenberg ferromagnets
for chiral symmetry breaking, ultracold fermionic quantum gases with
BEC-BCS crossover for two-color QCD at finite baryon density, the
very well studied spin models and their dualities for the
deconfinement transition and the electric fluxes in the pure gauge
theories  are some of the famous examples touched upon in these
lecture notes.      

The description of critical phenomena in statistical physics and
quantum field theory is today synonymous with the renormalization
group. Valuable introductory accounts of various aspects of the
renormalization group have been given at this $49^\mathrm{th}$ 
Schladming Winter School. I have presented some selected examples of
applications of basic renormalization group concepts towards a better
understanding of chiral dynamics at finite density, quark confinement, the
different phases of strongly interacting matter at extremes of
temperature or density, and the nature of the transitions between
them.             

Arguably, two of the most promising non-per\-tur\-ba\-tive tools to study
the QCD phase diagram at the moment are the functional  
renormalization group and lattice simulations, especially in
combination. These two complement each other very
well. Non-perturbative functional methods in general rely on
additional assumptions. These can be guided by intuition, derive from
fundamental requirements such as causality, locality or gauge
invariance, or they can be backed up by results from other approaches
as external input. This is what lattice simulations can provide where
they are possible without fermion sign problem. Therefore, it will be
extremely valuable to investigate and ultimately completely understand
the phase diagrams of QCD-like theories such as two-color QCD or the
$G_2$ gauge theory which don't have this problem.    

A lot of work in this direction has been done in two-color QCD as
reviewed in Section~\ref{QMDmodel}. Here we are naturally interested in
the functional renormalization group treatment which we have seen
to be well capable of describing the effects of finite baryon density
and the competing dynamics of collective mesonic and baryonic
fluctuations. The bosonic nature of the baryons in two-color QCD
with a BEC-BCS crossover provides interesting analogies, but  
I have argued that the impact of the finite baryon density on the
chiral transition reveals a genuine effect. There are reasons to believe
that there might not be enough chiral symmetry breaking left for a
chiral phase transition and a critical endpoint at finite baryon
density in the QCD phase diagram either. Suitable theories 
with fermionic baryons but without fermion sign problem seem to be the
logical next step to study at finite density. The $G_2$ gauge theory
is such a theory and the studies of its phase diagram have only just
begun. The challenge there will be to disentangle finite density
effects due to bosonic from those due to the fermionic baryons, an
ideal task for the next combined effort with the functional
renormalization group and lattice Monte-Carlo simulations.

The confinement problem is obviously a hard one. But a better
understanding on a fundamental level even in the simple $2+1$ 
dimensional SU$(N)$ gauge theories from exact results seems worthwhile
to work our way up from there. I found some of these results such as
the selfduality reflected in the center vortex and electric flux
ensembles quite enlightening and perhaps also unexpected. The models
may not be realistic but the results are quite powerful, and who
knows, perhaps we will eventually find that it wasn't wasted but that
combined vortices and disorder in the vacuum may be relevant when QCD
is embedded in the Standard Model.

\bigskip
\bigskip

\leftline{\bfseries Acknowledgements}

\medskip

I much enjoyed this $49^\mathrm{th}$ Schladming Winter School and
I would like to thank the organizers Natalia
and Reinhard Alkofer, Christian Fischer, Heimo Latal, Leopold
Mathelitsch, and of course the Director of the School, Bernd-Jochen
Schaefer. 

There are many colleagues and friends to whom I'm indebted for
collaborations and discussions. In relation to the material in these
lecture notes  I would like to mention especially Philippe de Forcrand,
Jeff Greensite, Kurt Langfeld, Axel Maas, Jan Pawlowski, Jochen Wambach, 
Bj\"orn Wellegehausen, and Andreas Wipf.

Finally, but most importantly I thank my students. Most of
the results from the last few years that I presented here, were
obtained together with Sam Edwards and Nils Strodthoff, in particular.
It has been my privilege and great pleasure to work with them. 

\smallskip

Financial support from the Helmholtz International Center for FAIR
within the LOEWE program of the State of Hesse, the Helmholtz
Association Grant No.~VH-NG-332, and the European Commission,
FP7-PEOPLE-2009-RG, Grant No.~249203 is greatly acknowledged.




\bigskip
\bigskip


\leftline{\bfseries References}

\medskip

\bibliographystyle{elsarticle-num}
\bibliography{Schladming_2011_LvS}







\end{document}